\documentclass[apj, onecolappendix, numberedappendix, appendixfloats]{emulateapj} 
\usepackage{amsmath}
\usepackage{soul}
\usepackage{natbib}
\newcommand*\meanbar[1]{\overline{#1}}

\newcommand{\noprint}[1]{}
\newcommand{\figsetstart}{{\bf Fig. Set} }
\newcommand{\figsetend}{}
\newcommand{\figsetgrpstart}{}
\newcommand{\figsetgrpend}{}
\newcommand{\figsetnum}[1]{{\bf #1.}}
\newcommand{\figsettitle}[1]{ {\bf #1} }
\newcommand{\figsetgrpnum}[1]{\noprint{#1}}
\newcommand{\figsetgrptitle}[1]{\noprint{#1}}
\newcommand{\figsetplot}[1]{\noprint{#1}}
\newcommand{\figsetgrpnote}[1]{\noprint{#1}}

\usepackage{mathtools}
\shorttitle{An ALMA Survey of H$_2$CO in Protoplanetary Disks}
\shortauthors{Pegues et al.}

\begin{document}

\title{An ALMA Survey of H$_2$CO in Protoplanetary Disks}


\author{Jamila Pegues$^1$, Karin I. \"Oberg$^1$, Jennifer B. Bergner$^{2,1}$, Ryan A. Loomis$^{3,1}$, Chunhua Qi$^1$, Romane Le Gal$^1$,
L. Ilsedore Cleeves$^4$, Viviana V. Guzm\'an$^{5,6}$, Jane Huang$^1$, Jes K. J{\o}rgensen$^7$,
Sean M. Andrews$^1$, Geoffrey A. Blake$^8$, John M. Carpenter$^5$, Kamber R. Schwarz$^{9}$, Jonathan P. Williams$^{10}$, David J. Wilner$^1$}
\affiliation{$^1$Center for Astrophysics $\mid$ Harvard \& Smithsonian, Cambridge, MA 02138, USA}
\affiliation{$^2$Department of Geophysical Sciences, University of Chicago, Chicago, IL 60637, USA}
\affiliation{$^3$National Radio Astronomy Observatory, Charlottesville, VA 22903, USA}
\affiliation{$^4$Department of Astronomy, University of Virginia, Charlottesville, VA 22904, USA}
\affiliation{$^5$Joint ALMA Observatory, Alonso de C{\'o}rdova 3107 Vitacura, Santiago, Chile}
\affiliation{$^6$Instituto de Astrof{\'i}sica, Pontificia Universidad Cat{\'o}lica de Chile, Av.~Vicu{\~n}a Mackenna 4860, 7820436 Macul, Santiago, Chile}
\affiliation{$^7$Niels Bohr Institute \& Centre for Star and Planet Formation, University of Copenhagen, {\O}ster Voldgade 5-7, DK-1350 Copenhagen K., Denmark}
\affiliation{$^8$Division of Geological \& Planetary Sciences, California Institute of Technology, Pasadena, CA
91125, USA}
\affiliation{$^{9}$Sagan Fellow at the Lunar and Planetary Laboratory, University of Arizona, Tucson, AZ 85721, USA}
\affiliation{$^{10}$Institute for Astronomy, University of Hawai'i at M\={a}noa, Honolulu, HI 96822, USA}


\begin{abstract}

H$_2$CO is one of the most abundant organic molecules in protoplanetary disks and can serve as a precursor to more complex organic chemistry. We present an ALMA survey of H$_2$CO towards 15 disks covering a range of stellar spectral types, stellar ages, and dust continuum morphologies.  H$_2$CO is detected towards 13 disks and tentatively detected towards a 14th.  We find both centrally-peaked and centrally-depressed emission morphologies, and half of the disks show ring-like structures at or beyond expected CO snowline locations.  Together these morphologies suggest that H$_2$CO in disks is commonly produced through both gas-phase and CO-ice-regulated grain-surface chemistry.  We extract disk-averaged and azimuthally-averaged H$_2$CO excitation temperatures and column densities for four disks with multiple H$_2$CO line detections.  The temperatures are between 20-50K, with the exception of colder temperatures in the DM Tau disk.  These temperatures suggest that H$_2$CO emission in disks is generally emerging from the warm molecular layer, with some contributions from the colder midplane.  Applying the same H$_2$CO excitation temperatures to all disks in the survey, we find that H$_2$CO column densities span almost three orders of magnitude ($\sim 5 \times 10^{11} - 5 \times 10^{14} \mathrm{cm}^{-2}$).  The column densities appear uncorrelated with disk size and stellar age, but Herbig Ae disks may have less H$_2$CO compared to T Tauri disks, possibly because of less CO freeze-out.  More H$_2$CO observations towards Herbig Ae disks are needed to confirm this tentative trend, and to better constrain under which disk conditions H$_2$CO and other oxygen-bearing organics efficiently form during planet formation.

\end{abstract}

   \keywords{astrochemistry, protoplanetary disks, ISM: molecules, radio lines: ISM}

\section{Introduction}
\label{sec_introduction}

Protoplanetary disks are the precursors of planetary systems.  These disks consist of gas and dust orbiting young stars.  Over time, the dust grains collide and stick together to form pebbles, eventually growing into planetesimals and planets~\citep[e.g.][and references therein]{cite_mordasinietal2008}.  The organic compositions of pebbles, planetesimals, and planets are therefore shaped by the organic chemistry of the gas and dust in their ancestral disks.  By studying this chemistry, we can ultimately model and predict the chemical formation environment of young planets~\citep[e.g.][]{cite_cridlandetal2016, cite_obergetal2016, cite_cridlandetal2017} and their potential ability to develop Earth-like life.

The small organic molecule formaldehyde (H$_2$CO) is expected to be important for the overall organic chemical budget of disks.  H$_2$CO is commonly and abundantly detected in disks~\citep[e.g.,][]{cite_dutreyetal1997, cite_aikawaetal2003, cite_thietal2004, cite_obergetal2010, cite_obergetal2011a}.  Based on cometary observations, H$_2$CO was also abundant in the solar nebula; in comets it is present at levels of $\sim$0.1-1\% with respect to water~\citep{cite_mummaetal2011}. There are several possible chemical origins of H$_2$CO in disks. H$_2$CO can form through neutral-neutral gas-phase chemistry~\citep[e.g.,][]{cite_fockenbergetal2002, cite_atkinsonetal2006}, and through grain-surface chemistry via CO ice hydrogenation~\citep[e.g.,][]{cite_hiraokaetal1994, cite_hiraokaetal2002, cite_watanabeetal2002, cite_hidakaetal2004, cite_watanabeetal2004, cite_fuchsetal2009}. Both processes can occur in disks, though likely in different locations, since neutral-neutral gas-phase chemical reactions increase with density and temperature towards the inner disk, while CO ice hydrogenation is only possible beyond the CO snowline.  In addition, H$_2$CO has been detected in ices around young stellar objects~\citep[e.g., detections/tentative detections by][]{cite_gibbetal2004, cite_pontoppidanetal2004, cite_boogertetal2008}, and some of the H$_2$CO observed in disks might be inherited from this preceding evolutionary stage~\citep[e.g.,][]{cite_drozdovskayaetal2014, cite_drozdovskayaetal2019}.  


Understanding H$_2$CO emission and its origins in disks is important for modeling more complex organic disk chemistry. When formed through CO-ice hydrogenation, H$_2$CO serves as a stepping-stone towards forming oxygen-bearing, complex organic molecules~\citep[known as COMs, and typically defined as unsaturated carbon-bearing molecules with five or more atoms; e.g.,][]{cite_dishoecketal2009}, especially methanol~\citep[CH$_3$OH; e.g.,][]{cite_hiraokaetal1994, cite_watanabeetal2002}.  COMs are expected to be rich in disks based on astrochemical disk models~\citep{cite_walshetal2014}.  Unfortunately, without the help of rare heating phenomena such as stellar outbursts~\citep[e.g.,][]{cite_hoffetal2018, cite_leeetal2019}, COM emission lines are predicted to be difficult to detect in disks~\citep{cite_henningetal2013, cite_walshetal2014}, due to their large partition functions and comparatively small abundances.  To date, the largest COMs detected towards disks are cyanoacetylene~\citep[HC$_3$N,][]{cite_chapillonetal2012}, methyl-cyanide~\citep[CH$_3$CN,][]{cite_obergetal2015}, methanol~\citep[CH$_3$OH,][]{cite_walshetal2016}, and formic acid~\citep[HCOOH,][]{cite_favreetal2018}, and only cyanoacetylene and methyl-cyanide have been detected in more than one disk~\citep{cite_bergneretal2018, cite_loomisetal2018}.  H$_2$CO, on the other hand, is readily detectable in disks.  The presence/absence of H$_2$CO emission, originating specifically from grain-surface chemistry, can thus indicate whether the disk hosts a rich organic grain-surface chemistry or not~\citep{cite_walshetal2014, cite_obergetal2017}.

The chemical origins of H$_2$CO in disks, and its potential to form more complex organic chemistry, can be addressed through spatially-resolved observations of H$_2$CO emission at millimeter and submillimeter wavelengths. To date, H$_2$CO has been spatially resolved with the Atacama Large Millimeter/submillimeter Array (ALMA) towards six disks: (1) the T Tauri disk Oph IRS 48~\citep{cite_mareletal2014}; (2) the T Tauri disk DM Tau~\citep{cite_loomisetal2015}; (3) the Herbig Ae disk HD 163296~\citep{cite_carneyetal2017, cite_guzmanetal2018}; (4) the T Tauri disk TW Hya~\citep{cite_obergetal2017}; (5) the T Tauri disk V4046 Sgr~\citep{cite_kastneretal2018}; and (6) the T Tauri disk DG Tau~\citep{cite_podioetal2019}.  These six studies show a variety of H$_2$CO morphologies, from centrally peaked (indicative of gas-phase formation) to ring-like in the outer disk (indicative of formation through CO ice hydrogenation). 
Two of these studies~\citep{cite_carneyetal2017, cite_guzmanetal2018} calculated H$_2$CO excitation temperatures of $>$20K and 24K, which are consistent with both formation pathways.  Three of these studies~\citep{cite_loomisetal2015, cite_carneyetal2017, cite_obergetal2017} determined that the detected H$_2$CO emission could be described by two distinct components of H$_2$CO: (1) a warm inner component produced through gas-phase chemistry, and (2) a cold outer component exterior to the CO snowline produced through grain-surface chemistry.  A larger survey of H$_2$CO emission towards disks is necessary to explore the typical distribution and chemical origins of H$_2$CO in protoplanetary disks, as well as their connections to stellar and disk characteristics.

In this work, we present an ALMA survey of H$_2$CO towards 15 protoplanetary disks.  We characterize the H$_2$CO morphologies, excitation temperatures, and column densities, evaluate their dependence on stellar and disk characteristics, and assess the chemical origins of H$_2$CO.  In Section~\ref{sec_data}, we describe the disk sample, the ALMA observations, and the data reduction process.  In Section~\ref{sec_methodology}, we present our methodology for visualizing and analyzing the imaged line emission.  In Section~\ref{sec_results}, we report the detections of H$_2$CO, the H$_2$CO emission morphologies, excitation temperatures, and column densities, and any trends relating the H$_2$CO to stellar and disk characteristics.  In Section~\ref{sec_discussion}, we discuss the results in the context of H$_2$CO formation chemistry and disk environment, and in Section~\ref{sec_summary} we summarize the key findings from our survey.


\section{Observational Details}
\label{sec_data}

\subsection{Disk Sample}
\label{sec_survey}

\begin{figure}
\resizebox{\hsize}{!}{
    \includegraphics{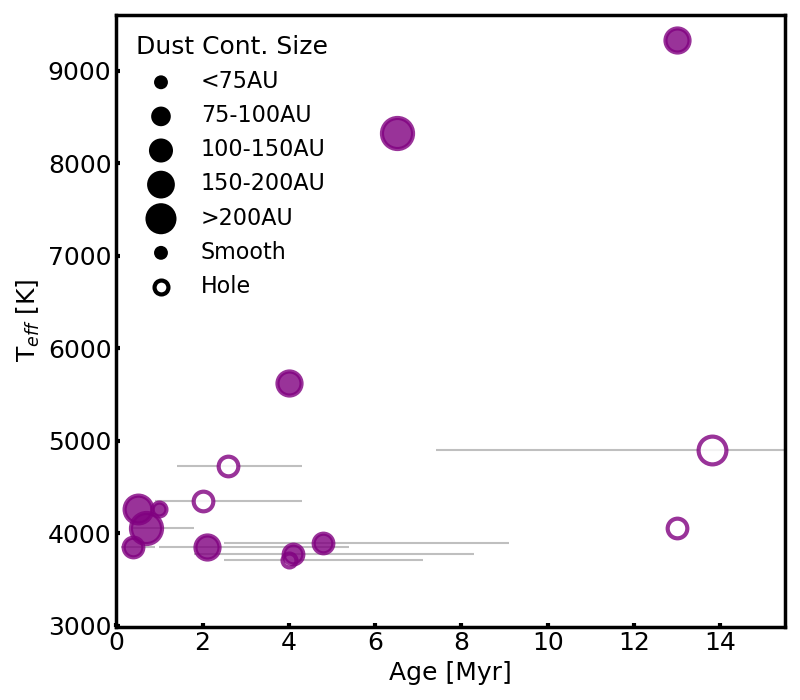}}
\caption{\textbf{Stellar Effective Temperature versus Stellar Age} for our disk sample (Table~\ref{table_majorstellarchar}).  Each point represents a disk, and the sizes denote the sizes of the submillimeter/millimeter dust continuum.  An empty vs. filled point marks whether or not the disk shows a large inner cavity in the submillimeter/millimeter dust continuum.  For disks where an age range rather than a single age is given in the literature, horizontal bars span that range.}
\label{fig_intro}
\end{figure}

Figure~\ref{fig_intro} and Table~\ref{table_majorstellarchar} summarize the stellar and disk characteristics of the 15 disks in our sample.  The sample was assembled from three main ALMA observing projects, with H$_2$CO lines from other ALMA observing projects added in as available.  The sample consists of well-studied T Tauri and Herbig Ae disks, and gas, dust, and temperature structures are available for most sources from previous studies. The host stars in the sample range in luminosity, mass, and age from $\sim$0.2-26$\sim$L$_\Sun$, $\sim$0.5-2.0$\sim$M$_\Sun$, and $\lesssim$1-15$\sim$Myrs, respectively.  All disks in the sample are relatively close by, $\sim$70-170$\sim$pc away.  Six disks in the sample are associated with the Taurus-Auriga region, five with the Upper Scorpius region, and one each with the Lupus and Ophiuchus regions.

The disks vary in dust morphology and size, but most are large.  Outer millimeter dust radii range from $<$75 AU to $\sim$200 AU (where we define the outer radius as the radius where the azimuthally-averaged continuum emission drops below 5\% of the peak).  Five of the disks are ``transition disks", i.e. have large inner holes in their submillimeter/millimeter dust continuum: DM Tau ~\citep{cite_marshetal1992}, GM Aur~\citep{cite_jensenetal1997}, J1604-2130~\citep{cite_mathewsetal2012}, LkCa 15~\citep{cite_berginetal2004}, and V4046 Sgr~\citep{cite_jensenetal1997}.

\begin{deluxetable*}{lllllllll}
\tablecaption{\textbf{Stellar and Disk Characteristics of the Sample.} \label{table_majorstellarchar}}
\tablehead{
Disk	& Stellar	& R.A.$^{[0]}$	& Dec.$^{[0]}$	& Dist  & Age$_*$	& L$_*$	& M$_*$	& T$_{eff}$ \\
	& Type	& [J2000] 	& [J2000]	& [pc]  & [Myr]	& [L$_\Sun$]	& [M$_\Sun$]	& [K]
}
\startdata
AS 209$^{[1]}$             & K5           & 16:49:15.3             & -14:22:09.0            & 121            & 1.0            & 1.41                  & 0.83                 & 4266                  \\
CI Tau$^{[2]}$             & K7           & 04:33:52.0             & +22:50:29.8            & 159            & 0.7 (0.4-1.8)  & 1.20                  & 0.66                 & 4060                  \\
DM Tau$^{[2]}$             & M1           & 04:33:48.7             & +18:10:9.7             & 145            & 4.0 (2.5-7.1)  & 0.24                  & 0.53                 & 3705                  \\
DO Tau$^{[2]}$             & M0           & 04:38:28.6             & +26:10:49.1            & 140            & 0.4 (0.1-0.9)  & 1.40                  & 0.45                 & 3850                  \\
GM Aur$^{[2]}$             & K5.5$^{[3]}$      & 04:55:11.0             & +30:21:59.0            & 159            & 2.6 (1.4-4.3)  & 1.60                  & 1.30                 & 4730                  \\
HD 143006$^{[1]}$          & G7           & 15:58:36.9             & -22:57:15.5            & 165            & 4.0            & 3.80                  & 1.78                 & 5623                  \\
HD 163296$^{[1]}$          & A1           & 17:56:21.3             & -21:57:22.5            & 101            & 13             & 17                    & 2.04                 & 9333                  \\
IM Lup$^{[1]}$             & K5           & 15:56:09.2             & -37:56:06.5            & 158            & 0.5            & 2.57                  & 0.89                 & 4266                  \\
J16042165-2130284$^{*[2]}$ & K2           & 16:04:21.6             & -21:30:28.9            & 149            & 13.8 (7.4-32)  & 0.62                  & 1.11                 & 4900                  \\
J16090075-1908526$^{*[2]}$ & K9           & 16:09:00.7             & -19:08:53.1            & 137            & 4.8 (2.5-9.1)  & 0.32                  & 0.68                 & 3890                  \\
J16123916-1859284$^{*[2]}$ & M0.5         & 16:12:39.2             & -18:59:28.9            & 138            & 4.1 (1.8-8.3)  & 0.29                  & 0.56                 & 3775                  \\
J16142029-1906481$^{*[2]}$ & M0           & 16:14:20.3             & -19:06:48.5            & 143            & 2.1 (1.0-5.4)  & 0.46                  & 0.60                 & 3850                  \\
LkCa 15$^{[2]}$            & K5           & 04:39:17.8             & +22:21:03.1            & 158            & 2.0 (0.9-4.3)  & 1.04                  & 1.03                 & 4350                  \\
MWC 480$^{[2]}$            & A4.5$\pm$2     & 04:58:46.3             & +29:50:36.6            & 161            & 6.5            & 25                    & 1.84                 & 8330                  \\
V4046 Sgr$^{**[4]}$   & K5,K7$^{[5]}$  & 18:14:10.5             & -32:47:35.3            & 73             & 13 & 0.86 & 1.75     & 4350, 4060
\enddata
\tablecomments{{R.A. and Dec. (Columns 3 and 4) are from GAIA~\citep[e.g.,][]{cite_gaiaetal2016a, cite_gaiaetal2016b, cite_gaiaetal2018}.  All other table values are taken from the literature.  For AS 209, HD 143006, HD 163296, and IM Lup, we use values as updated by [2] using new GAIA distances.  For disks denoted by [1], we use values calculated/scaled using the same methodology of [2] unless otherwise individually referenced.  For V4046 Sgr, we take all values except for the stellar type from [4], which assumed a distance of 73pc.  Stellar ages (Column 6) are given as ranges in parentheses (which account for error) when available.  $*$: The disks J16042165-2130284, J16090075-1908526, J16123916-1859284, and J16142029-1906481 are referred to respectively as J1604-2130, J1609-1908, J1612-1859, and J1614-1906 in subsequent figures, tables, and text.  $**$: V4046 Sgr is a protoplanetary disk orbiting a binary star system.  The individual stellar effective temperatures and total stellar luminosity and mass are reported.}
\textit{[0]~\cite{cite_gaiaetal2016a, cite_gaiaetal2016b, cite_gaiaetal2018}, [1]~\cite{cite_andrewsetal2018}, [2] Calculated using the same methodology as~\cite{cite_andrewsetal2018}, [3]~\cite{cite_espaillatetal2010}, [4]~\cite{cite_rosenfeldetal2012}, [5]~\cite{cite_quastetal2000}.}}
\end{deluxetable*}

%
\begin{deluxetable*}{lllllll}
\tablecaption{\textbf{Observed H$_2$CO Lines in the Sample.} \label{table_linechar}}
\tablehead{
H$_2$CO & Freq.	& A$_{ul}$	& E$_{u}$	&   g$_u$ & ALMA  & Observed	\\
Line    & [GHz]	& [$10^{-4}$s$^{-1}$]	& [K]   &	& Project Code  & Disks	}
\startdata
p-3$_{03}$-2$_{02}$ & 218.222 & 2.818 & 20.956 & 7 & 2016.1.00627.S  & CI Tau, DM Tau, DO Tau \\
									& & & & &	2015.1.00964.S  & HD 143006, J1604-2130, J1609-1908,
									J1612-1859, J1614-1906 \\
p-3$_{22}$-2$_{21}$ & 218.476 & 1.571 & 68.094 & 7 & 2015.1.00964.S  & HD 143006, J1604-2130, J1609-1908, 
 J1612-1859, J1614-1906 \\
                    & &       &       &       & 2013.1.00226.S  & AS 209, HD 163296, IM Lup, 
                    LkCa 15, MWC 480, V4046 Sgr \\
p-3$_{21}$-2$_{20}$ & 218.760 & 1.578 & 68.111 & 7 & 2015.1.00964.S  & HD 143006, J1604-2130, J1609-1908, 
 J1612-1859, J1614-1906 \\
                    & &       &       &       & 2013.1.00226.S  & AS 209, HD 163296, IM Lup, 
                    LkCa 15, MWC 480, V4046 Sgr \\
o-4$_{14}$-3$_{13}$ & 281.527 & 5.884 & 45.570 & 27 & 2015.1.00657.S  & LkCa 15, MWC 480 \\
p-4$_{04}$-3$_{03}$ & 290.623 & 6.903 & 34.904 & 9 & 2015.1.00678.S  & AS 209, DM Tau, GM Aur, IM Lup, V4046 Sgr \\
									& & & & & 2012.1.00681.S  & HD 163296 \\
									& & & & &	2015.1.00657.S  & LkCa 15, MWC 480 \\
p-4$_{23}$-3$_{22}$ & 291.238 & 5.211 & 82.071 & 9 & 2015.1.00657.S  & LkCa 15, MWC 480 \\
p-4$_{22}$-3$_{21}$ & 291.948 & 5.249 & 82.122 & 9 & 2015.1.00657.S  & LkCa 15, MWC 480 \\
o-4$_{13}$-3$_{12}$ & 300.837 & 7.178 & 47.887 & 27 & 2015.1.00657.S  & LkCa 15, MWC 480 \\
o-5$_{15}$-4$_{14}$ & 351.769 & 12.02 & 62.452 & 33 & 2011.0.00629.S  & DM Tau, MWC 480 
\enddata
\tablecomments{{`o-' and `p-' denote ortho and para lines, respectively.  All frequencies, Einstein coefficients (A$_{ul}$), upper energy levels (E$_u$), and degeneracies (g$_u$) were obtained from the Cologne Database for Molecular Spectroscopy~\citep[CDMS;][]{cite_cdms2016}.}}
\end{deluxetable*}

\subsection{ALMA Observations}
\label{sec_observations}

The core of our study contains observations of the H$_2$CO 3$_{03}$-2$_{02}$ and 4$_{04}$-3$_{03}$ lines towards eight disks each (both lines are observed towards one disk, DM Tau).  These observations are complemented by the H$_2$CO 5$_{15}$-4$_{14}$ line towards two disks; the H$_2$CO 3$_{22}$-2$_{21}$ and 3$_{21}$-2$_{20}$ lines towards 11 disks each; and the H$_2$CO 4$_{14}$-3$_{13}$, 4$_{13}$-3$_{12}$, 4$_{23}$-3$_{22}$, and 4$_{22}$-3$_{21}$ lines towards two disks each.  When detected, these additional lines are used to determine H$_2$CO excitation temperatures and column densities.

In total, nine H$_2$CO lines were considered, providing a total of 48 H$_2$CO observations across the 15 disks.  All H$_2$CO lines and their molecular characteristics, ALMA observing projects, and corresponding disks are listed in Table~\ref{table_linechar}.  The ALMA observing projects are described and referenced briefly in Table~\ref{table_obsmain}.

We also consider complementary dust continuum and C$^{18}$O observations towards all disks except GM Aur.  All CO observations were taken from ALMA observing projects 2013.1.00226.S, 2015.1.00964.S, and 2016.1.00627.S.

\begin{deluxetable*}{llllll}
\tabletypesize{\scriptsize}
\tablecaption{\textbf{ALMA Observing Project Details and References.} \label{table_obsmain}}
\startdata
\tablehead{
ALMA Project  & On-Source & Num. & Baseline & Largest & References \\
Code, Band & Int. Times & of Ant. & Ranges & Ang. &  \\
 & [min] & & [m] & Scale ["] &
}
2016.1.00627.S, 6    & 22-26                    & 47                 & 15-704        & 4-5 & \cite{cite_bergneretal2019} \\
2015.1.00964.S, 6    & 14-21                     & 48                 & 15-704   & 5-10 & \cite{cite_bergneretal2019}    \\
2015.1.00678.S, 7     & 20                       & 41                 & 20-640      & 2.6-4.2$^*$ & \cite{cite_kastneretal2018, cite_qietal2019} \\
2015.1.00657.S, 7     & 23-61                    & 39-44              & 15-650      & 5 & Loomis et al. (subm.) \\
2013.1.00226.S, 6     & 12-21                    & 31-37              & 20-784      & 6 & \cite{cite_huangetal2017} \\ 
2012.1.00681.S, 7     & 45                       & 32                 & 18-650      & 2.9$^*$ & \cite{cite_qietal2015} \\
2011.0.00629.S, 7     & 35                       & 23                 & 17.5-380    & 6 & \cite{cite_loomisetal2015}
\enddata
\tablecomments{$^*$: The H$_2$CO 4$_{04}$-3$_{03}$ line towards DM Tau, HD 163296, and V4046 Sgr was observed with largest angular scales of 2".6, 2".9, and 4".2, respectively.  It is possible that we are missing extended H$_2$CO emission for these large disks.}
\end{deluxetable*}

\subsection{Data Reduction}
\label{sec_reduction}


The ALMA/NAASC staff performed the initial calibration (including bandpass, flux, and phase calibration) for each observation.  The flux calibration for HD 163296 in ALMA observing projects 2012.1.00681.S and 2013.1.00226.S was further optimized by~\cite{cite_qietal2015} and~\cite{cite_huangetal2017}, respectively.  As the datasets were observed at different times across the past decade, we note that this initial calibration was non-uniform across the sample.  After initial calibration, we used the \textsc{Common Astronomy Software Applications} package (CASA) version 4.7.2 to self-calibrate and image each targeted line.  We self-calibrated using the continuum emission from all available spectral windows, after removing any strong lines.  Solution intervals were 10-100 seconds.  We used intervals within this range that maximized the number of solutions with signal-to-noise ratios $\geq$2.  We carried out 2 iterations of phase calibration and 1 iteration of amplitude calibration for each line, and we used a Briggs weighting scheme and a robust value of 0.5 during the self-calibration process for all disks.
There were a few exceptions to this approach.  For one observation of the H$_2$CO 4$_{13}$-3$_{12}$ line towards MWC 480, we used 3 iterations of phase calibration and 2 iterations of amplitude calibration to improve phase and amplitude solutions.  We did not self-calibrate the H$_2$CO 4$_{04}$-3$_{03}$ line towards DM Tau, or the H$_2$CO 3$_{03}$-2$_{02}$ line towards J1612-1859, due to low continuum peaks for these sources.


After self-calibration, we subtracted out the continuum using CASA's \texttt{uvcontsub} function and imaged each targeted line using CASA's \texttt{clean} algorithm.  We initially used a Briggs weighting scheme and a robust value of 0.5 for \texttt{clean}ing for all disks. We created the \texttt{clean}ing masks by hand to cover the emission and little else in each channel.  For the weaker lines (e.g., H$_2$CO 3$_{22}$-2$_{21}$), we recycled \texttt{clean} masks from the stronger lines available.  The weaker H$_2$CO lines that were tentatively-detected or non-detected with robust=0.5 were reimaged with robust=2 to maximize sensitivity.

We did not image the two weaker H$_2$CO 3$_{22}$-2$_{21}$ and 3$_{21}$-2$_{20}$ lines for the disks J1609-1908, J1612-1859, and J1614-1906, because the emission from the much brighter H$_2$CO 3$_{03}$-2$_{02}$ line was already very faint for these three disks.


\section{Image Analysis}
\label{sec_methodology}

\begin{figure*}
\resizebox{\hsize}{!}{
    \includegraphics[trim=35pt 125pt 4pt 25pt, clip]{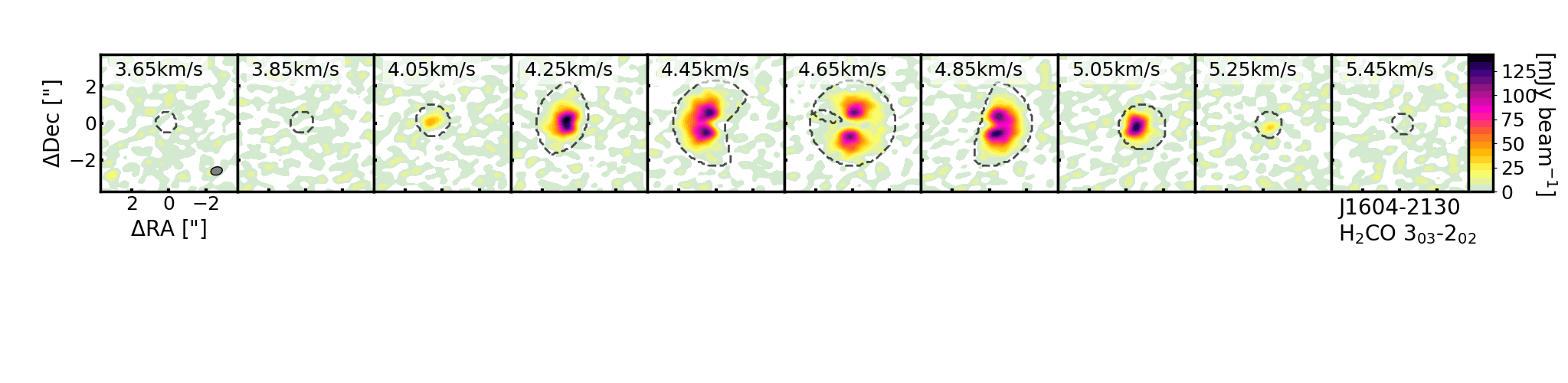}}
\caption{\textbf{H$_2$CO 3$_{03}$-2$_{02}$ Channel Map and Keplerian Masks for J1604-2130.}   The colorbars have been clipped to exclude values below zero.  The outlines of the calculated Keplerian masks are drawn with dashed lines.}
\label{fig_exseries}
\end{figure*}

Our basic image products were image cubes, which are visualized as channel maps. Figure~\ref{fig_exseries} displays the channel map for the H$_2$CO 3$_{03}$-2$_{02}$ line towards J1604-2130; channel maps for all detected lines and sources are shown in Figure Sets A1-A3 in the Appendix.
We generated Keplerian masks~\citep[e.g.,][]{cite_rosenfeldetal2013a, cite_yenetal2016, cite_salinasetal2017} to extract the line emission within each channel for each disk, ultimately reducing the noise incorporated into the extracted emission.  
We generated each set of Keplerian masks in four steps\footnote{The Keplerian mask package is available on GitHub (\url{https://github.com/jpegues/kepmask}).  Version 1.0.0 was used for this paper and is available at https://doi.org/10.5281/zenodo.3382082.}: (1) calculate each disk's line-of-sight velocities within the disk's coordinate system, (2) incorporate thermal and turbulent broadening using the approach of~\cite{cite_yenetal2016}, (3) mask the pixels within each velocity channel that fall within the broadened velocity bins, and (4) convolve the masks with the average beam size.  For this survey, we empirically fit for the position and inclination angles using the masks, C$^{18}$O emission towards each disk, and a grid-search algorithm ($^{12}$CO emission was used for the three disks where C$^{18}$O was not detected, and H$_2$CO emission was used for the one disk where no CO was observed).  The Keplerian masks calculated for the H$_2$CO 3$_{03}$-2$_{02}$ line towards J1604-2130 are overplotted in the channel map of Figure~\ref{fig_exseries}.  
All parameters used to generate the Keplerian masks for the sample are given in Table~\ref{table_kepmaskparams} in the Appendix. 

We created a velocity-integrated emission map (a.k.a. a moment-0 map) for each line and disk pair by summing up the emission within the Keplerian masks across the velocity channels in the data cube. To estimate the noise of each velocity-integrated emission map, it is important to account for how each pixel in the image appears in a different number of Keplerian masks, depending on the pixel's location.  We therefore generated an `rms map' for each line and disk, which calculates the rms at each pixel \textit{through} all masks.  Each rms map has the same dimensions as the velocity-integrated emission map and holds different rms values for each pixel.  For each line and disk, we randomly sampled 1000 rms maps across channels with no emission, and then took the standard deviation at each pixel.  We used the median of the standard deviations as the estimate of the line emission map's noise ($\sigma_{map}$).  See~\cite{cite_bergneretal2018} for the full methodology for generating rms maps.

Velocity-integrated fluxes for each line and disk were calculated by summing over the velocity-integrated emission maps pixel-by-pixel. Noise was estimated similarly to the velocity-integrated maps, i.e. from the standard deviation of 1000 random samples. To create a radial emission profile for each line, we azimuthally averaged each velocity-integrated emission map.  In this process, we deprojected the disks and divided them into 0.1" rings about the host star, and then averaged the emission within each ring.  Following~\cite{cite_bergneretal2018}, we estimated the noise within each ring as ($\sigma_{map}/\sqrt{N}$), where $N$ is the number of independent measurements in this ring, taken to be the number of pixels in the ring divided by the beam area (in pixels).  For rings within the beam's circumference, we fixed $N$ to be the beam circumference (in pixels) divided by the beam area.
Finally, we summed the emission in the Keplerian masks as a function of velocity channel to extract a spectrum for each line and disk (shown in Figure~\ref{fig_series32and43spec} in the Appendix).


\section{Results}
\label{sec_results}

\subsection{Detection Statistics}
\label{sec_analysis_detections}

\begin{deluxetable*}{lllllllllll}
\tablecaption{\textbf{H$_2$CO Emission Flux Measurements.} \label{table_h2coemfluxes}}
\tablehead{
Disk	& H$_2$CO  & Flux$^*$	& Peak Flux$^*$    & Int. Vel.   & Chan.  & rms$^{**}$	& Beam  & Robust \\
	& Line 	& [mJy	&  [mJy beam$^{-1}$  & Range  & Width   & [mJy	& Size (P.A.) & Value  & \\
	&  	& $\times$ km s$^{-1}$]	&  $\times$ km s$^{-1}$]  & [km s$^{-1}$]  &  [km s$^{-1}$]  & beam$^{-1}$]	&  &
}
\startdata
AS 209     & 4$_{04}$-3$_{03}$            & 580 $\pm$ 17    & 19 $\pm$ 2.6        & 0.9 - 8.1        & 0.4              & 2.9            & 0.31" x 0.22" (-69.84$^{\circ}$) & 0.5                       \\
           & 3$_{22}$-2$_{21}$            & $<$82  & 30 $\pm$ 7.2        & 0.9 - 8.1        & 0.4              & 7.5            & 0.61" x 0.56" (-69.47$^{\circ}$) & 2.0                       \\
           & 3$_{21}$-2$_{20}$            & $<$80  & 21 $\pm$ 6.4        & 0.9 - 8.1        & 0.4              & 6.6            & 0.61" x 0.56" (-70.17$^{\circ}$) & 2.0                       \\
CI Tau     & 3$_{03}$-2$_{02}$            & 420 $\pm$ 12    & 28 $\pm$ 3.1        & 2.6 - 9.0        & 0.2              & 4.8            & 0.67" x 0.46" (30.53$^{\circ}$)  & 0.5                       \\
DM Tau     & 4$_{04}$-3$_{03}$            & 337 $\pm$ 28    & 20 $\pm$ 3.1        & 3.6 - 8.4        & 0.3              & 5.5            & 0.59" x 0.51" (-81.65$^{\circ}$)$^\dagger$ & 0.5                 \\
           & 3$_{03}$-2$_{02}$            & 462 $\pm$ 17    & 24 $\pm$ 2.7        & 3.6 - 8.6        & 0.2              & 4.9            & 0.66" x 0.50" (35.24$^{\circ}$)$^\dagger$  & 0.5                 \\
           & 5$_{15}$-4$_{14}$            & 256 $\pm$ 37    & 39 $\pm$ 6.6        & 3.6 - 8.6        & 0.2              & 13             & 0.77" x 0.60" (-25.76$^{\circ}$)$^\dagger$ & 0.5               \\
DO Tau     & 3$_{03}$-2$_{02}$            & 85 $\pm$ 17     & 36 $\pm$ 3.7        & 3.4 - 8.6        & 0.4              & 3.4            & 0.90" x 0.58" (32.19$^{\circ}$)  & 2.0                       \\
GM Aur     & 4$_{04}$-3$_{03}$            & 1074 $\pm$ 17   & 44 $\pm$ 3.6        & -0.1 - 11.3      & 0.3              & 3.1            & 0.29" x 0.19" (-2.69$^{\circ}$)  & 0.5                       \\
HD 143006  & 3$_{03}$-2$_{02}$            & 161 $\pm$ 9.4   & 32 $\pm$ 3.0        & 5.4 - 10.2       & 0.2              & 4.7            & 0.68" x 0.47" (-74.10$^{\circ}$) & 0.5                       \\
           & 3$_{22}$-2$_{21}$            & $<$30$^\times$  & 8.2 $\pm$ 2.3       & 5.4 - 10.2       & 0.2              & 4.0            & 0.87" x 0.57" (-71.56$^{\circ}$) & 2.0                       \\
           & 3$_{21}$-2$_{20}$            & $<$32           & 17 $\pm$ 3.2        & 5.4 - 10.2       & 0.2              & 4.2            & 0.87" x 0.57" (-71.74$^{\circ}$) & 2.0                       \\
HD 163296  & 4$_{04}$-3$_{03}$            & 942 $\pm$ 43    & 51 $\pm$ 5.9        & -1.4 - 13.0      & 0.4              & 6.9            & 0.57" x 0.52" (86.68$^{\circ}$)$^\dagger$  & 0.5                  \\
           & 3$_{22}$-2$_{21}$            & $<$132 & $<$30               & -1.4 - 13.0      & 0.4              & 3.8            & 0.69" x 0.59" (67.20$^{\circ}$)  & 2.0                       \\
           & 3$_{21}$-2$_{20}$            & 119 $\pm$ 40          & 47 $\pm$ 9.8        & -1.4 - 13.0      & 0.4              & 3.4            & 0.72" x 0.59" (70.98$^{\circ}$)  & 2.0                       \\
IM Lup     & 4$_{04}$-3$_{03}$            & 1063 $\pm$ 25   & 45 $\pm$ 4.3        & 0.2 - 8.6        & 0.2              & 7.4            & 0.45" x 0.39" (70.99$^{\circ}$)$^\dagger$  & 0.5                 \\
           & 3$_{22}$-2$_{21}$            & $<$55$^\times$  & 12 $\pm$ 3.5        & 0.6 - 8.6        & 0.4              & 3.7            & 0.67" x 0.49" (-76.11$^{\circ}$) & 2.0                       \\
           & 3$_{21}$-2$_{20}$            & $<$45$^\times$  & 9.1 $\pm$ 3.0       & 0.6 - 8.6        & 0.4              & 3.2            & 0.66" x 0.48" (-76.35$^{\circ}$) & 2.0                       \\
J1604-2130 & 3$_{03}$-2$_{02}$            & 821 $\pm$ 6.4   & 60 $\pm$ 1.6        & 3.6 - 5.4        & 0.2              & 4.5            & 0.63" x 0.44" (-76.97$^{\circ}$) & 0.5                       \\
           & 3$_{22}$-2$_{21}$            & 157 $\pm$ 7.1   & 18 $\pm$ 1.5        & 3.6 - 5.4        & 0.2              & 4.6            & 0.63" x 0.44" (-77.13$^{\circ}$) & 0.5                       \\
           & 3$_{21}$-2$_{20}$            & 151 $\pm$ 6.6   & 18 $\pm$ 1.3        & 3.6 - 5.4        & 0.2              & 4.7            & 0.63" x 0.44" (-77.51$^{\circ}$) & 0.5                       \\
J1609-1908 & 3$_{03}$-2$_{02}$            & 43 $\pm$ 13     & 36 $\pm$ 4.2        & -1.6 - 9.2       & 0.4              & 3.3            & 0.81" x 0.53" (-72.33$^{\circ}$) & 2.0                       \\
J1612-1859 & 3$_{03}$-2$_{02}$            & $<$26$^\times$  & $<$12               & -0.8 - 10.4      & 0.4              & 3.0            & 0.81" x 0.53" (-71.66$^{\circ}$) & 2.0                       \\
J1614-1906 & 3$_{03}$-2$_{02}$            & $<$30           & 25 $\pm$ 5.2        & -3.0 - 10.6      & 0.4              & 3.3            & 0.80" x 0.53" (-72.36$^{\circ}$) & 2.0                       \\
LkCa 15    & 4$_{04}$-3$_{03}$            & 662 $\pm$ 24    & 20 $\pm$ 2.8        & 2.1 - 10.2       & 0.3              & 3.3            & 0.26" x 0.20" (-14.57$^{\circ}$) & 0.5                       \\
           & 3$_{22}$-2$_{21}$            & $<$77           & 16 $\pm$ 3.3        & 2.1 - 10.2       & 0.3              & 4.6            & 0.64" x 0.51" (13.70$^{\circ}$)  & 2.0                       \\
           & 3$_{21}$-2$_{20}$            & $<$52           & 15 $\pm$ 2.8        & 2.1 - 10.2       & 0.3              & 3.5            & 0.64" x 0.50" (14.42$^{\circ}$)  & 2.0                       \\
           & 4$_{14}$-3$_{13}$            & 1199 $\pm$ 16   & 124 $\pm$ 3.2       & 1.9 - 10.7       & 1.1              & 2.2            & 0.82" x 0.47" (-36.05$^{\circ}$) & 0.5                       \\
           & 4$_{13}$-3$_{12}$            & 1342 $\pm$ 33   & 169 $\pm$ 5.7       & 1.9 - 10.7       & 1.1              & 3.9            & 0.80" x 0.48" (-37.21$^{\circ}$) & 0.5                       \\
           & 4$_{23}$-3$_{22}$            & $<$62           & 29 $\pm$ 4.0        & 1.9 - 10.7       & 1.1              & 2.3            & 0.99" x 0.58" (-40.33$^{\circ}$) & 2.0                       \\
           & 4$_{22}$-3$_{21}$            & 37 $\pm$ 20           & 32 $\pm$ 3.6        & 1.9 - 10.7       & 1.1              & 2.2            & 0.99" x 0.58" (-40.35$^{\circ}$) & 2.0                       \\
MWC 480    & 4$_{04}$-3$_{03}$            & 239 $\pm$ 21    & 74 $\pm$ 8.3        & -1.4 - 11.8      & 1.1              & 3.5            & 1.02" x 0.83" (-3.58$^{\circ}$)  & 0.5                       \\
           & 3$_{22}$-2$_{21}$            & 36 $\pm$ 30           & 55 $\pm$ 9.8        & -1.2 - 11.2      & 0.4              & 4.0            & 0.76" x 0.50" (13.92$^{\circ}$)  & 2.0                       \\
           & 3$_{21}$-2$_{20}$            & 55 $\pm$ 22           & 60 $\pm$ 8.3        & -1.2 - 11.2      & 0.4              & 3.0            & 0.76" x 0.50" (14.31$^{\circ}$)  & 2.0                       \\
           & 4$_{14}$-3$_{13}$            & 297 $\pm$ 16    & 49 $\pm$ 4.9        & -1.4 - 11.8      & 1.1              & 2.1            & 0.90" x 0.46" (-28.87$^{\circ}$) & 0.5                       \\
           & 4$_{13}$-3$_{12}$            & 452 $\pm$ 32    & 86 $\pm$ 8.4        & -1.4 - 11.8      & 1.1              & 3.6            & 0.87" x 0.47" (-28.96$^{\circ}$) & 0.5                       \\
           & 4$_{23}$-3$_{22}$            & $<$59$^\times$  & $<$16               & -1.4 - 11.8      & 1.1              & 2.1            & 1.08" x 0.59" (-32.09$^{\circ}$) & 2.0                       \\
           & 4$_{22}$-3$_{21}$            & $<$61           & 30 $\pm$ 6.0        & -1.4 - 11.8      & 1.1              & 2.1            & 1.08" x 0.59" (-32.10$^{\circ}$) & 2.0                       \\
           & 5$_{15}$-4$_{14}$            & 396 $\pm$ 54    & 104 $\pm$ 13        & -1.2 - 11.2      & 0.4              & 12             & 0.88" x 0.54" (-22.89$^{\circ}$)$^\dagger$ & 0.5                 \\
V4046 Sgr  & 4$_{04}$-3$_{03}$            & 1218 $\pm$ 40   & 33 $\pm$ 6.3        & -4.2 - 10.2      & 0.3              & 5.8            & 0.54" x 0.45" (-67.39$^{\circ}$)$^\dagger$ & 0.5                 \\
           & 3$_{22}$-2$_{21}$            & $<$69  & $<$19               & -4.2 - 10.2      & 0.3              & 3.9            & 0.86" x 0.54" (-84.12$^{\circ}$) & 2.0                       \\
           & 3$_{21}$-2$_{20}$            & $<$67           & 14 $\pm$ 3.6        & -4.2 - 10.2      & 0.3              & 3.4            & 0.85" x 0.53" (-84.18$^{\circ}$) & 2.0                      
\enddata
\tablecomments{{*: The velocity-integrated fluxes (Column 3) were measured within the radii of the Keplerian masks (Table~\ref{table_kepmaskparams} in the Appendix)}.  The 3$\sigma$ upper limit is given for tentative and non-detections, and the upper limits of non-detections are additionally marked with a $^\times$.  The peak fluxes (Column 4) are the peaks of the velocity-integrated emission maps; note the difference in unit compared to the velocity-integrated fluxes.  The uncertainty in each peak flux is the standard deviation of the peaks across 1000 random samples.  The 3$\sigma$ upper limit is given whenever the peak flux is less than 3$\sigma$ itself.  Uncertainties do not include absolute flux uncertainties.  $**$: The channel rms was estimated as the standard deviation of 1000 1"-by-1" random samples.  $\dagger$: The following UV tapers were applied to reduce the effects of small-scale noise in the image: 0.25" for the H$_2$CO 4$_{04}$-3$_{03}$ line towards IM Lup and V4046 Sgr, 0.40" for the 5$_{15}$-4$_{14}$ line towards MWC 480, and 0.50" for the 4$_{04}$-3$_{03}$ line towards HD 163296.  UV tapers of 0.25", 0.50", and 0.60" were applied to the 3$_{03}$-2$_{02}$ (0.25"), 5$_{15}$-4$_{14}$ (0.50"), and 4$_{04}$-3$_{03}$ (0.60") lines towards DM Tau for the same reason, as well as to make the beam sizes comparable for these lines.}
\end{deluxetable*}

 We define a \textit{detection} as a line that fulfills all of the following criteria:

\begin{enumerate}
	\item Peak emission is $\geq 3\sigma$ in the velocity-integrated emission map
	\item Emission within the Keplerian masks is $\geq 3\sigma$ in at least 4 velocity channels of the channel map
\end{enumerate}

We define a \textit{tentative detection} as a line that fulfills at least one of the above criteria, and a \textit{non-detection} as a line that fails all criteria.  Based on these criteria, we detect the H$_2$CO 3$_{03}$-2$_{02}$ line towards 6/8 disks.  We tentatively detect this line towards the disk J1614-1906, and do not detect it towards the disk J1612-1859.  We detect the H$_2$CO 4$_{04}$-3$_{03}$ line towards 8/8 disks.  
Altogether, we report detections of H$_2$CO towards 13/15 disks.  Of these 13 disks, we detect one line towards 8 disks, two lines towards 1 disk, three lines towards 2 disks, four lines towards 1 disk, and six lines towards 1 disk.  Table~\ref{table_h2coemfluxes} lists the fluxes for all detected H$_2$CO lines and flux upper limits for all tentative and non-detected H$_2$CO lines.  The fluxes for the associated continuum and C$^{18}$O emission are in Tables~\ref{table_h2cocontfluxes} and~\ref{table_coemfluxes}, respectively, in the Appendix.


\subsection{Spatial Distributions of H$_2$CO Emission}
\label{sec_analysis_morphologies}


\begin{figure*}[!htbp]
\centering
\resizebox{0.95\hsize}{!}{
\centering
    \includegraphics{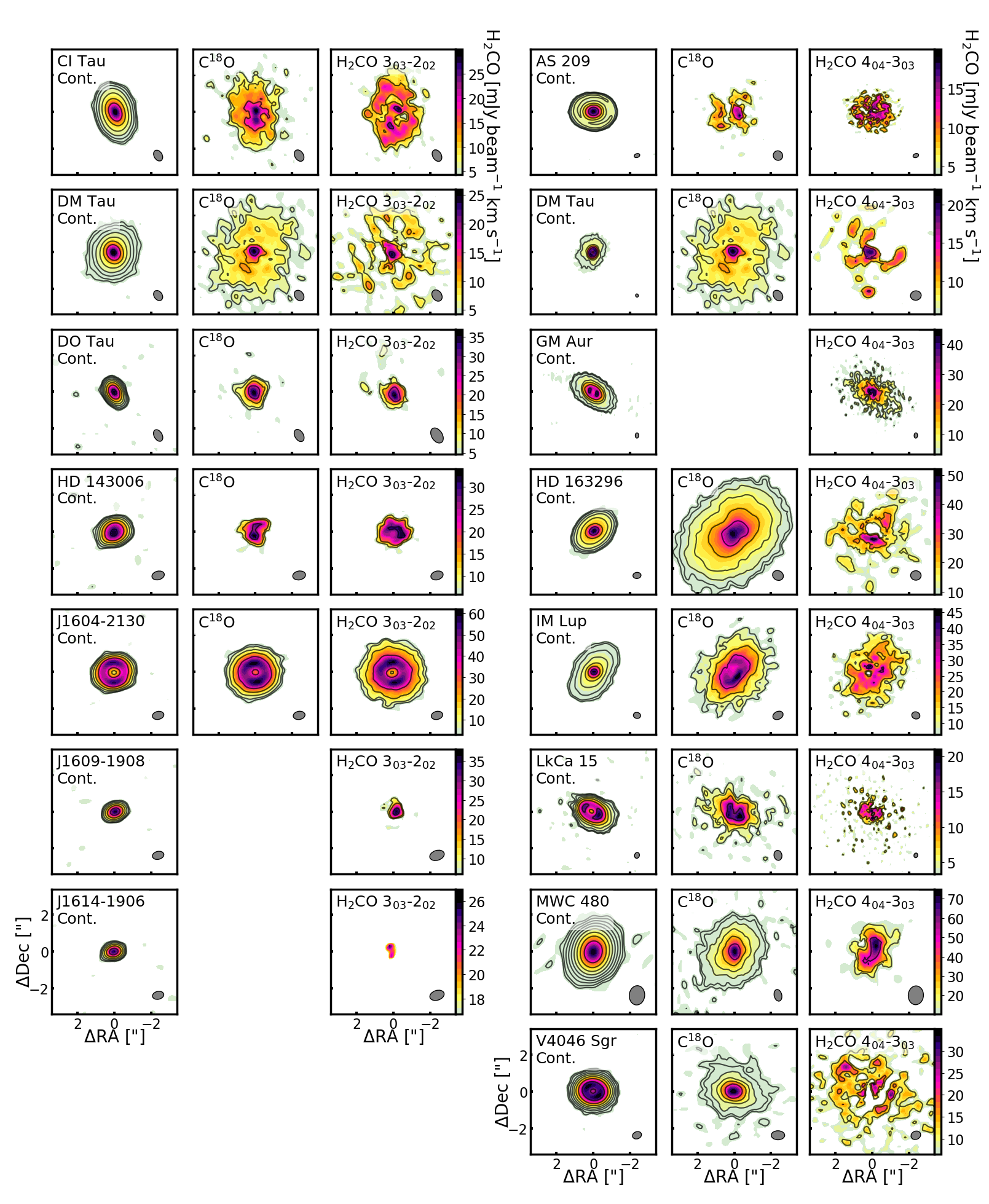}}
\caption{\textbf{H$_2$CO 3$_{03}$-2$_{02}$ and 4$_{04}$-3$_{03}$ Velocity-Integrated Emission Maps}. The left and right groups of columns correspond to all detected/tentatively-detected H$_2$CO 3$_{03}$-2$_{02}$ (3-2) and 4$_{04}$-3$_{03}$ (4-3) lines, respectively.  In each group, each row shows emission for a different disk.  The first column in each group shows the dust continuum at wavelengths of either 1.3mm (left group) or 1.0mm (right group), the second shows the velocity-integrated emission maps above $2\sigma_{map}$ for C$^{18}$O, and the third shows the H$_2$CO velocity-integrated emission maps above $2\sigma_{map}$.  Colorbars are provided for the H$_2$CO emission maps to the right of each group.  The contour lines for all subplots are [$3\sigma, 5\sigma, 10\sigma, 20\sigma$...]; for the continuum emission $\sigma$ is the continuum rms, while for the integrated emission maps $\sigma$ is the corresponding $\sigma_{map}$.  Beams are drawn in the lower right corners.  The H$_2$CO 3-2 line is tentatively detected towards J1614-1906.  All other H$_2$CO 3-2 and 4-3 lines are detected.  C$^{18}$O is not observed towards GM Aur and is not detected towards J1609-1908 and J1614-1906.}
\label{fig_series32and43mom0}
\end{figure*}

\begin{figure}[!htbp]
\centering
\resizebox{0.9\hsize}{!}{
\centering
    \includegraphics[trim=7.5pt 0pt 7.5 5pt, clip]{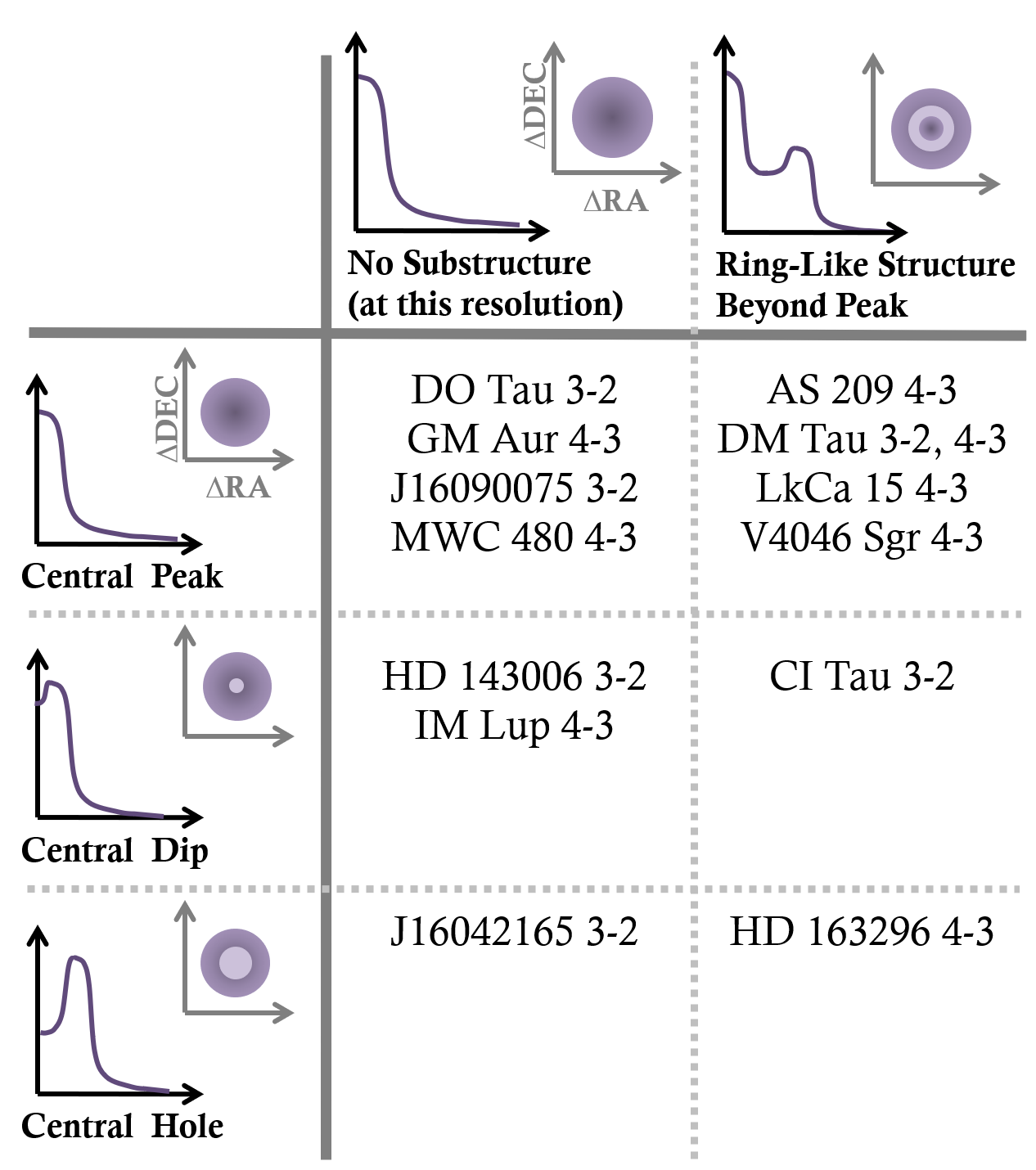}}
\caption{\textbf{H$_2$CO 3$_{03}$-2$_{02}$ and 4$_{04}$-3$_{03}$ Grid of Morphologies.}  The rows classify the H$_2$CO detections by the shape of the central emission at/inwards of the emission peak.  The columns classify the detections based on the substructure beyond the central emission when it is significant in both the integrated emission maps and radial profiles.}
\label{fig_morphologies}
\end{figure}

\begin{figure*}[!htbp]
\centering
\resizebox{0.85\hsize}{!}{
    \includegraphics[trim=10pt 5pt 55pt 5pt, clip]{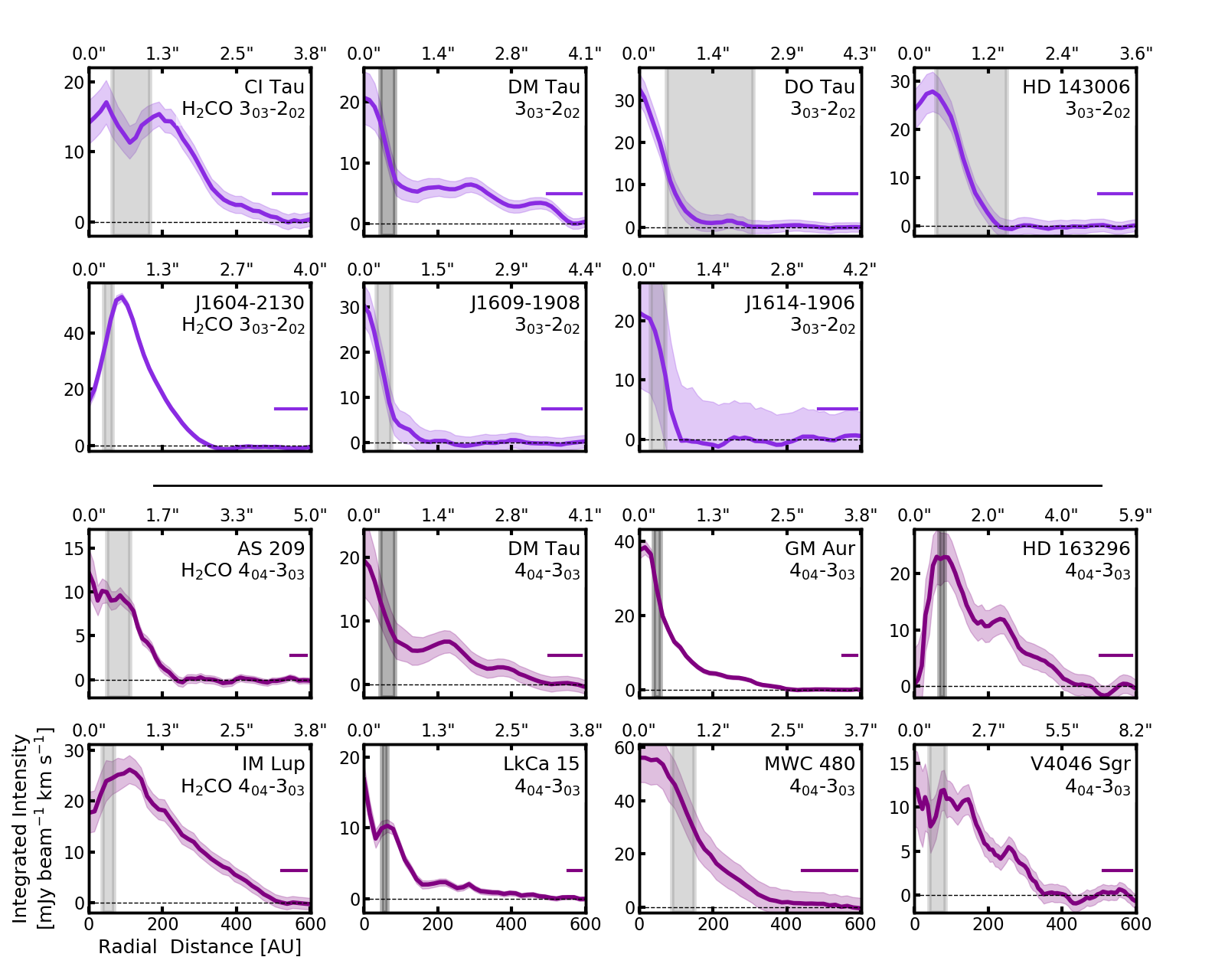}}
\caption{\textbf{H$_2$CO 3$_{03}$-2$_{02}$ and 4$_{04}$-3$_{03}$ Radial Profiles.}  H$_2$CO 3$_{03}$-2$_{02}$ (3-2) and 4$_{04}$-3$_{03}$ (4-3) lines are depicted in light and dark purple, respectively, above and below the black line.  DM Tau appears twice because both lines are observed towards DM Tau.  The H$_2$CO 3-2 line is tentatively detected towards J1614-1906.  All other H$_2$CO 3-2 and 4-3 lines are detected.  The shading represents the 1$\sigma$ uncertainties, which do not include absolute flux uncertainties.  Proposed locations for the midplane CO snowlines of the disks, extracted from the literature and listed in Table~\ref{table_cosnowlocs} (Section~\ref{sec_discussion_cosnowline}), are overplotted with gray vertical shading.  The light gray or dark gray regions indicate the snowline locations were derived from either disk temperature models or N$_2$H$^+$ fits, respectively.  The average half-power beam widths are represented by the horizontal bars in the lower right corners.}
\label{fig_series32and43prof}
\end{figure*}

\begin{figure*}[!htbp]
\centering
\resizebox{0.85\hsize}{!}{
    \includegraphics[trim=5pt 5pt 55pt 5pt, clip]{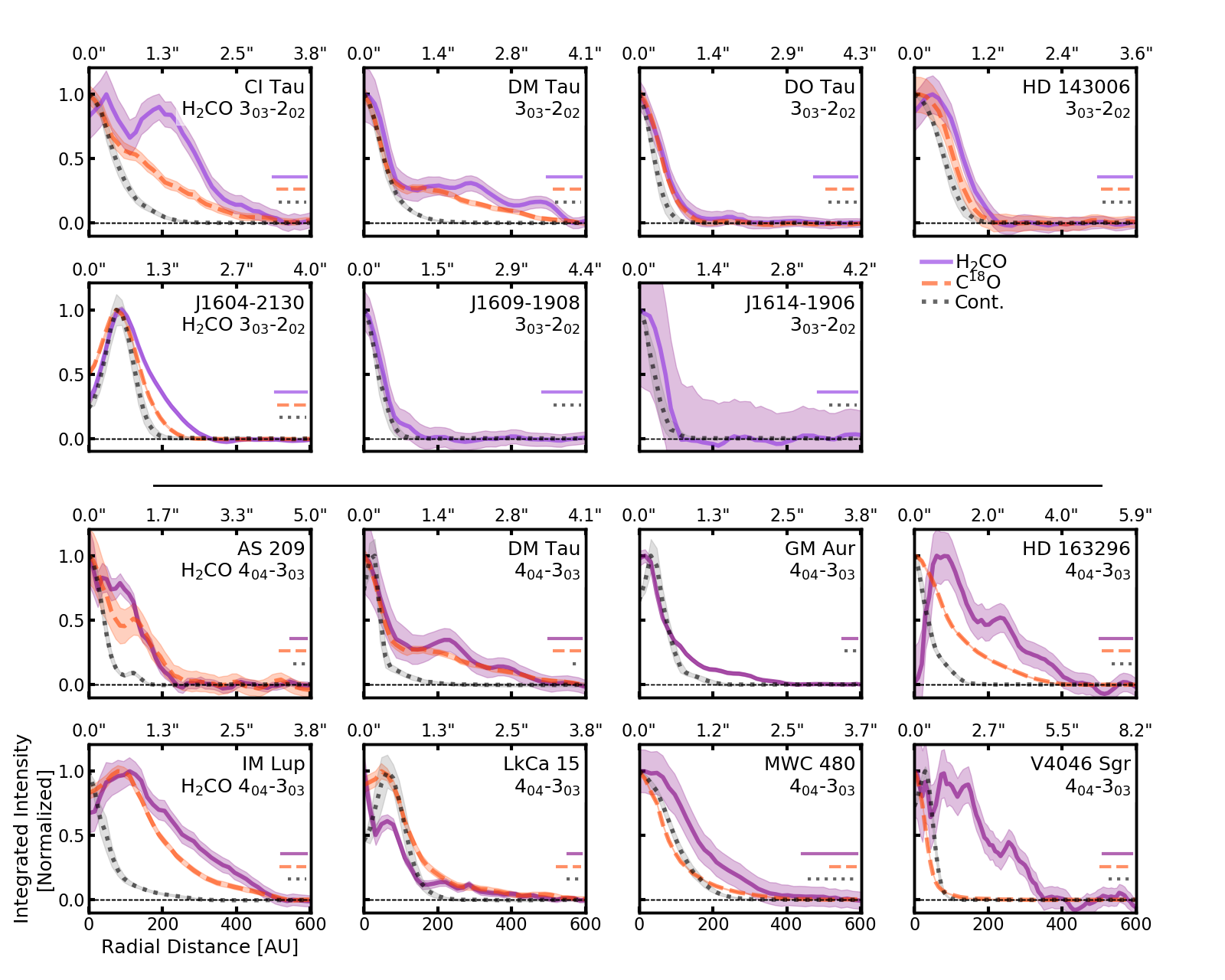}}
\caption{\textbf{Same as Figure~\ref{fig_series32and43prof}, Normalized with C$^{18}$O (orange) and Dust Continuum (gray) Radial Profiles for Comparison.}   C$^{18}$O is not detected towards J1609-1908 and J1614-1906, and we have no CO observations for GM Aur.}
\label{fig_series32and43line}
\end{figure*}

Figure~\ref{fig_series32and43mom0} presents emission maps for all observed H$_2$CO 3$_{03}$-2$_{02}$ and 4$_{04}$-3$_{03}$ lines, associated dust continuum, and C$^{18}$O line emission.  Figure~\ref{fig_morphologies} summarizes the emission morphologies of the disks detected in H$_2$CO.  Figure~\ref{fig_series32and43prof} shows the radial profiles of all H$_2$CO 3$_{03}$-2$_{02}$ and 4$_{04}$-3$_{03}$ lines.  We see a variety of H$_2$CO spatial structures between the 13 detected disks.  The majority (8/13) of the disks are centrally peaked in H$_2$CO.  Three disks have central dips (central depressions that are $\sim$1-3$\sigma_{peak}$ deep) in H$_2$CO, while the last two disks have central holes (central depressions that are $\gtrsim$3$\sigma_{peak}$ deep) in H$_2$CO.
About half (6/13) of the disks show outer rings or plateaus in H$_2$CO emission.  DM Tau, for example, has rings in its H$_2$CO 4$_{04}$-3$_{03}$ emission and a plateau with suggestive ring-like substructure in its H$_2$CO 3$_{03}$-2$_{02}$ emission. 

Figure~\ref{fig_series32and43line} compares the H$_2$CO radial profiles with the dust continuum and C$^{18}$O radial profiles.  The majority (10/13) of the disks show H$_2$CO emission beyond the edges of the dust continuum (defined as the radial extent at which the azimuthally-averaged continuum emission is 5\% of its peak).  Of the five disks that show inner holes in the dust continuum at our resolution, only one disk, J1604-2130, shows a similar inner hole in the H$_2$CO emission.  Dust and H$_2$CO emission substructure thus do not necessarily coincide.
The C$^{18}$O and H$_2$CO emission morphologies are more similar than the dust and H$_2$CO: seven disks have similar radial extents in H$_2$CO and C$^{18}$O, and five have similar outer slopes in H$_2$CO and C$^{18}$O emission. However, there are also distinct differences between the H$_2$CO and C$^{18}$O emission morphologies, indicating that H$_2$CO substructure is set by chemistry as well as by the disk gas distribution.

A priori, one possible explanation for the central H$_2$CO depressions observed is dust opacity, i.e. optically thick dust blocking out molecular emission. C$^{18}$O emission profiles provide an avenue to test this. We do see C$^{18}$O depressions in three disks with H$_2$CO inner dips or holes: IM Lup, J1604-2130, and LkCa 15 (Figure~\ref{fig_series32and43line}). However, for two of these disks (J1604-2130 and LkCa 15) the C$^{18}$O depressions coincide with central holes in the dust continuum and are therefore explained by holes in gas and dust rather than by dust opacity. This leaves IM Lup as the only case where the central H$_2$CO depression is likely due to dust opacity, at least in part. For all disks except IM Lup, optically thick dust is likely not a major cause for the H$_2$CO emission substructure in the sample.



\subsection{Rotational Diagram Analysis}
\label{sec_analysis_trotfits}

\begin{figure*}[!htbp]
\centering
\resizebox{0.8\hsize}{!}{
	\centering
    \includegraphics[trim=5pt 0pt 40pt 5pt, clip]{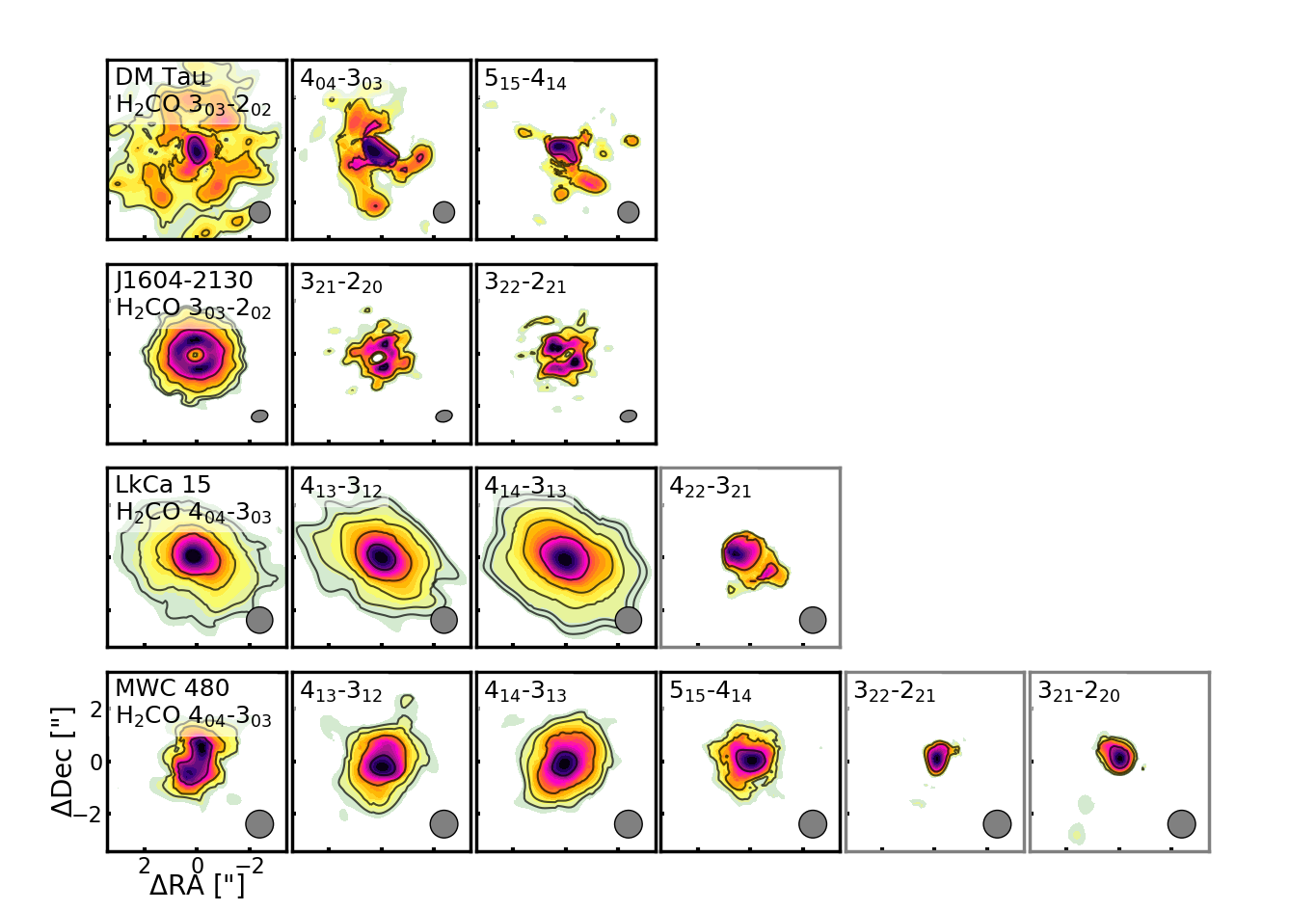}}
\caption{\textbf{Velocity-Integrated Emission Maps for Disks with Multiple H$_2$CO Detections.}  Each row plots emission for a different disk, and each subplot gives the velocity-integrated emission map above $2\sigma_{map}$ of a different line of H$_2$CO.  Each subplot has an individual color scaling.  The contour lines are [$3\sigma_{map}, 5\sigma_{map}, 10\sigma_{map}, 20\sigma_{map}$...].  Beams are drawn in the lower right corners.  All lines have been smoothed and circularized using CASA's \textsc{imsmooth} function so that the beams are identical across each disk.  Subplots outlined in black show the lines included in the rotational diagrams of Section~\ref{sec_analysis_trotfits}.  Subplots outlined in gray show the lines that were excluded from the diagrams due to low signal-to-noise in the outer disk regions.}
\label{fig_multimom0}
\end{figure*}

\begin{figure*}[!htbp]
\resizebox{\hsize}{!}{
    \includegraphics[trim=7.5pt 0pt 65pt 5pt, clip]{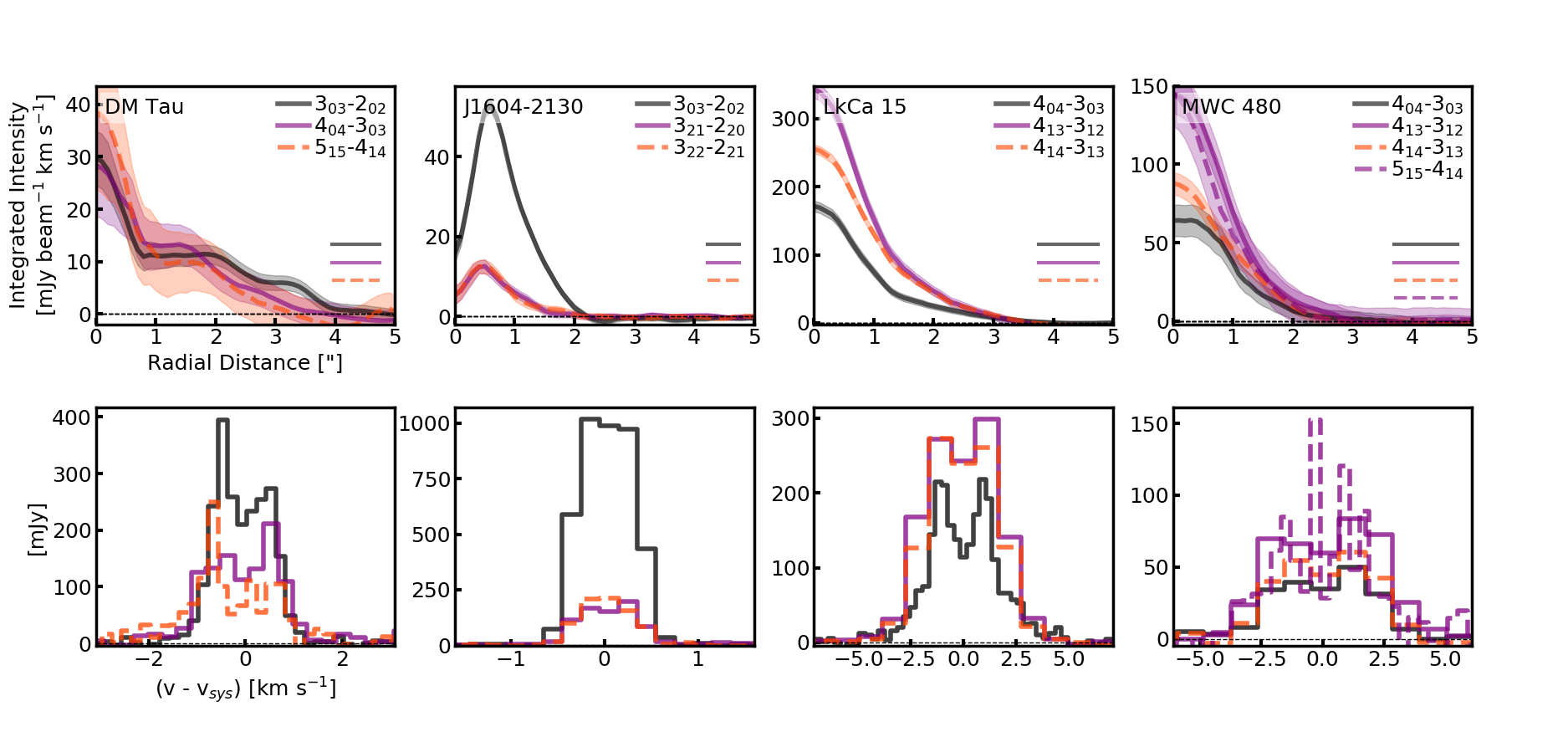}}
\caption{\textbf{Radial Profiles and Spectra for Lines used in Rotational Diagrams.}  The top and bottom rows show radial profiles and spectra, respectively, for the H$_2$CO lines used in the rotational diagrams (Section~\ref{sec_analysis_trotfits}).  Each column corresponds to a different disk, and each line style and color depict a different H$_2$CO line, as listed in the legends in the top row.  The shading represents the 1$\sigma$ uncertainties.  The average half-power beam widths are represented by the horizontal bars in the lower right corners of the top row.}
\label{fig_multigrid}
\end{figure*}

Three or more H$_2$CO lines are detected towards four disks: DM Tau, J1604-2130, LkCa 15, and MWC 480.  The velocity-integrated emission maps are shown in Figure~\ref{fig_multimom0}, while the radial profiles and spectra are shown in Figure~\ref{fig_multigrid}.  For these four disks, we fit for the area-averaged H$_2$CO excitation temperatures ($\meanbar{T_{ex}}$) and column densities ($\meanbar{N_{tot}}$) using the rotational diagram method~\citep[e.g.][]{cite_blakeetal1987, cite_goldsmithetal1999}.  This method assumes that the lines are in local thermodynamic equilibrium (LTE), that all lines originate from the same disk environment, and that the emission fills the beam.  LTE should be a reasonable approximation, since the H$_2$CO critical densities are lower than typical disk densities in the midplane and warm molecular disk layers (see Figure~\ref{fig_critdens} in the Appendix). It is a reasonable first approximation that the emission fills the beam, since the H$_2$CO emission is always resolved; however unresolved emission substructure may be present.

Following~\cite{cite_loomisetal2018}, we construct opacity-corrected rotational diagrams by relating $\meanbar{T_{ex}}$ and $\meanbar{N_{tot}}$ to the area-averaged column density $\meanbar{N_u}$ and degeneracy $g_u$ of each line's upper state:

\begin{align}
\ln \bigg( \frac{ \meanbar{N_u}}{g_u} \bigg) &= \ln \bigg( \frac{\meanbar{N_u^{thin}} C_{\tau_{ul}}}{g_u} \bigg) \nonumber \\
    &= \ln \bigg[ \bigg( \frac{4\pi (\meanbar{\int S dv})}{A_{ul} \Omega_{a} hc} \bigg) \bigg( \frac{\tau_{ul}}{1 - e^{-\tau_{ul}}} \bigg) \bigg] \nonumber \\
    &= \bigg[ \ln \bigg( \frac{\meanbar{N_{tot}}}{Q\{\meanbar{T_{ex}}\}} \bigg) - \frac{E_u}{\meanbar{T_{ex}}} \bigg] ,
\label{eq_fitline}
\end{align}

\noindent where $\meanbar{N_u^{thin}}$ is the area-averaged column density if the line is actually optically thin, $\meanbar{\int S dv}$ is the area-averaged velocity-integrated flux (in units of spectral flux density $\times$ velocity), $A_{ul}$ is the Einstein coefficient, and $\Omega_{a}$ is the solid angle of area over which $\meanbar{\int S dv}$ is averaged.  Next, $h$ is the Planck constant, $c$ is the speed of light, $Q\{T\}$ is the partition function at temperature $T$~\citep[interpolated from $T$ and $Q\{T\}$ values from CDMS;][]{cite_cdms}, and $E_u$ is the upper energy level.  Finally, $C_{\tau_{ul}} = \tau_{ul}/(1 - e^{-\tau_{ul}})$ is the opacity correction factor~\citep[e.g.,][]{cite_goldsmithetal1999}, where the opacity of the line $\tau_{ul}$ can be calculated as:

\begin{align}
\tau_{ul} &= \bigg( \frac{A_{ul} c^3}{8\pi \nu^3 \Delta v_{l}} \bigg) N_u (e^{h\nu/(k_B T_{ex})} - 1) \nonumber \\
 &= \bigg( \frac{A_{ul} c^3}{8\pi \nu^3 \Delta v_{l}} \bigg) N_u^{thin} C_{\tau_{ul}} (e^{h\nu/(k_B T_{ex})} - 1) \nonumber
 \end{align}
 
 \begin{align}
& (1 - e^{-\tau_{ul}}) = \bigg( \frac{A_{ul} c^3}{8\pi \nu^3 \Delta v_{l}} \bigg) N_u^{thin} (e^{h\nu/(k_B T_{ex})} - 1) \nonumber \\
& \tau_{ul} - \frac{\tau_{ul}^2}{2} + \frac{\tau_{ul}^3}{6} \approx \bigg( \frac{A_{ul} c^3}{8\pi \nu^3 \Delta v_{l}} \bigg) N_u^{thin} (e^{h\nu/(k_B T_{ex})} - 1) ,
\label{eq_tau}
\end{align}

\noindent where $(1 - e^{-\tau_{ul}}) \approx \tau_{ul} - \frac{\tau_{ul}^2}{2} + \frac{\tau_{ul}^3}{6}$, with accuracy within 99\%, 95\%, and 90\% at $\tau_{ul}$ of 0.6, 1.0, and 1.25 respectively.  $\nu$ is the line frequency, $k_B$ is the Boltzmann constant, and $\Delta v_{l}$ is the intrinsic line width, assumed to be a combination of thermal and turbulent broadening~\citep[e.g.,][]{cite_flahertyetal2015}:

\begin{equation}
\Delta v_{l}\approx\sqrt{\bigg( \sqrt{\frac{2k_B T_{ex}}{m_{H_2CO}}} \bigg)^2 + \bigg( t_0 \sqrt{ \frac{k_B T_{ex}}{\mu m_H}} \bigg)^2},
\label{eq_linewidth}
\end{equation}

\noindent where $m_{H_2CO}$ and $m_H$ are the masses of H$_2$CO and H respectively, $\mu=2.37$ is the assumed mean molecular weight of the disk, and $t_0 \approx 0.01$ is the assumed prefactor for the turbulent contribution~\citep[e.g.,][]{cite_flahertyetal2015}.

Combining Equations~\ref{eq_fitline} and~\ref{eq_tau}, we initially solve for the opacity-uncorrected $\meanbar{T_{ex}}$ and $\meanbar{N_{tot}}$ using a linear fit assuming $C_{\tau_{ul}}=1$, and then we correct the rotational diagram for the opacity using an iterative approach: (1) solve for each $\tau_{ul}$ using Equation~\ref{eq_tau}, (2) recalculate each $\meanbar{N_{u}}/g_u$ using Equation~\ref{eq_fitline}, (3) redo the linear fit to the rotational diagram (the last line of Equation~\ref{eq_fitline}), and use the fit to (4) recalculate the values for $\meanbar{T_{ex}}$ and $\meanbar{N_{tot}}$ from the fit's slope and y-intercept, respectively.  We repeat this process until we reach convergence, defined as when the difference in the $\tau_{ul}$ values and $\meanbar{T_{ex}}$ between two iterations is at most 0.001\% (or, in the case of the uncertain spike in excitation temperature for LkCa 15, until 5000 iterations have passed).

\begin{deluxetable*}{llll}
\tablecaption{\textbf{Disk-Averaged H$_2$CO Excitation Temperatures and Column Densities.}
\label{table_trotfitarea}}
\tablehead{
Disk	& $\meanbar{T_{ex}}$	& $\meanbar{N_{tot}}$ & $\tau_{ul}$ (H$_2$CO Line) \\
	& [K]	& [$10^{12}$ cm$^{-2}$] & 
}
\startdata
DM Tau  & $11 \pm 0.39$ & $2.4 \pm 0.34$    & $0.85 \pm 0.20$ (3$_{03}$-2$_{02}$), $0.35 \pm 0.08$ (4$_{04}$-3$_{03}$), $0.19 \pm 0.06$ (5$_{15}$-4$_{14}$) \\
J1604-2130 & $37 \pm 3.6$ & $21 \pm 3.2$    & $0.44 \pm 0.06$ (3$_{03}$-2$_{02}$), $0.07 \pm 0.01$ (3$_{21}$-2$_{20}$), $0.07 \pm 0.01$ (3$_{22}$-2$_{21}$)  \\
LkCa 15   & $29 \pm 8.8$ & $3.1 \pm 1.4$  & $0.14 \pm 0.04$ (4$_{04}$-3$_{03}$), $0.26 \pm 0.08$ (4$_{13}$-3$_{12}$), $0.26 \pm 0.07$ (4$_{14}$-3$_{13}$)  \\
MWC 480   & $21 \pm 2.2$ & $1.9 \pm 0.47$    & $0.16 \pm 0.03$ (4$_{04}$-3$_{03}$), $0.32 \pm 0.07$ (4$_{13}$-3$_{12}$), $0.22 \pm 0.04$ (4$_{14}$-3$_{13}$), $0.19 \pm 0.05$ (5$_{15}$-4$_{14}$)
\enddata
\tablecomments{Values were derived from the linear fits of Figure~\ref{fig_trotfitarea}.  Errors and fits include 10\% absolute flux uncertainties, added in quadrature to the velocity-integrated flux uncertainties. }
\end{deluxetable*}

\begin{figure*}[!htbp]
\centering
\resizebox{\hsize}{!}{
\centering
    \includegraphics[trim=10pt 5pt 5pt 5pt, clip]{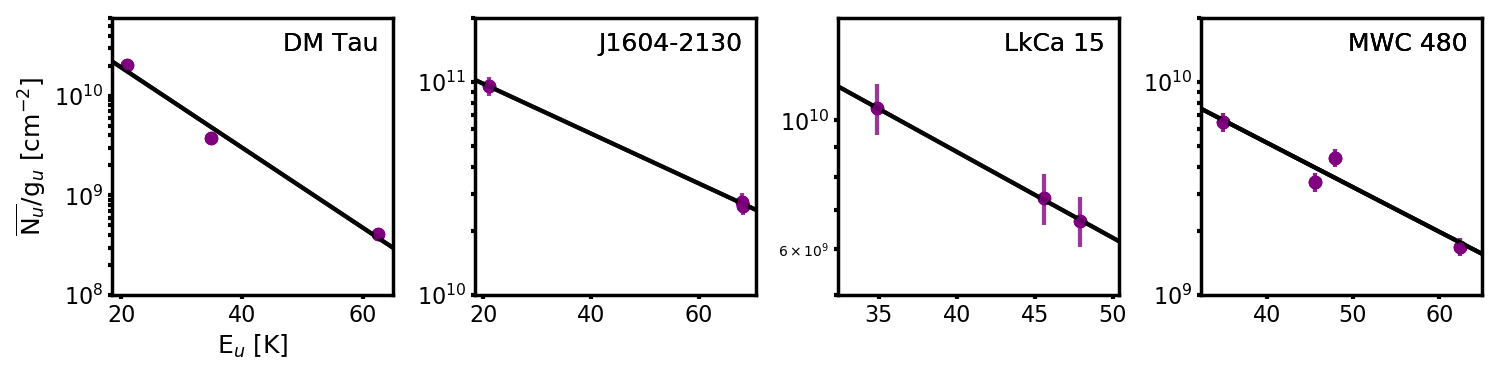}}
\caption{\textbf{Disk-Averaged Rotational Diagrams.}  The purple points represent the $\meanbar{N_u}/g_u$ values extracted from the observations, while the black lines indicate the best linear fits taking into account calculated line opacities.  Disk areas extend out to where at least one H$_2$CO line has azimuthally-averaged intensities above 1$\sigma$.  Errorbars are 1$\sigma$ and include 10\% absolute flux uncertainties, added in quadrature to the velocity-integrated flux uncertainties.}
\label{fig_trotfitarea}
\end{figure*}

\begin{figure*}[!htbp]
\resizebox{\hsize}{!}{
    \includegraphics[trim=10pt 5pt 5pt 5pt, clip]{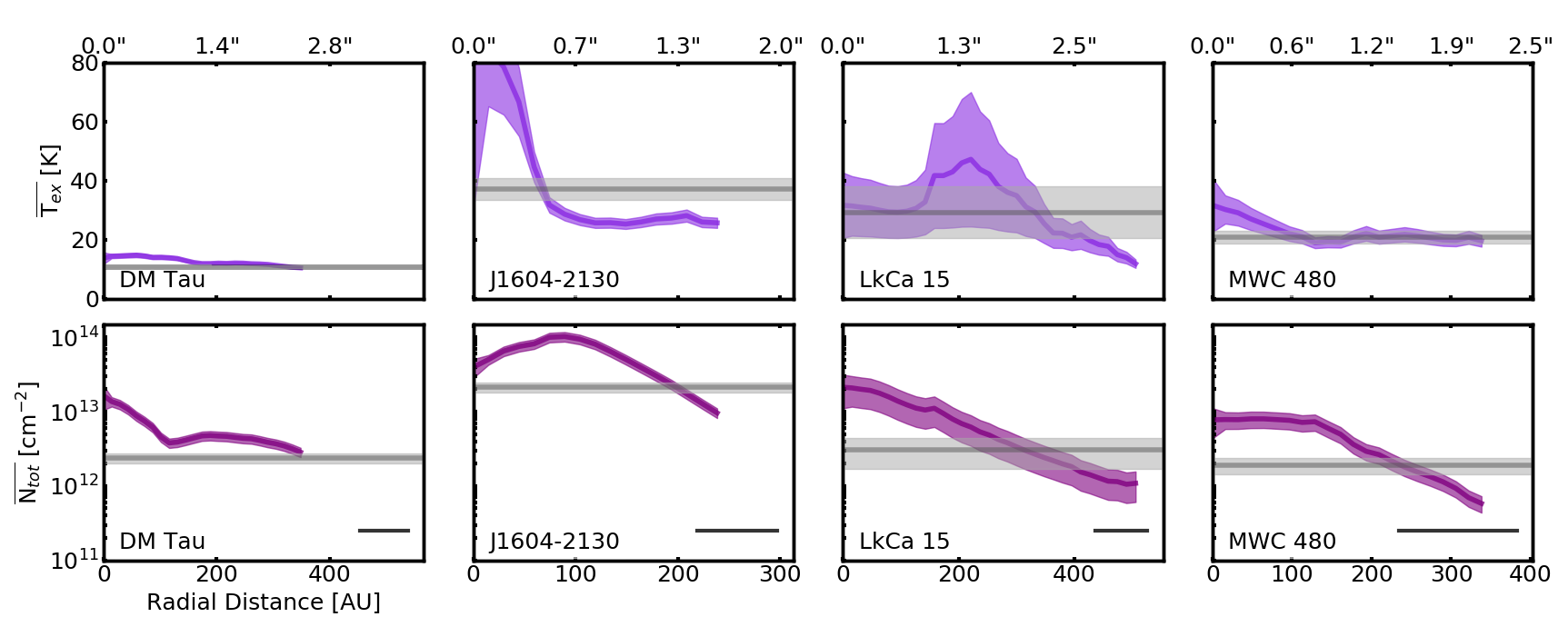}}
\caption{\textbf{Radial H$_2$CO Excitation Temperatures and Column Densities.}  The top and bottom rows show H$_2$CO excitation temperatures and column densities, respectively, as a function of radial distance from the host star.  The shaded regions give the 1$\sigma$ errors in the fits and include 10\% absolute flux uncertainties, added in quadrature to the flux uncertainties within each radial ring around the host star.   The horizontal gray lines show the disk-averaged excitation temperatures and column densities from Table~\ref{table_trotfitarea}.  All profiles are estimated out to where at least three H$_2$CO lines have azimuthally-averaged intensities above 1$\sigma$.  The subplots' x-axes are extended to show the disk area over which the disk-averaged temperatures and column densities are measured, out to where at least one H$_2$CO line has azimuthally-averaged intensities above 1$\sigma$.  The average half-power beam widths are represented by the horizontal bars in the lower right corners of the bottom panels.}
\label{fig_trotfitrad}
\end{figure*}

Figure~\ref{fig_trotfitarea} displays the disk-averaged rotational diagrams.  The disk area is measured out to where at least one H$_2$CO line has azimuthally-averaged intensities above 1$\sigma$.  Table~\ref{table_trotfitarea} lists the extracted disk-averaged excitation temperatures and column densities.  The disk-averaged excitation temperatures span $\sim$11-37K.  The disk-averaged column densities span a little more than an order of magnitude: J1604-2130 has the largest column density at $2.1 \times 10^{13}$cm$^{-2}$, while MWC 480 has the lowest at $1.9 \times 10^{12}$cm$^{-2}$.  Notably the energy ranges covered by the H$_2$CO lines for each disk are similar, implying that these results are not due to excitation effects from using different H$_2$CO lines.

For each disk we also construct azimuthally-averaged rotational diagrams. For each fit, we extract the integrated flux for each line from within the same narrow (0.1") rings of the deprojected disk around the host star.  We then construct a rotational diagram for each ring, assuming that each ring has uniform temperature and column density (see~\cite{cite_loomisetal2018} for more details).
Figure~\ref{fig_trotfitrad} displays the azimuthally-averaged excitation temperatures. In all disks except J1604-2130, the radially-resolved temperature profiles are close to consistent with a constant temperature.  The excitation temperature increase towards J1604-2130 conincides with the inner dust edge, and may be due to direct radiation from the host star.  Otherwise, the H$_2$CO excitation temperature across disks are $\sim$20-50$\sim$K, with the exception of DM Tau, where most H$_2$CO appears to be at $\sim$10-15$\sim$K. We note that DM Tau is an M-star with a massive disk, and it is therefore perhaps unsurprising that it hosts the coldest H$_2$CO.  

Figure~\ref{fig_trotfitrad} also displays the azimuthally-averaged H$_2$CO column densities.  The column density profiles generally vary between the four disks, closely following their H$_2$CO line emission profiles. Notably the observed H$_2$CO line emission plateau towards DM Tau seems to be due to a near-flat column density profile beyond $\sim$100 AU in this disk.  Interestingly, H$_2$CO has the smallest column density towards MWC 480, the only disk of these four that is irradiated by a Herbig Ae star.


\subsection{Radial H$_2$CO Column Density Profiles across the Sample}
\label{sec_analysis_coldens}

\begin{figure*}[!htbp]
\resizebox{\hsize}{!}{
    \includegraphics[trim=20pt 20pt 15pt 20pt, clip]{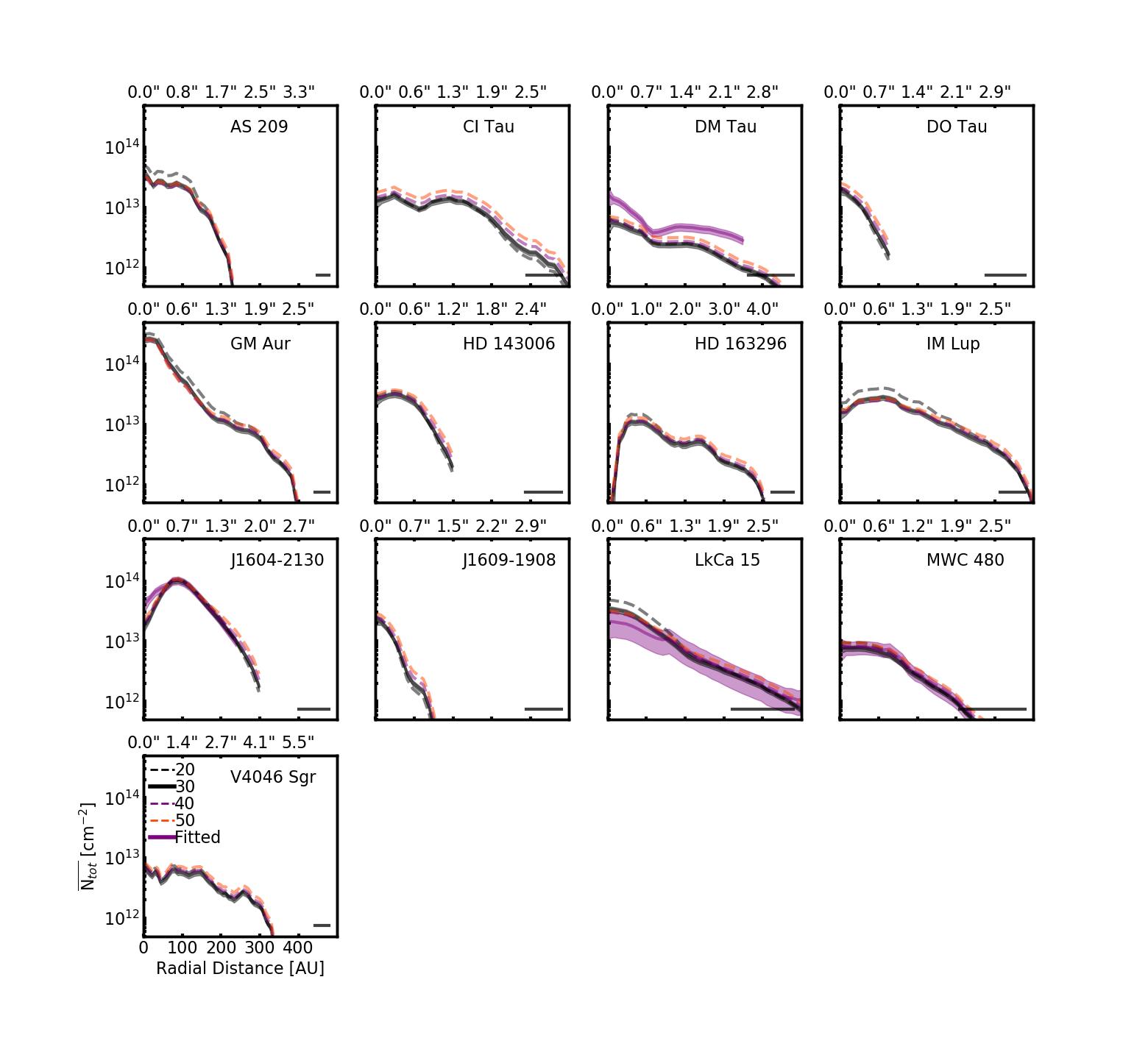}}
\caption{\textbf{Radial H$_2$CO Column Density Profiles for the Disk Sample.}  The panels show azimuthally-averaged H$_2$CO column densities for the disks with detected H$_2$CO 3$_{03}$-2$_{02}$ and 4$_{04}$-3$_{03}$ lines, as a function of radial distance from the host star.  Colors and line styles indicate different assumed excitation temperatures (listed in the bottom panel).  Fitted column densities from Figure~\ref{fig_trotfitrad} are overplotted in solid purple.  Errors, which include 10\% absolute flux uncertainties, are shown as shaded regions for the column densities at 30K and for the fitted column densities.   The average half-power beam widths are represented by the horizontal bars in the lower right corners.}
\label{fig_colestrad}
\end{figure*}

For each disk where a single H$_2$CO line was detected, we estimate a range of azimuthally-averaged column density profiles using Equation~\ref{eq_fitline} for H$_2$CO excitation temperatures between 20-50K.
Figure~\ref{fig_colestrad} displays the resulting H$_2$CO column density profiles, demonstrating that they are quite insensitive to excitation temperature assumptions in this range.  Across the disks, estimated column densities span nearly three orders of magnitude, from $\sim5 - 5000 \times 10^{11}$cm$^{-2}$.   The highest column densities are seen at about 25 AU and 100 AU for the disks GM Aur and J1604-2130, respectively; notably, these locations are near the inner edges of the holes in the disks' dust continuum.  The lowest column densities are seen for the disks HD 163296, MWC 480, and V4046 Sgr, which present low column densities across their entire disks.  Notably HD 163296 and MWC 480 are the only Herbig Ae disks in the sample, and V4046 Sgr is a very old disk.



\begin{figure*}[!htbp]
\centering
\resizebox{0.49\hsize}{!}{
    \includegraphics[trim=10pt 0pt 12pt 0pt, clip]{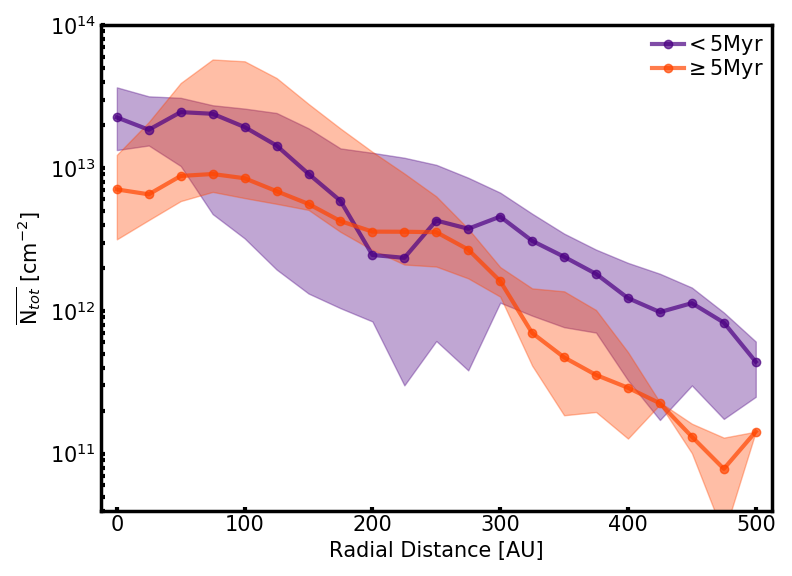}}
\resizebox{0.49\hsize}{!}{
    \includegraphics[trim=10pt 0pt 12pt 0pt, clip]{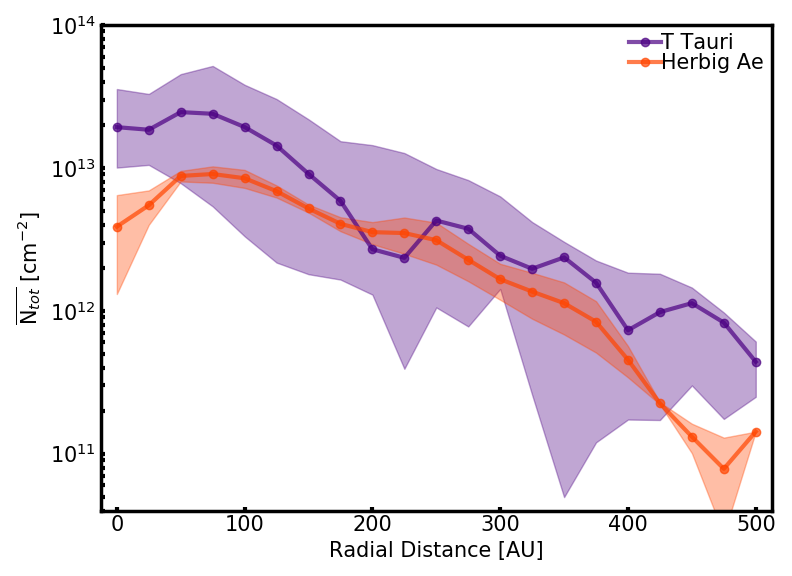}}
\caption{\textbf{Radial H$_2$CO Column Density Profiles Grouped by Stellar Characteristics}.  The panels plot statistical summaries of the azimuthally-averaged H$_2$CO column densities as a function of radial distance from the host star, assuming an excitation temperature of 30K (from Figure~\ref{fig_colestrad}).  The statistics are computed at intervals of 25 AU, and only for disks that are within the bounds of the Keplerian masks at that radius.  The dots indicate the median values for each group and are connected by solid lines to guide the eye.  The shaded regions span the 16$^{th}$-84$^{th}$ percentile ranges.  Colors group the column densities by stellar age on the left and stellar type on the right (from Table~\ref{table_majorstellarchar}).}
\label{fig_colvsallh2co}
\end{figure*}

Figure~\ref{fig_colvsallh2co} plots the median radial H$_2$CO column density profiles from Figure~\ref{fig_colestrad}, grouped by stellar age on the left and stellar type (T Tauri or Herbig Ae) on the right.  The right-hand panel of Figure~\ref{fig_colvsallh2co} shows that the median Herbig Ae H$_2$CO column densities are generally lower than the T Tauri medians.  Note that this trend is tentative, since there are only two Herbig Ae disks in the sample, and there is overlap between the T Tauri and Herbig Ae distributions.  The left-hand panel shows that the median H$_2$CO column densities for the old disks ($\geq$5Myr) are similar to the Herbig Ae medians, because both Herbig Ae disks are $\geq$5Myr.  However, given that the $\geq$5Myr disk J1604-2130 hosts some of the largest H$_2$CO column densities in the survey, it seems more likely that the tentative trend is explained by stellar type rather than stellar age.  There is no trend in H$_2$CO column density with dust continuum size (not shown).

\section{Discussion}
\label{sec_discussion}

\subsection{H$_2$CO Formation Chemistry}
\label{sec_discussion_chemistry}

Interpretations of our observed H$_2$CO morphologies and profiles rely on an understanding of the possible astrochemical origins of H$_2$CO in disks, which we briefly review here.

Astrochemical laboratory experiments and models have shown that H$_2$CO can form efficiently from both gas-phase and grain-surface chemistry.  In the gas phase, H$_2$CO can form through multiple pathways, including the neutral-neutral reaction~\citep[e.g.][]{cite_fockenbergetal2002, cite_atkinsonetal2006}:

\vspace{\baselineskip}
\centerline{
\noindent CH$_3$ + O $\longrightarrow$ H$_2$CO + H
}
\vspace{\baselineskip}

Under laboratory conditions from $\sim$290-930K, the H$_2$CO yield from this reaction is roughly constant~\citep[e.g.,][]{cite_fockenbergetal2002}.  Astrochemical disk modeling has shown that at low temperatures ($<$100K), this pathway is the dominant source of gas-phase H$_2$CO formation~\citep{cite_mareletal2014, cite_loomisetal2015}, and that gas-phase production of H$_2$CO in general is most efficient in the warm, dense inner regions of the disk~\citep{cite_loomisetal2015}.

On grain surfaces, H$_2$CO can form through the sequential hydrogenation of CO ice, which also produces CH$_3$OH if the hydrogenation continues~\citep[e.g.][]{cite_hiraokaetal1994, cite_hiraokaetal2002, cite_watanabeetal2002, cite_hidakaetal2004, cite_watanabeetal2004, cite_fuchsetal2009}:

\vspace{\baselineskip}
\centerline{
\noindent CO $\xrightarrow{\text{+H}}$ HCO $\xrightarrow{\text{+H}}$ H$_2$CO $\xrightarrow{\text{+H}}$ H$_2$COH $\xrightarrow{\text{+H}}$ CH$_3$OH
}
\vspace{\baselineskip}

We expect CO hydrogenation to produce H$_2$CO efficiently beyond the CO snowline.  However, since H$_2$CO grain-surface production is possible as long as some CO is on the grains long enough to react with atomic hydrogen, H$_2$CO grain-surface production likely also occurs somewhat interior to the CO snowline~\citep{cite_obergetal2017}.

H$_2$CO is less volatile than CO; the desorption energies of H$_2$CO off of different icy surfaces range from 3300-3700K~\citep{cite_nobleetal2012}, while the desorption energies of $^{13}$CO range from 900-1600K~\citep{cite_fayolleetal2016}.  Therefore H$_2$CO freezes out at temperatures higher than $\sim$80K, which is interior to the CO snowline at smaller radial distances in the disks than we are sensitive to.  However, frozen-out H$_2$CO beyond the H$_2$CO snowline can still be released into the gas via non-thermal desorption~\citep[e.g.][]{cite_walshetal2014, cite_loomisetal2015, cite_obergetal2017, cite_feraudetal2019}.


\subsection{H$_2$CO and the CO Snowline}
\label{sec_discussion_cosnowline}

\begin{deluxetable}{lll}[hb!]
\tablecaption{\textbf{CO Midplane Snowline Locations.}
\label{table_cosnowlocs}}
\tablehead{
Disk	& CO Midpl.	& Reference \\
	& Snowline 	&   \\
	& Range 	& \\
	& [AU] & 
}
\startdata
AS 209      & 50-111                     & \cite{cite_bergneretal2019}                \\
CI Tau      & 64-166                     & \cite{cite_bergneretal2019}                \\
DM Tau      & 45-85                      & \cite{cite_qietal2019}                     \\
DO Tau      & 76-310                     & \cite{cite_bergneretal2019}                \\
GM Aur      & 40-58                      & \cite{cite_qietal2019}                     \\
HD 143006   & 59-251                     & \cite{cite_bergneretal2019}                \\
HD 163296   & 69-83                      & \cite{cite_qietal2015}                   \\
IM Lup      & 38-69                      & \cite{cite_bergneretal2019}                \\
J1604-2130 & 43-64                      & \cite{cite_bergneretal2019}                \\
J1609-1908 & 36-73                      & \cite{cite_bergneretal2019}                \\
J1612-1859 & 36-78                      & \cite{cite_bergneretal2019}                \\
J1614-1906 & 31-71                      & \cite{cite_bergneretal2019}                \\
LkCa 15     & 48-64                      & \cite{cite_qietal2019}                     \\
MWC 480     & 90-150$^*$                    & Loomis et al. (subm.)        \\
V4046 Sgr   & 40-85                      & \cite{cite_bergneretal2019}               
\enddata
\tablecomments{{All locations have been scaled to new GAIA distances (Table~\ref{table_majorstellarchar}).  $*$: The CO snowline was estimated as 120 AU by Loomis et al. (subm.); here we broaden the snowline by $\pm$30 AU to account for the resolution of the data.}}
\end{deluxetable}

To constrain H$_2$CO formation pathways in our survey, we compare the observed H$_2$CO emission profiles with estimated locations of CO snowlines at the disk midplane.  Studies have inferred CO snowlines for disks using CO isotopologues and chemical tracers of CO freeze-out, and found freeze-out temperatures between $\sim$17-26$\sim$K~\citep[e.g.][]{cite_mathewsetal2013, cite_qietal2013, cite_qietal2015, cite_obergetal2015b, cite_schwarzetal2016, cite_pinteetal2018}.  For DM Tau, GM Aur, HD 163296, and LkCa 15, we use the CO snowline locations fitted using N$_2$H$^+$ as a tracer by~\cite{cite_qietal2015} and~\cite{cite_qietal2019}.  For the nine other disks with detected H$_2$CO, we estimate midplane snowlines at locations corresponding to a similar temperature range of 17-25K, extracted from the disk toy models of~\cite{cite_bergneretal2019} and Loomis et al. (subm.).  Note that these model-based estimates are very uncertain, and in at least two cases (AS 209 and V4046 Sgr) they overestimate the CO snowline location compared to the location based on N$_2$H$^+$~\citep{cite_qietal2019}.  Three disks DO Tau, HD 143006, and J1609-1908 have CO snowline ranges that are too broad to be informative, and we exclude them from further analysis.

Figure~\ref{fig_series32and43prof} compares the H$_2$CO profiles and the inferred CO snowline locations.  There is significant H$_2$CO emission both interior and exterior to the CO snowline of all disks.  A majority (6/10) of the H$_2$CO radial profiles peak interior to the CO snowline, but have rings or plateaus in H$_2$CO at or beyond the CO snowline.  Two of the remaining disks, IM Lup and J1604-2130, have H$_2$CO radial profiles that peak beyond the CO snowline.  One of the last two disks, MWC 480, may show more substructure at higher resolution.
The simplest explanation for these H$_2$CO morphologies is that H$_2$CO is produced through both gas-phase and grain-surface chemistry in disks.  The substantial central H$_2$CO emission found interior to the CO snowline implies an inner component of H$_2$CO produced in the gas phase and/or due to sublimation of inward-moving grains.  The rings and local enhancements in H$_2$CO emission found at or beyond the CO snowline are best explained by grain-surface production, since the efficiency of H$_2$CO gas-phase formation is expected to decrease with disk radius.

It is possible that H$_2$CO ices inherited from the preceding protostellar stage are contributing to the observed H$_2$CO emission in disks.  While we would expect inherited H$_2$CO ices to mainly contribute to emission in the inner disk, where H$_2$CO desorption would be efficient, we cannot exclude a smaller contribution in the outer disk, which may arise from the mixing of midplane ices into the warm upper layers of the disk.  However, there should be no relationship between this emission and the CO snowline, which suggests that inheritance alone cannot explain the observed H$_2$CO emission morphologies.

Our results thus indicate that the two-component H$_2$CO model discussed by~\cite{cite_loomisetal2015},~\cite{cite_carneyetal2017}, and~\cite{cite_obergetal2017} for three specific disks, containing one warm inner component and one cold component exterior to the CO snowline, likely holds for disks generally.  Furthermore, it is likely that H$_2$CO is commonly produced through both gas-phase and grain-surface chemistry in disks.

We expect that the H$_2$CO formed through CO hydrogenation beyond the CO snowline will coincide with the formation of complex oxygen-bearing organics, particularly methanol (CH$_3$OH), as these molecules can form along the same hydrogenation pathway as H$_2$CO~\citep[e.g.,][]{cite_hiraokaetal1994, cite_watanabeetal2002}. However, since there is also some contribution from gas-phase chemistry to the total H$_2$CO budget at these radii, it is difficult to assess how much of these COMs are present. Detailed modeling, as well as a few benchmarking disks with both H$_2$CO and COM detections, are needed to quantify the relationship between H$_2$CO and COM production in disks. 
In the meantime, we conclude that a majority of disks do have an active organic ice chemistry, and that COM production should be expected in the outer regions of most protoplanetary disks.

The exception to this conclusion may be disks around Herbig Ae stars, because their warmer temperatures allow for limited CO freeze-out. Our survey tentatively shows that T Tauri disks have more H$_2$CO than Herbig Ae disks, despite how the Herbig Ae disks present the highest C$^{18}$O flux densities. This finding is consistent with the single-dish survey of H$_2$CO in disks conducted by~\cite{cite_guilloteauetal2013}, which showed that warmer disks tended to have less H$_2$CO emission.  It is also consistent with the Submillimeter Array (SMA) survey conducted by~\cite{cite_obergetal2010}, which noted that the lower detection rates towards Herbig Ae disks could be because their midplanes are too warm for a cold CO-ice based chemistry.  
Therefore our results could be explained by a small or non-existing production of H$_2$CO through CO ice hydrogenation in Herbig Ae disks.  More spatially-resolved observations of H$_2$CO towards Herbig Ae disks are needed to further evaluate this tentative trend.

\subsection{Relationship between H$_2$CO and Dust Substructure}
\label{sec_discussion_cont}

\cite{cite_carneyetal2017} and \cite{cite_obergetal2017} note three ways that H$_2$CO could be affected by a hole or edge in the dust continuum: (1) through enhanced temperatures due to greater irradiation in regions with fewer solids~\citep{cite_cleevesetal2011, cite_cleeves2016}; (2) through increased photodesorption of H$_2$CO off the grains from the increased irradiation~\citep{cite_obergetal2015b}; and/or (3) through increased CO photodissociation in the upper layers, since more CO in the upper layers could produce more hydrocarbon radicals and atomic oxygen in the gas, and then possibly allow for more efficient gas-phase production of H$_2$CO~\citep{cite_carneyetal2017, cite_obergetal2017}.  
Five of the disks in our survey (DM Tau, GM Aur, J1604-2130, LkCa 15, and V4046 Sgr) have inner holes in the submillimeter/millimeter dust continuum that are detected at our spatial resolution.  For these five disks, we see no universal trend in H$_2$CO emission morphologies in the inner disks.  Four of the five disks show centrally-peaked H$_2$CO emission despite the central depletions in dust continuum.  The final disk, J1604-2130, does show an inner hole in the H$_2$CO emission.  Notably J1604-2130 also has a large inner hole in its C$^{18}$O emission \citep[see also][]{cite_zhangetal2014, cite_mareletal2015}.  We also do not see a consistent trend in H$_2$CO emission morphologies at the outer dust continuum edges, but four of the disks show clear bumps in H$_2$CO emission right at those edges. It therefore appears that, while dust substructure does influence the distribution of H$_2$CO in disks, there is no single relationship between the two.



\subsection{H$_2$CO Excitation Temperatures}
\label{sec_discussion_temperatures}

The H$_2$CO excitation temperature should reflect the temperature of the vertical layer in the disk within which the gas resides, as long as the densities are high enough for LTE.  T Tauri disk models predict that beyond $\sim$50 AU (the scale we are most sensitive to), the midplane will be 10-30K, the molecular layer 20-50K, and the atmosphere 50-500K ~\citep[e.g.,][]{cite_walshetal2010}.  Herbig Ae disks will have warmer temperatures at the same radial distances, while low-mass M-star disks will have colder temperatures. In the four disks where we could measure H$_2$CO excitation temperatures, three are consistent with the warm molecular layer, and one (DM Tau) with the disk midplane, indicating that the vertical distribution of H$_2$CO may vary between disks.

Our radially-resolved rotational diagram analysis is largely consistent with a constant radial temperature per disk in the outer disks beyond 100-200 AU.  This suggests that for three of the four disks, H$_2$CO is consistently emitting from the molecular layer at/above CO freeze-out temperatures.  This is in accordance with model predictions, which find the most abundant H$_2$CO in the molecular layer above 20K~\citep{cite_walshetal2010}.  However, since error bars are currently large (typically 10$\sim$K or more), we need more precise observations in order to better constrain radial H$_2$CO excitation temperatures.


\section{Summary}
\label{sec_summary}

We have conducted an ALMA survey of H$_2$CO towards 15 protoplanetary disks.  We have analyzed the H$_2$CO morphologies, excitation temperatures, and column densities, and their relationships to stellar and disk characteristics.  We summarize our main findings below:

\begin{enumerate}
	\item We report a total of 26 H$_2$CO detections towards 13/15 disks, consisting of one line towards 8 disks, two lines towards 1 disk, three lines towards 2 disks, four lines towards 1 disk, and six lines towards 1 disk.  We report an additional 11 tentative H$_2$CO detections across 7/15 disks. 
	\item Over half of the disks have centrally-peaked H$_2$CO emission, while the rest have central dips or holes in H$_2$CO emission.  About half of the disks show rings or a plateau in H$_2$CO emission in the outer disk.
    \item We measure disk-averaged and azimuthally-averaged H$_2$CO excitation temperatures for four disks with multiple H$_2$CO line detections.  Three of the four disks have disk-averaged excitation temperatures of 21-37K, while the fourth disk, DM Tau, is colder at 11K.  The azimuthally-averaged temperatures show an increase at the inner dust edge of J1604-2130, but otherwise close to constant radial temperatures for the disks.
    \item In addition to the disks we could directly model, we estimate radial H$_2$CO column densities for all disks assuming excitation temperatures of 20-50K.  The column densities between and across the disks of our survey span three orders of magnitude, from $\sim5-5000 \times 10^{11} \mathrm{cm}^{-2}$.  The highest H$_2$CO column densities are found near the inner dust continuum edges of the disks GM Aur and J1604-2130.  The lowest H$_2$CO column densities are found across the Herbig Ae disks HD 163296 and MWC 480, as well as in the old disk V4046 Sgr.
    \item Based on simple estimates of CO midplane snowline locations, significant H$_2$CO emission is present both interior and exterior to the CO snowline in all disks, with ring-like structures in the emission occurring beyond the CO snowline in half of the disks.  This suggests that H$_2$CO is commonly formed through both gas-phase and grain-surface pathways in protoplanetary disks, and that a majority of disks present an active organic ice chemistry that is likely also producing COMs. The low H$_2$CO column densities toward the two Herbig Ae disks in the sample may be due to a smaller degree of CO ice chemistry in these warmer disks.
\end{enumerate}

\acknowledgments

This paper has benefited from discussions with Alyssa Goodman, Dimitar Sasselov, and David Wilner.

Jamila Pegues gratefully acknowledges the support of the National Science Foundation (NSF) Graduate Research Fellowship through Grant Numbers DGE1144152 and DGE1745303.
Karin \"Oberg gratefully acknowledges the support of the Simons Foundation through a Simons Collaboration on the Origins of Life (SCOL) PI Grant (Number 321183).
Jane Huang gratefully acknowledges support from the NSF Graduate Research Fellowship under Grant No. DGE-1144152.
Geoffrey Blake gratefully acknowledges support from the NSF (Grant AST-1514670) and NASA (Grant NNX16AB48G).
Jes K. J{\o}rgensen gratefully acknowledges support from the European Research Council (ERC) under the European Union’s Horizon 2020 research and innovation programme through ERC Consolidator Grant "S4F" (Grant Agreement No. 646908).
Kamber Schwarz, a Sagan Fellow, gratefully acknowledges the support of NASA through Hubble Fellowship Program Grant HST-HF2-51419.001, awarded by the Space Telescope Science Institute, which is operated by the Association of Universities for Research in Astronomy, Inc., for NASA, under contract NAS5-26555.

This paper makes use of the following ALMA data:
\vspace{-3pt}
\begin{itemize}
\itemsep-0.5em
    \item ADS/JAO.ALMA\#2011.0.00629.S
    \item ADS/JAO.ALMA\#2012.1.00681.S
    \item ADS/JAO.ALMA\#2013.1.00226.S
    \item ADS/JAO.ALMA\#2015.1.00657.S
    \item ADS/JAO.ALMA\#2015.1.00678.S
    \item ADS/JAO.ALMA\#2015.1.00964.S
    \item ADS/JAO.ALMA\#2016.1.00627.S.
\end{itemize}
\vspace{-3pt}

 ALMA is a partnership of ESO (representing its member states), NSF (USA) and NINS (Japan), together with NRC (Canada), MOST and ASIAA (Taiwan), and KASI (Republic of Korea), in cooperation with the Republic of Chile. The Joint ALMA Observatory is operated by ESO, AUI/NRAO and NAOJ.  The National Radio Astronomy Observatory is a facility of the National Science Foundation operated under cooperative agreement by Associated Universities, Inc.
 
This work has made use of data from the European Space Agency (ESA) mission
{\it Gaia} (\url{https://www.cosmos.esa.int/gaia}), processed by the {\it Gaia}
Data Processing and Analysis Consortium (DPAC,
\url{https://www.cosmos.esa.int/web/gaia/dpac/consortium}). Funding for the DPAC
has been provided by national institutions, in particular the institutions
participating in the {\it Gaia} Multilateral Agreement.

All computer code used for this research was written in Python (version 2.7).  All plots were generated using Python's Matplotlib package~\citep{cite_matplotlib}.  This research also made use of Astropy (\url{http://www.astropy.org}), a community-developed core Python package for Astronomy, and the NumPy~\citep{cite_numpy} and SciPy~\citep{cite_scipy} Python packages. 


\appendix


\setcounter{section}{1}
\setcounter{figure}{0}
\setcounter{table}{0}


\figsetstart
\label{figset_h2comain}
\figsetnum{A1}
\figsettitle{Channel maps of detected H$_2$CO 3$_{03}$-2$_{02}$ and 4$_{04}$-3$_{03}$ above 2$\sigma$.}

\figsetgrpstart
\figsetgrpnum{A1.1}
\figsetgrptitle{}
\figsetplot{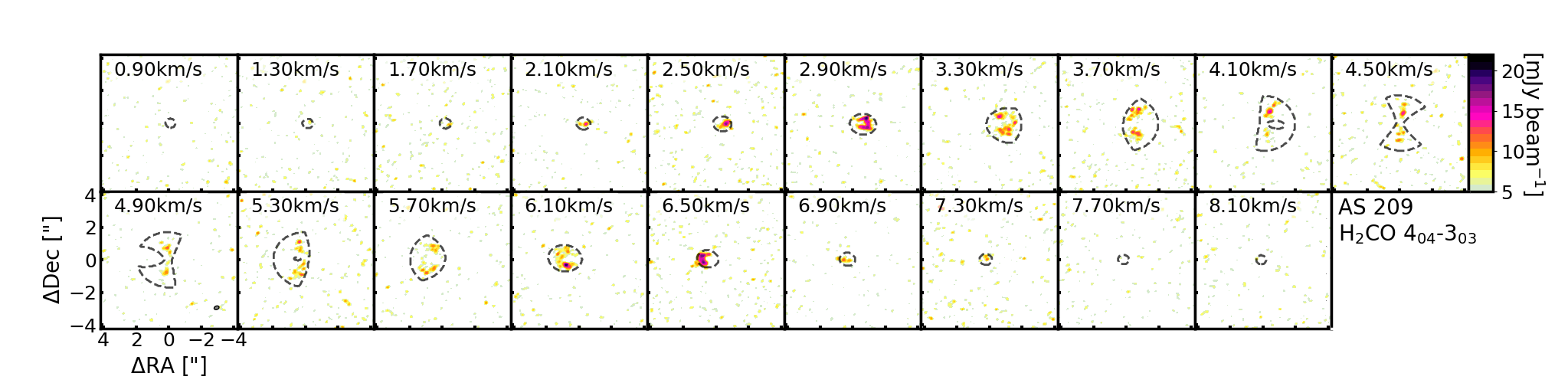}
\figsetgrpnote{H$_2$CO 4$_{04}$-3$_{03}$ towards AS 209 above 2$\sigma$.}
\figsetgrpend

\figsetgrpstart
\figsetgrpnum{A1.2}
\figsetgrptitle{}
\figsetplot{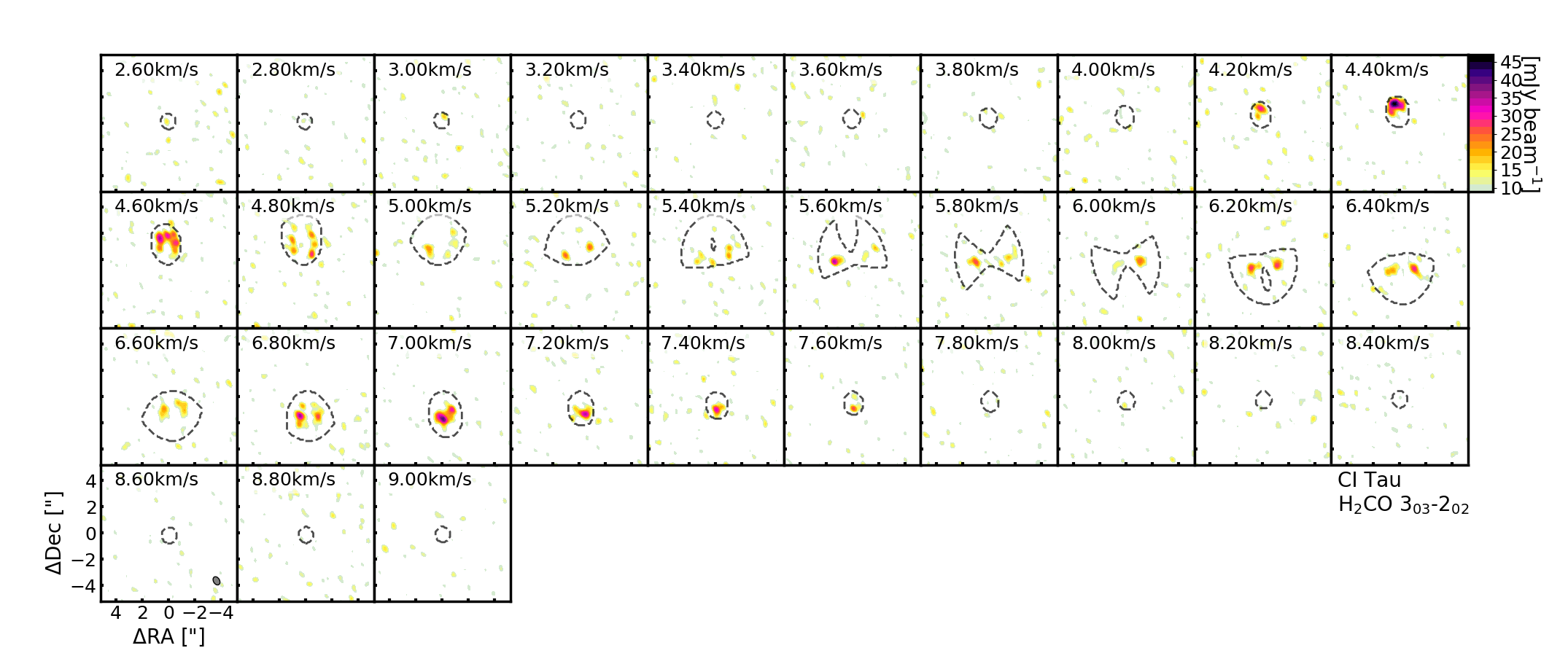}
\figsetgrpnote{H$_2$CO 3$_{03}$-2$_{02}$ towards CI Tau above 2$\sigma$.}
\figsetgrpend

\figsetgrpstart
\figsetgrpnum{A1.3}
\figsetgrptitle{}
\figsetplot{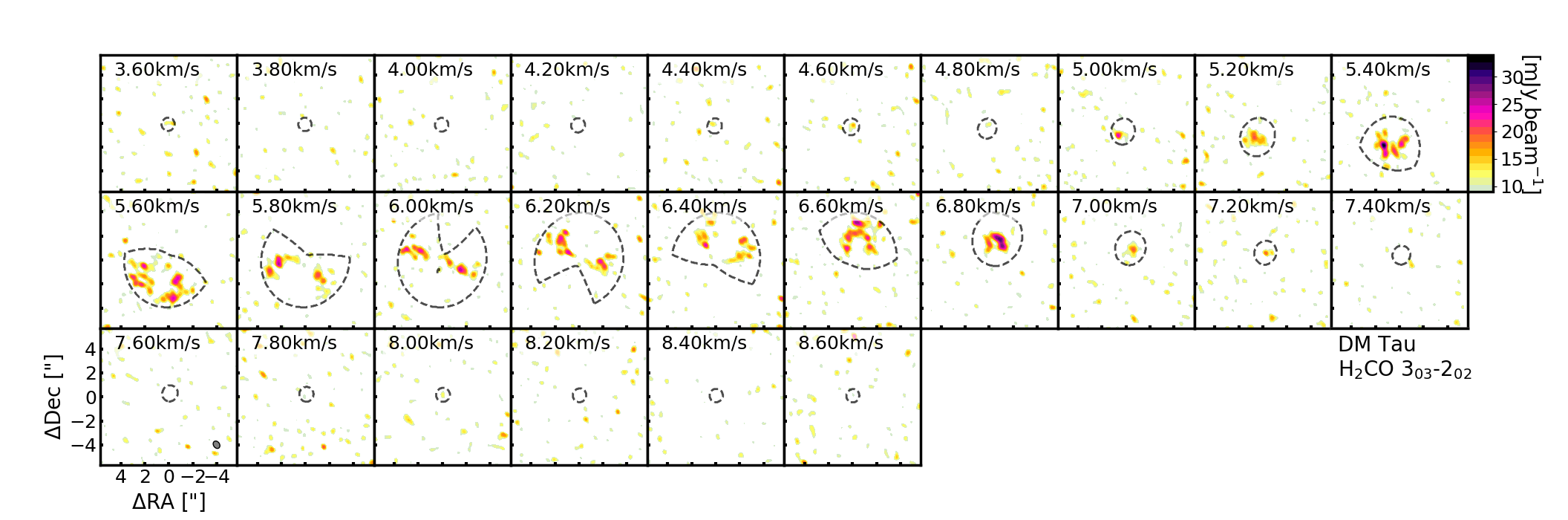}
\figsetgrpnote{H$_2$CO 3$_{03}$-2$_{02}$ towards DM Tau above 2$\sigma$.}
\figsetgrpend

\figsetgrpstart
\figsetgrpnum{A1.4}
\figsetgrptitle{}
\figsetplot{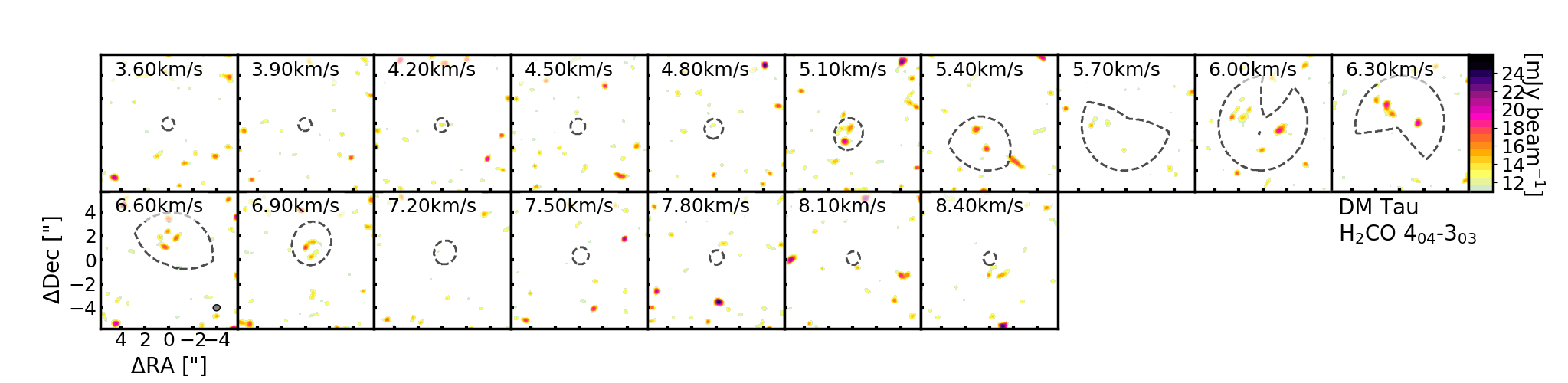}
\figsetgrpnote{H$_2$CO 4$_{04}$-3$_{03}$ towards DM Tau above 2$\sigma$.}
\figsetgrpend

\figsetgrpstart
\figsetgrpnum{A1.5}
\figsetgrptitle{}
\figsetplot{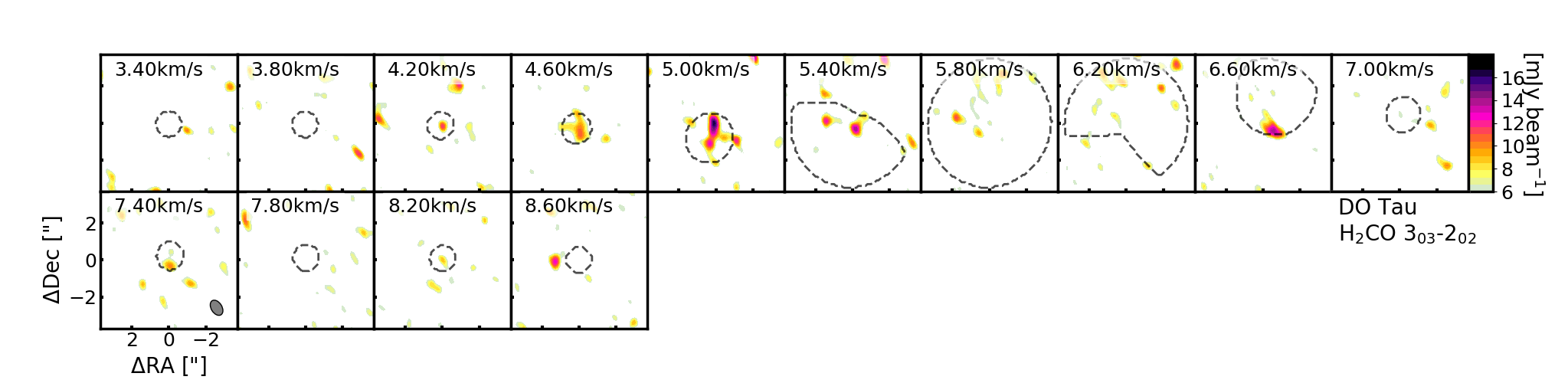}
\figsetgrpnote{H$_2$CO 3$_{03}$-2$_{02}$ towards DO Tau above 2$\sigma$.}
\figsetgrpend

\figsetgrpstart
\figsetgrpnum{A1.6}
\figsetgrptitle{}
\figsetplot{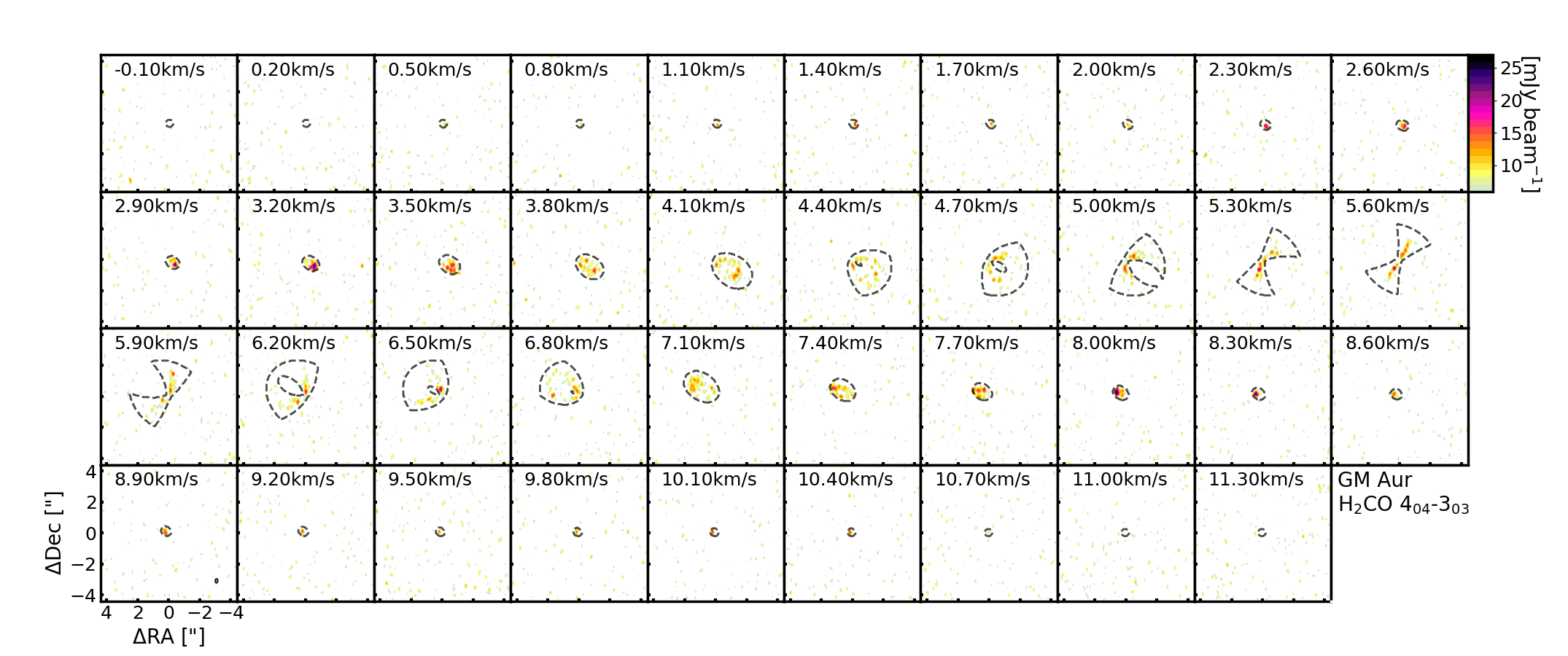}
\figsetgrpnote{H$_2$CO 4$_{04}$-3$_{03}$ towards GM Aur above 2$\sigma$.}
\figsetgrpend

\figsetgrpstart
\figsetgrpnum{A1.7}
\figsetgrptitle{}
\figsetplot{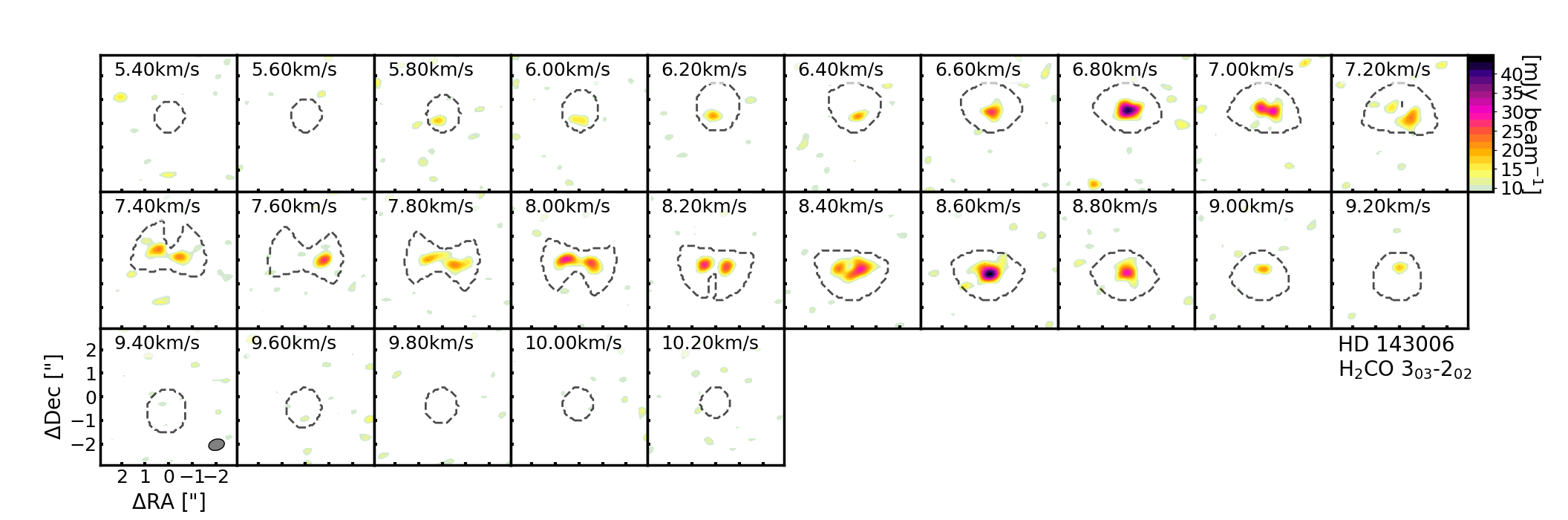}
\figsetgrpnote{H$_2$CO 3$_{03}$-2$_{02}$ towards HD 143006 above 2$\sigma$.}
\figsetgrpend

\figsetgrpstart
\figsetgrpnum{A1.8}
\figsetgrptitle{}
\figsetplot{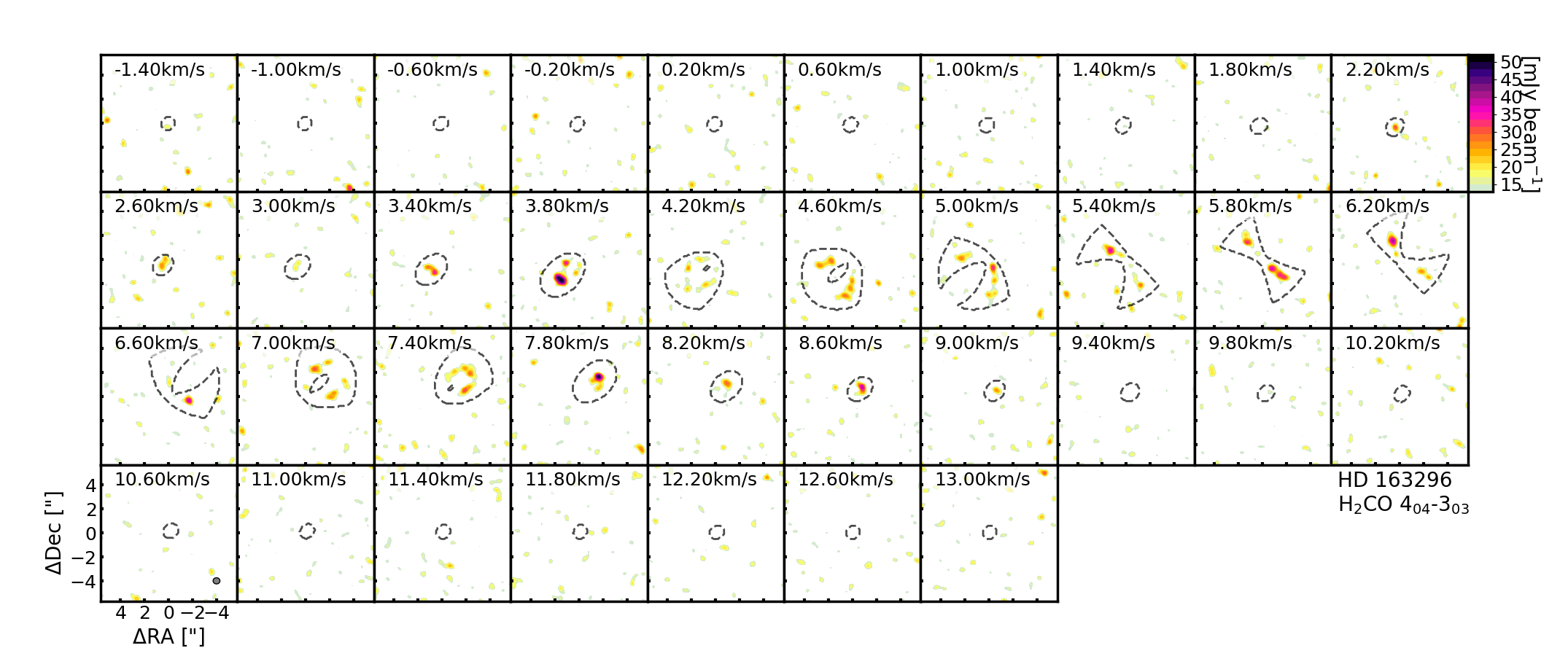}
\figsetgrpnote{H$_2$CO 4$_{04}$-3$_{03}$ towards HD 163296 above 2$\sigma$.}
\figsetgrpend

\figsetgrpstart
\figsetgrpnum{A1.9}
\figsetgrptitle{}
\figsetplot{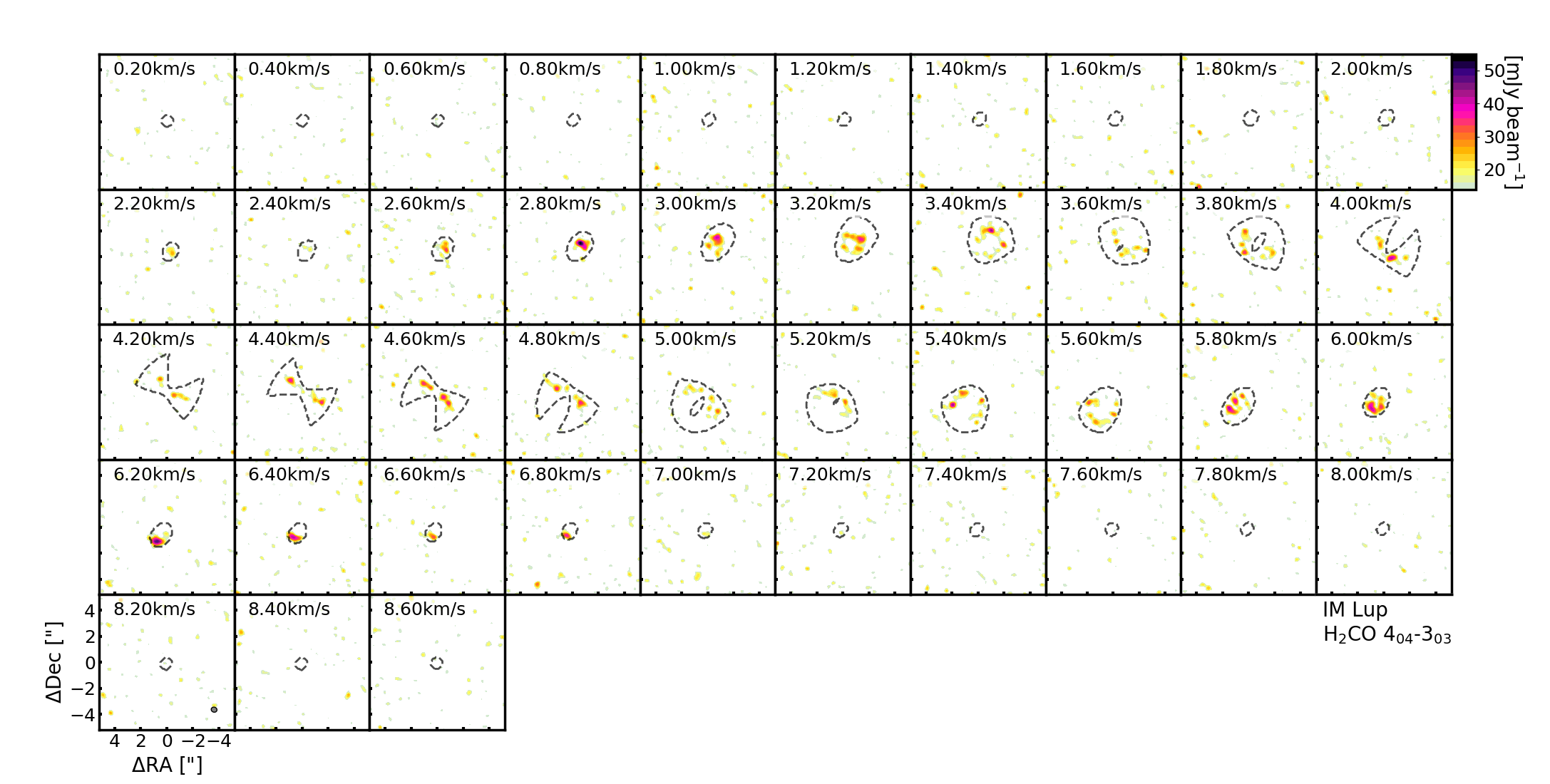}
\figsetgrpnote{H$_2$CO 4$_{04}$-3$_{03}$ towards IM Lup above 2$\sigma$.}
\figsetgrpend

\figsetgrpstart
\figsetgrpnum{A1.10}
\figsetgrptitle{}
\figsetplot{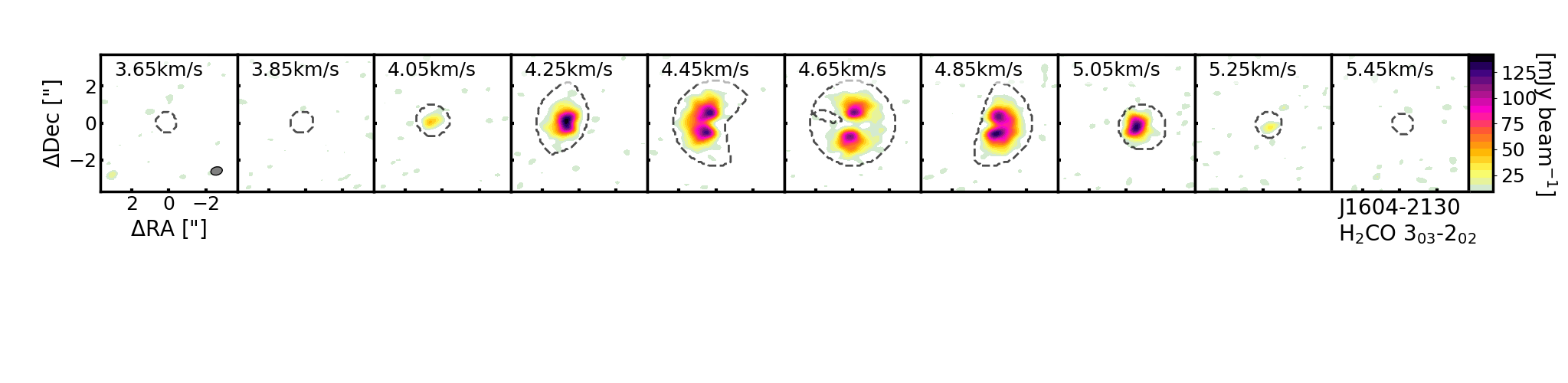}
\figsetgrpnote{H$_2$CO 3$_{03}$-2$_{02}$ towards J1604-2130 above 2$\sigma$.}
\figsetgrpend

\figsetgrpstart
\figsetgrpnum{A1.11}
\figsetgrptitle{}
\figsetplot{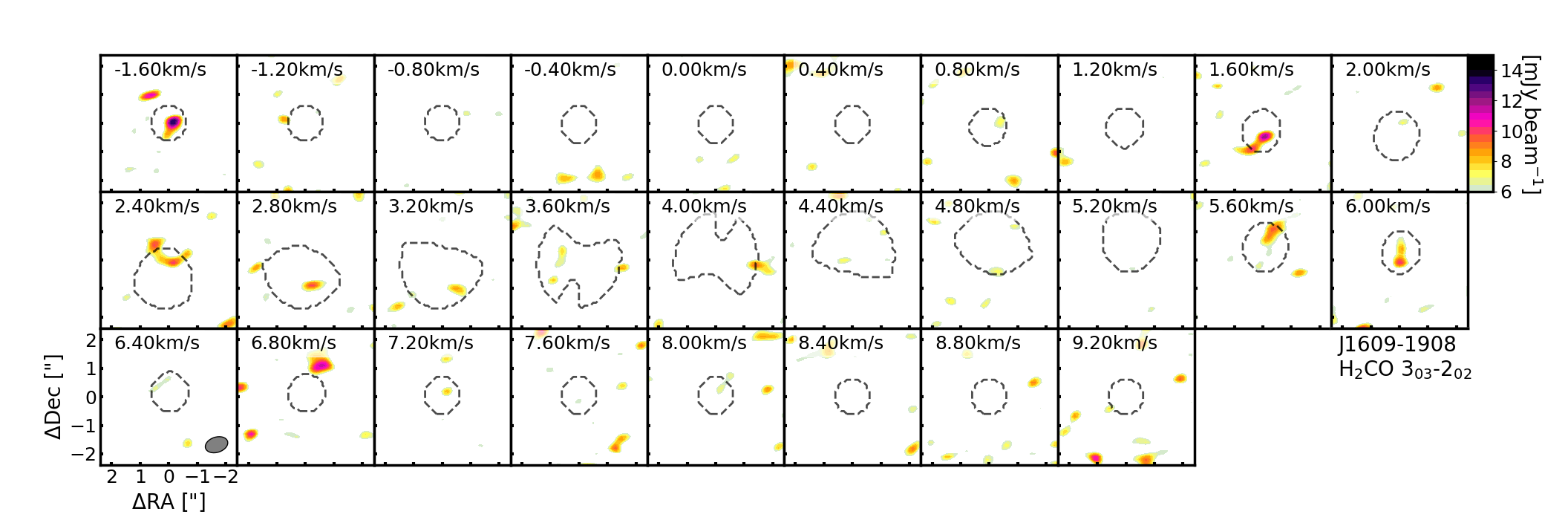}
\figsetgrpnote{H$_2$CO 3$_{03}$-2$_{02}$ towards J1609-1908 above 2$\sigma$.}
\figsetgrpend


\figsetgrpstart
\figsetgrpnum{A1.12}
\figsetgrptitle{}
\figsetplot{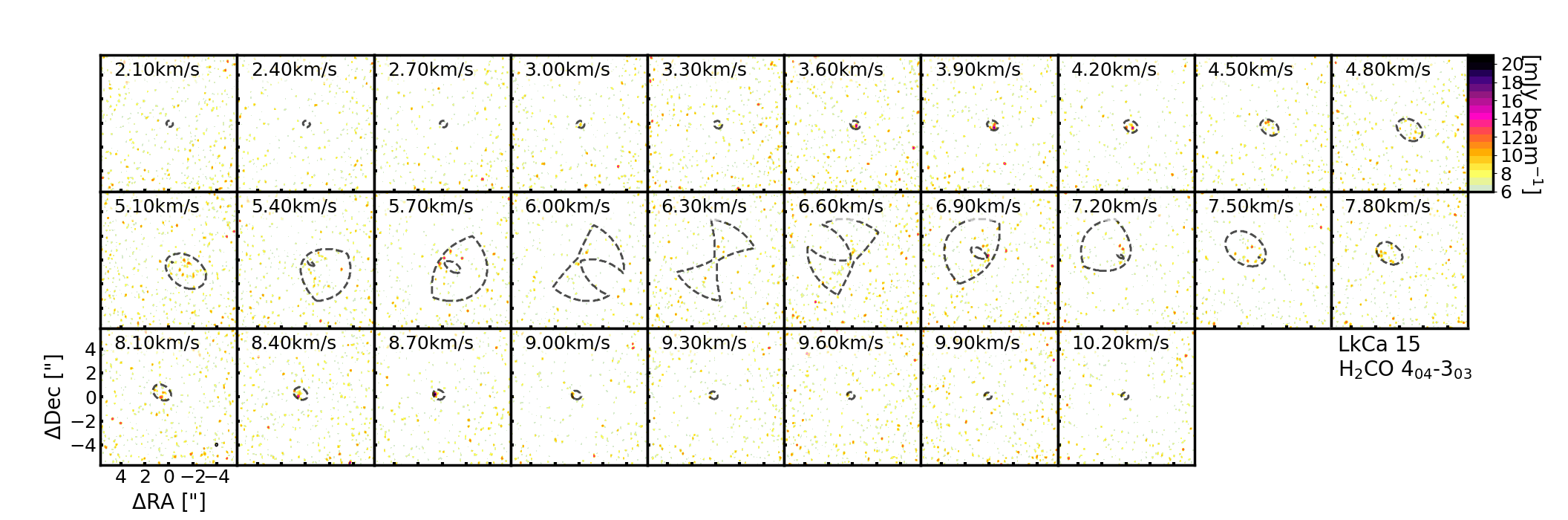}
\figsetgrpnote{H$_2$CO 4$_{04}$-3$_{03}$ towards LkCa 15 above 2$\sigma$.}
\figsetgrpend

\figsetgrpstart
\figsetgrpnum{A1.13}
\figsetgrptitle{}
\figsetplot{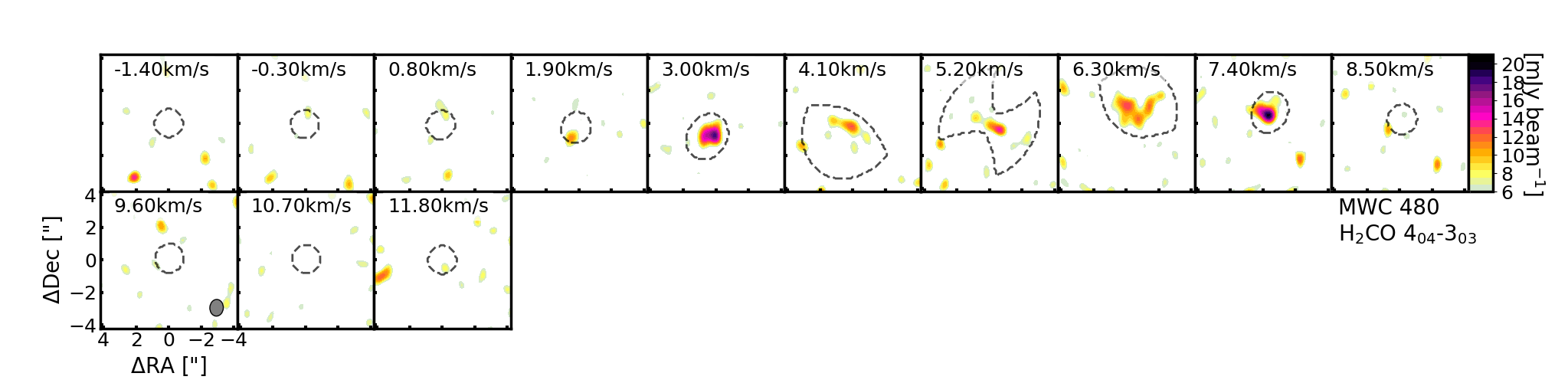}
\figsetgrpnote{H$_2$CO 4$_{04}$-3$_{03}$ towards MWC 480 above 2$\sigma$.}
\figsetgrpend

\figsetgrpstart
\figsetgrpnum{A1.14}
\figsetgrptitle{}
\figsetplot{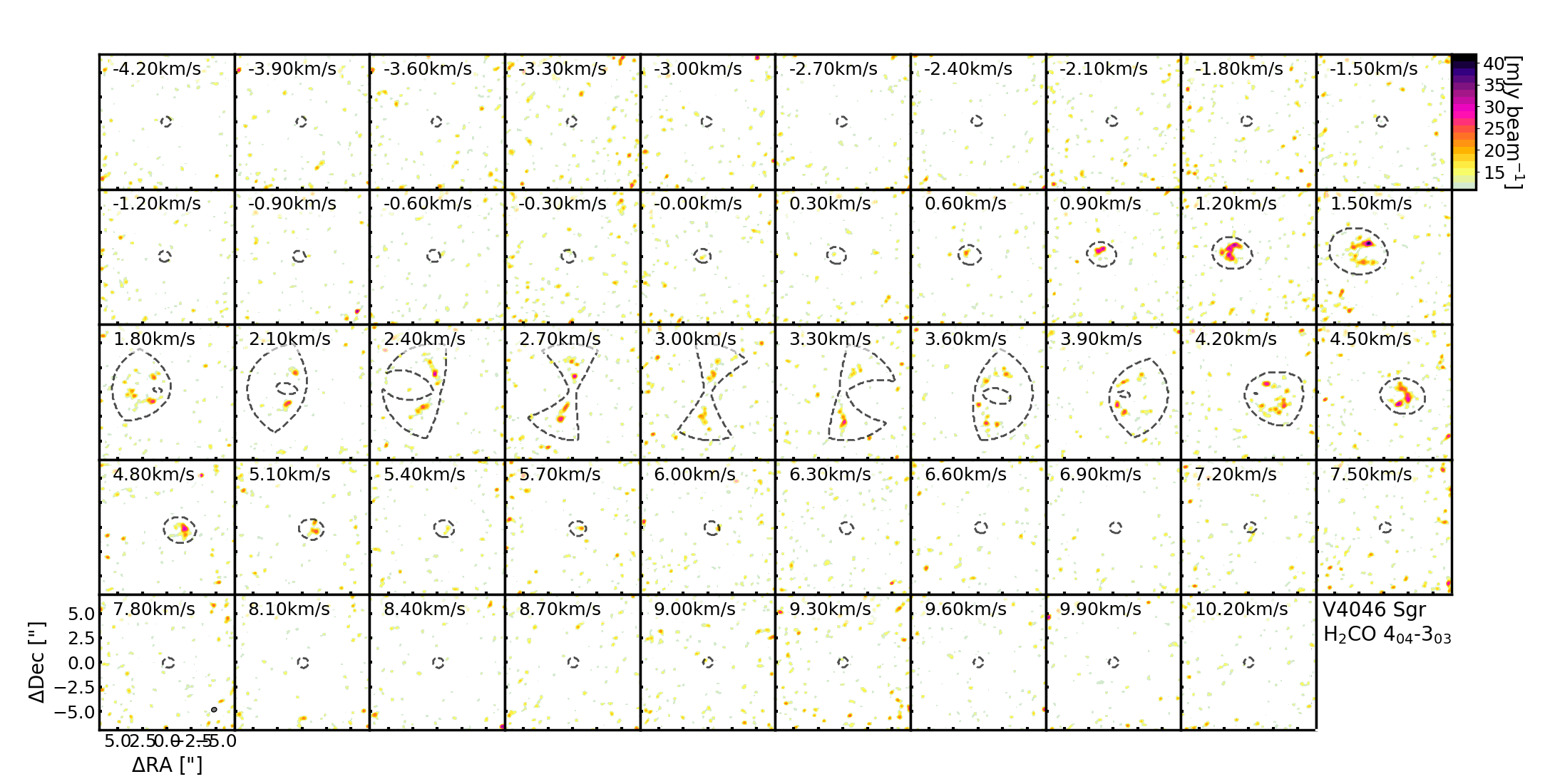}
\figsetgrpnote{H$_2$CO 4$_{04}$-3$_{03}$ towards V4046 Sgr above 2$\sigma$.}
\figsetgrpend

\figsetend


%

\figsetstart
\label{figset_c18o}
\figsetnum{A2}
\figsettitle{Channel maps of detected C$^{18}$O above 2$\sigma$.}

\figsetgrpstart
\figsetgrpnum{A2.1}
\figsetgrptitle{}
\figsetplot{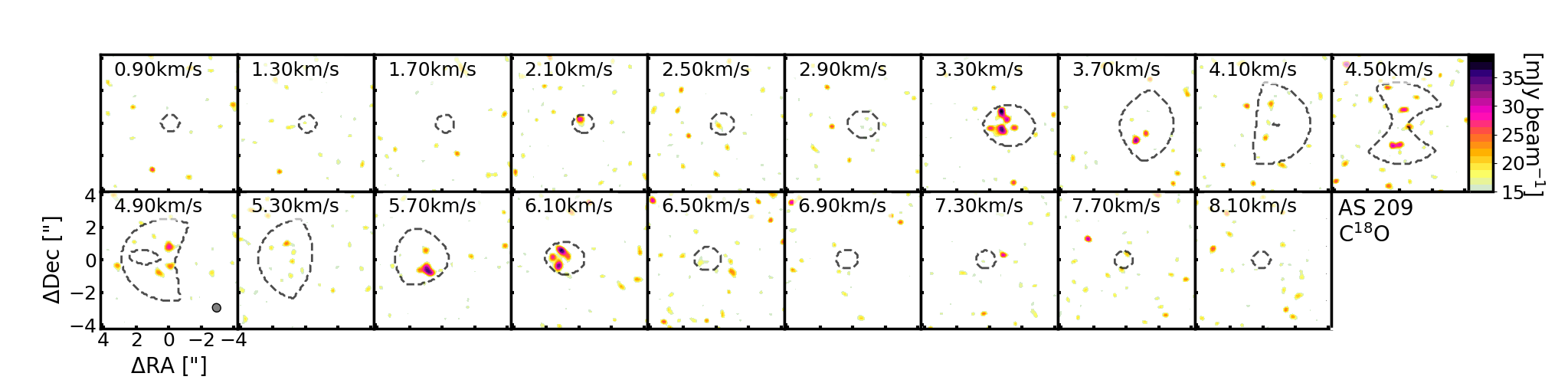}
\figsetgrpnote{C$^{18}$O towards AS 209 above 2$\sigma$.}
\figsetgrpend

\figsetgrpstart
\figsetgrpnum{A2.2}
\figsetgrptitle{}
\figsetplot{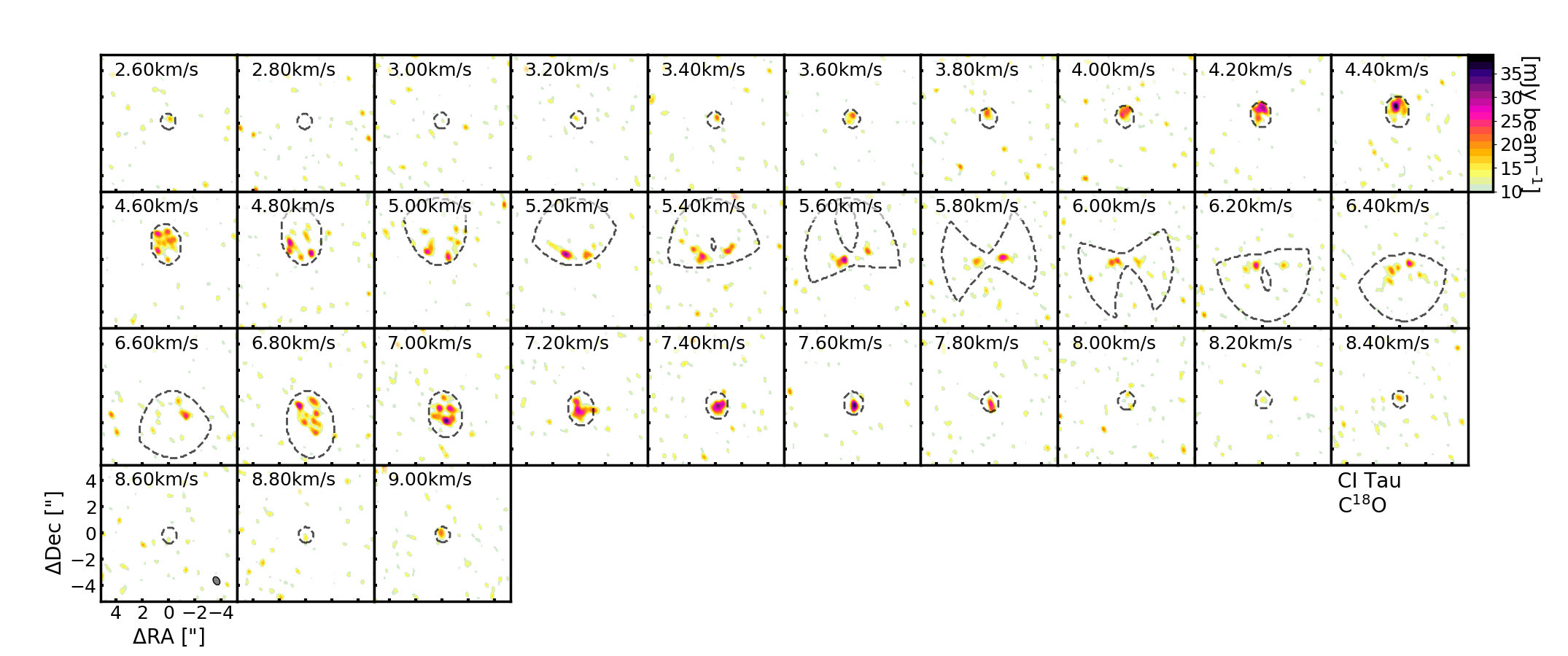}
\figsetgrpnote{C$^{18}$O towards CI Tau above 2$\sigma$.}
\figsetgrpend

\figsetgrpstart
\figsetgrpnum{A2.3}
\figsetgrptitle{}
\figsetplot{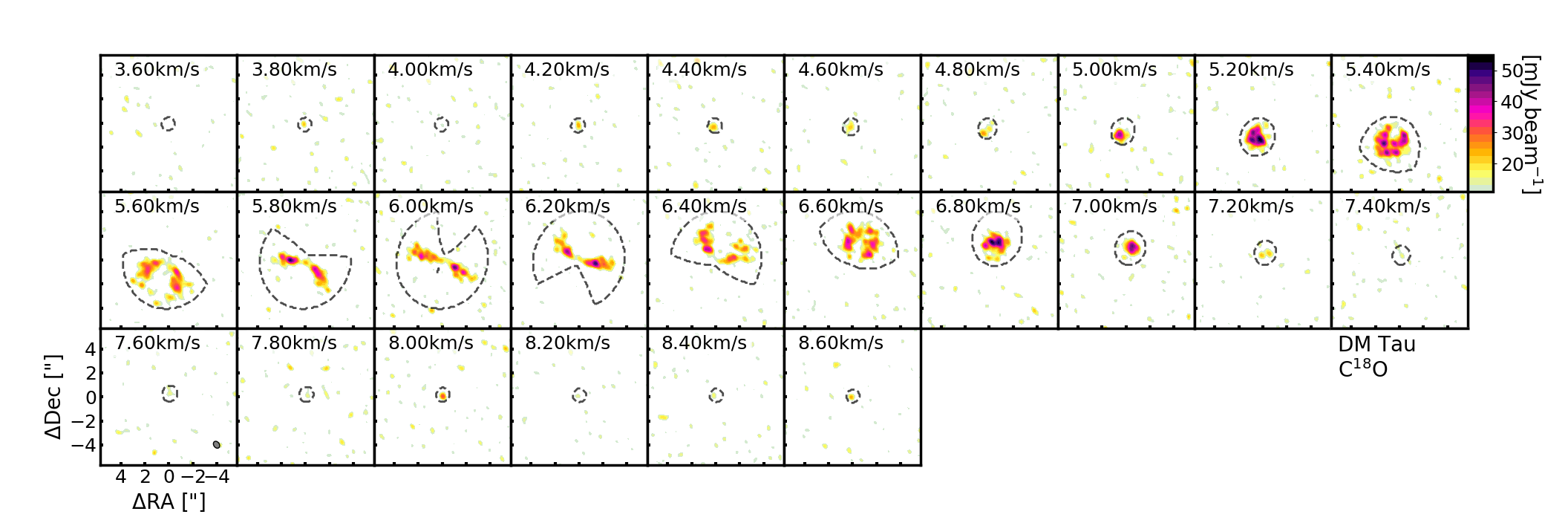}
\figsetgrpnote{C$^{18}$O towards DM Tau above 2$\sigma$.}
\figsetgrpend

\figsetgrpstart
\figsetgrpnum{A2.4}
\figsetgrptitle{}
\figsetplot{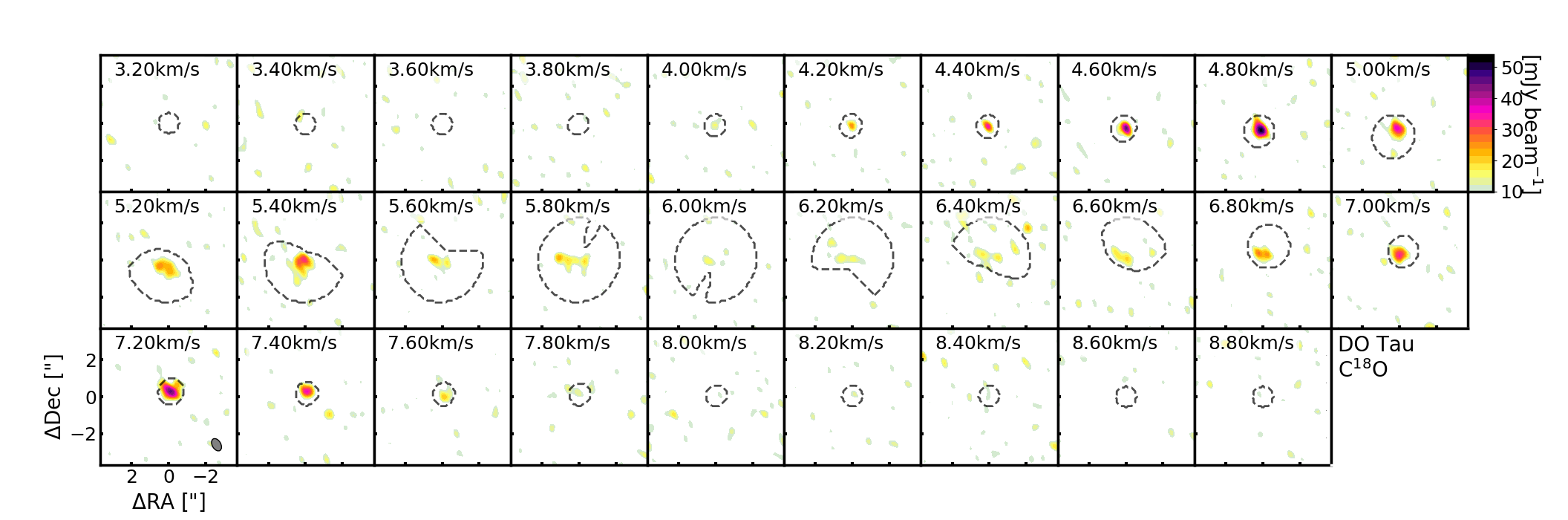}
\figsetgrpnote{C$^{18}$O towards DO Tau above 2$\sigma$.}
\figsetgrpend

\figsetgrpstart
\figsetgrpnum{A2.5}
\figsetgrptitle{C$^{18}$O towards HD 143006 above 2$\sigma$.}
\figsetplot{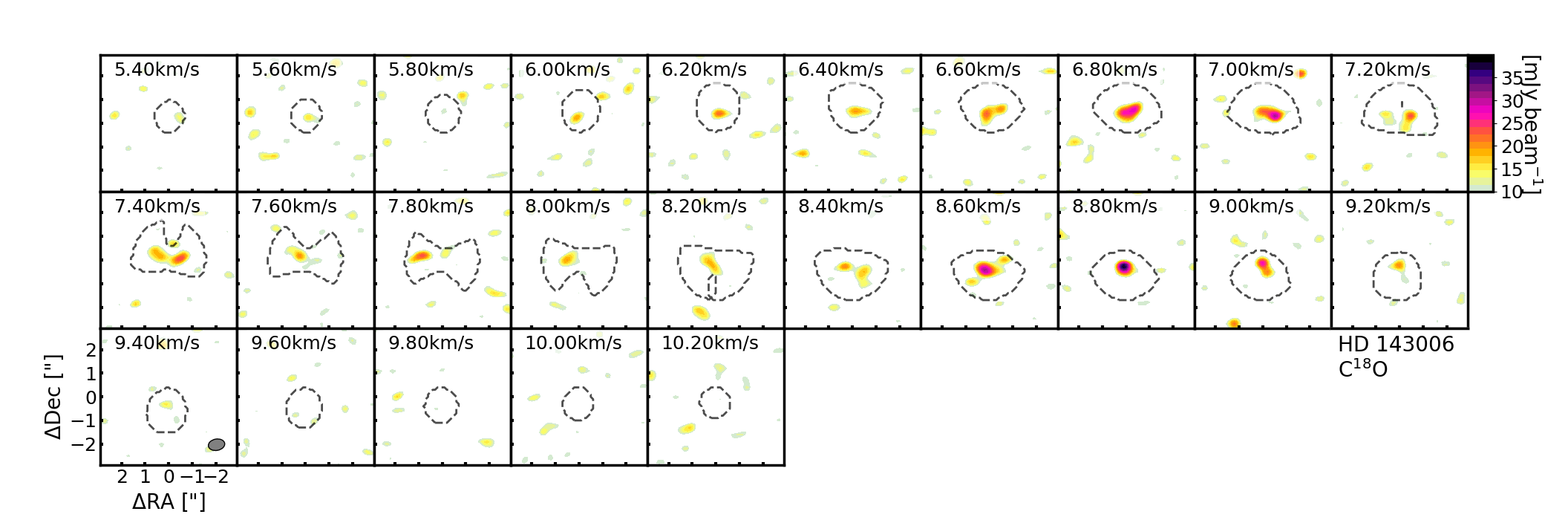}
\figsetgrpnote{ }
\figsetgrpend

\figsetgrpstart
\figsetgrpnum{A2.6}
\figsetgrptitle{}
\figsetplot{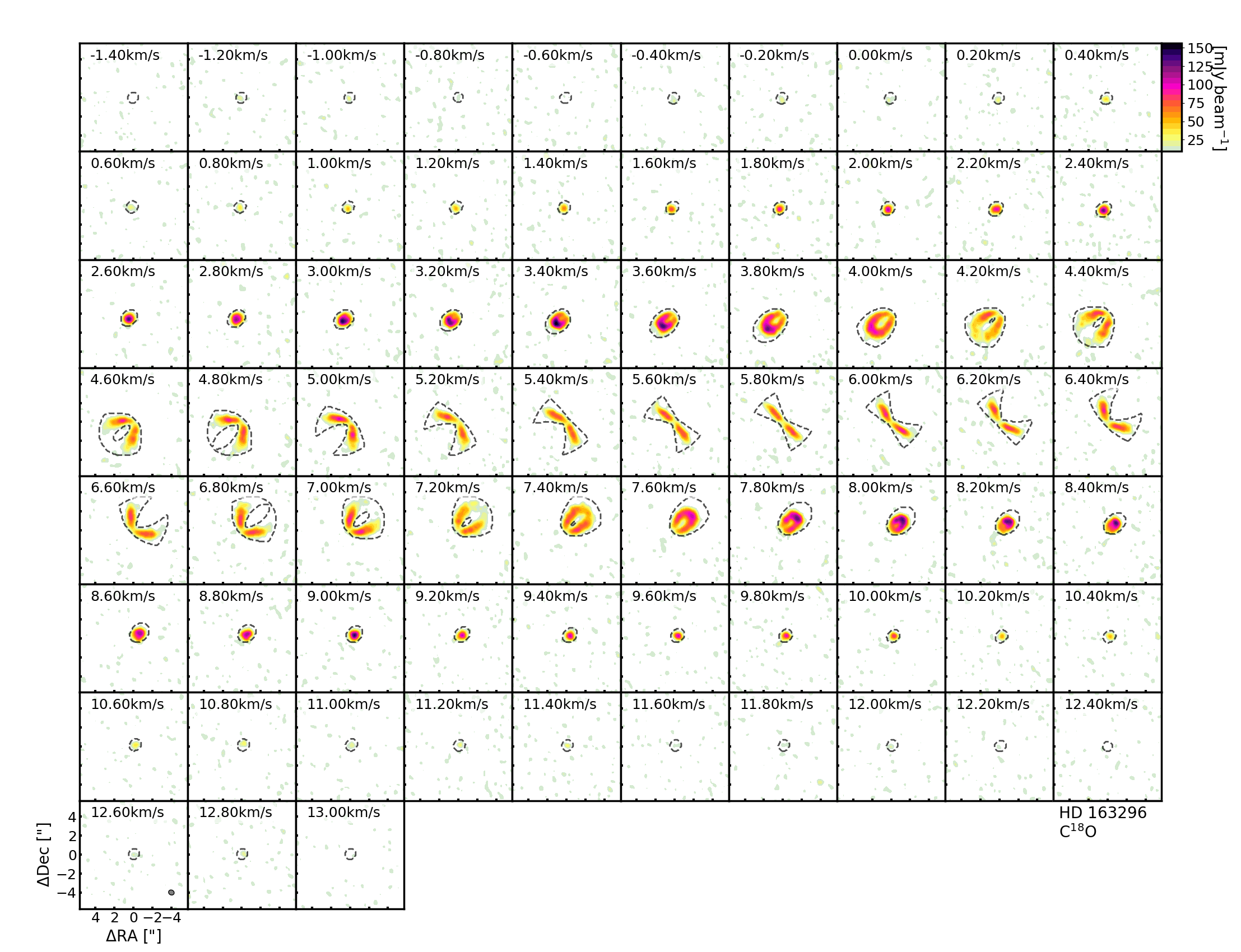}
\figsetgrpnote{C$^{18}$O towards HD 163296 above 2$\sigma$.}
\figsetgrpend

\figsetgrpstart
\figsetgrpnum{A2.7}
\figsetgrptitle{}
\figsetplot{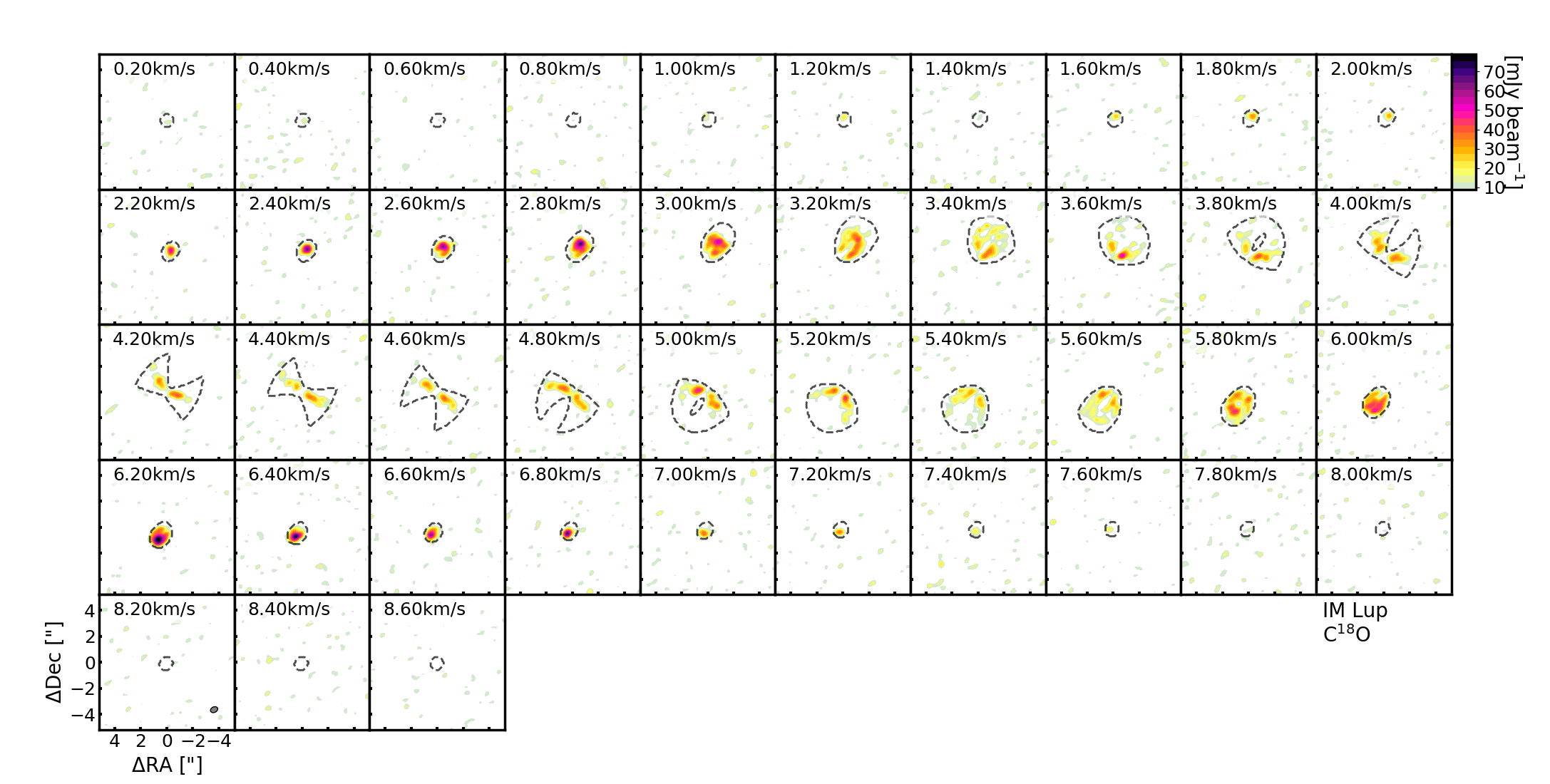}
\figsetgrpnote{C$^{18}$O towards IM Lup above 2$\sigma$.}
\figsetgrpend

\figsetgrpstart
\figsetgrpnum{A2.8}
\figsetgrptitle{}
\figsetplot{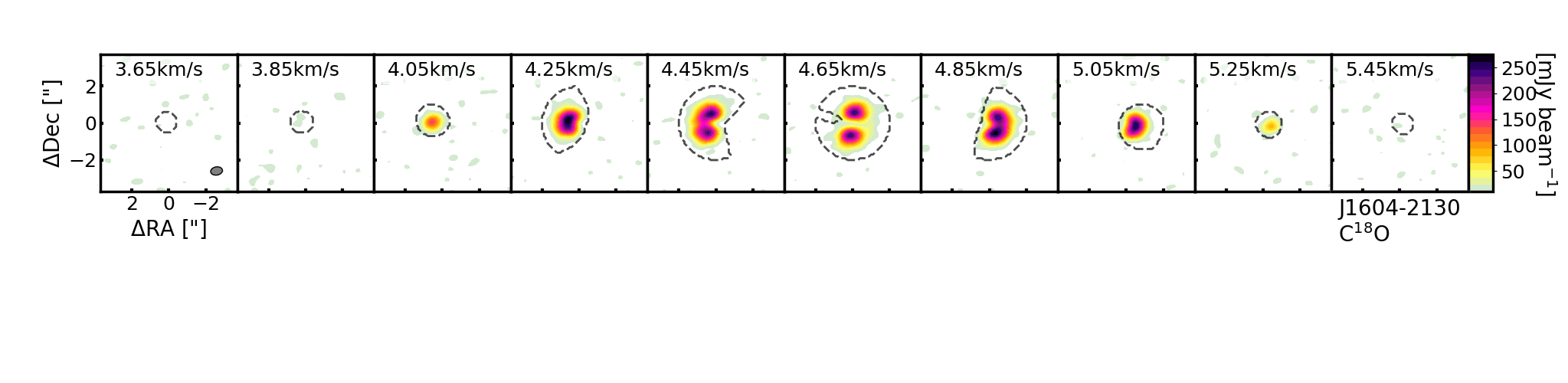}
\figsetgrpnote{C$^{18}$O towards J1604-2130 above 2$\sigma$.}
\figsetgrpend

\figsetgrpstart
\figsetgrpnum{A2.9}
\figsetgrptitle{}
\figsetplot{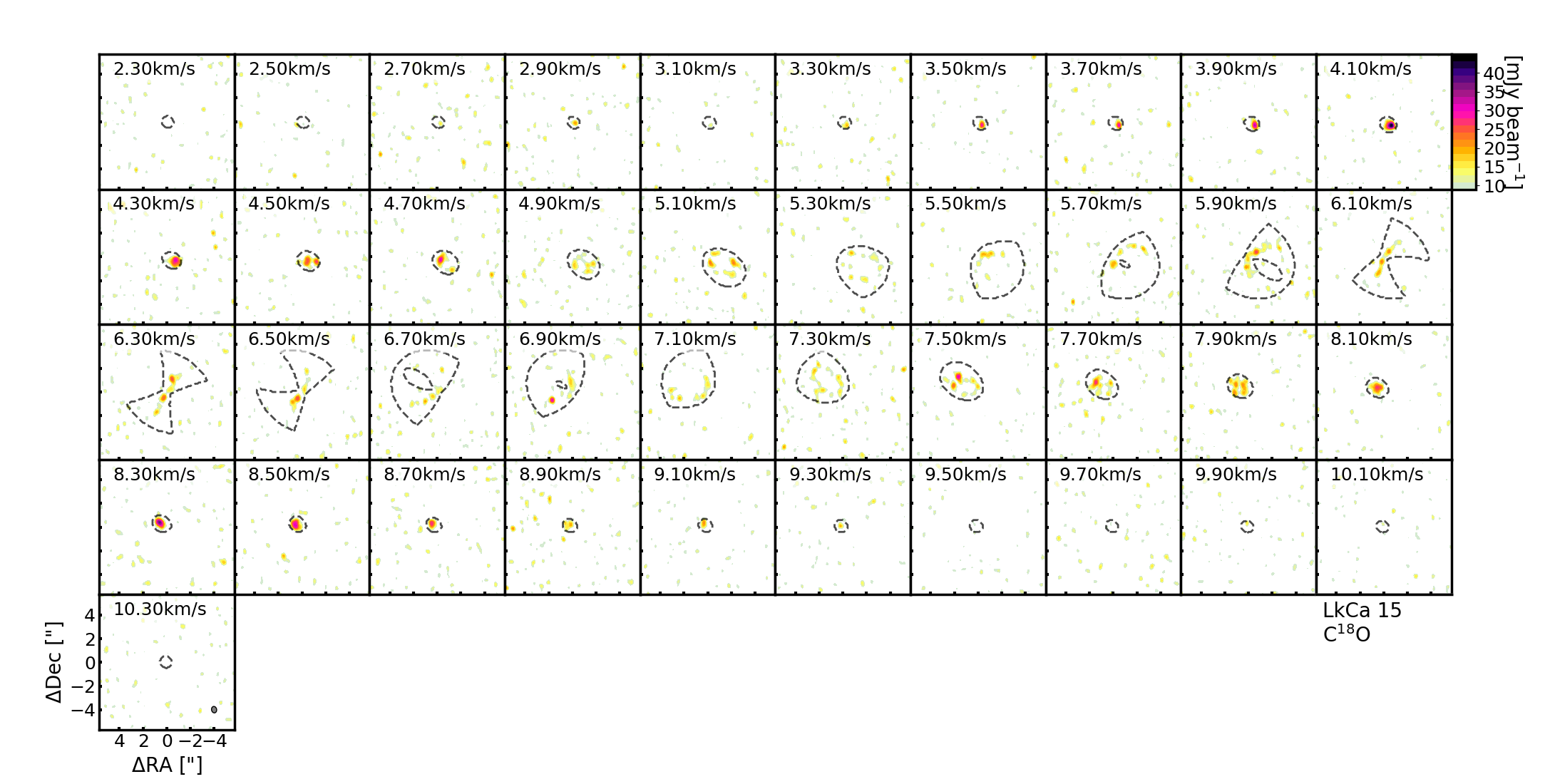}
\figsetgrpnote{C$^{18}$O towards LkCa 15 above 2$\sigma$.}
\figsetgrpend

\figsetgrpstart
\figsetgrpnum{A2.10}
\figsetgrptitle{}
\figsetplot{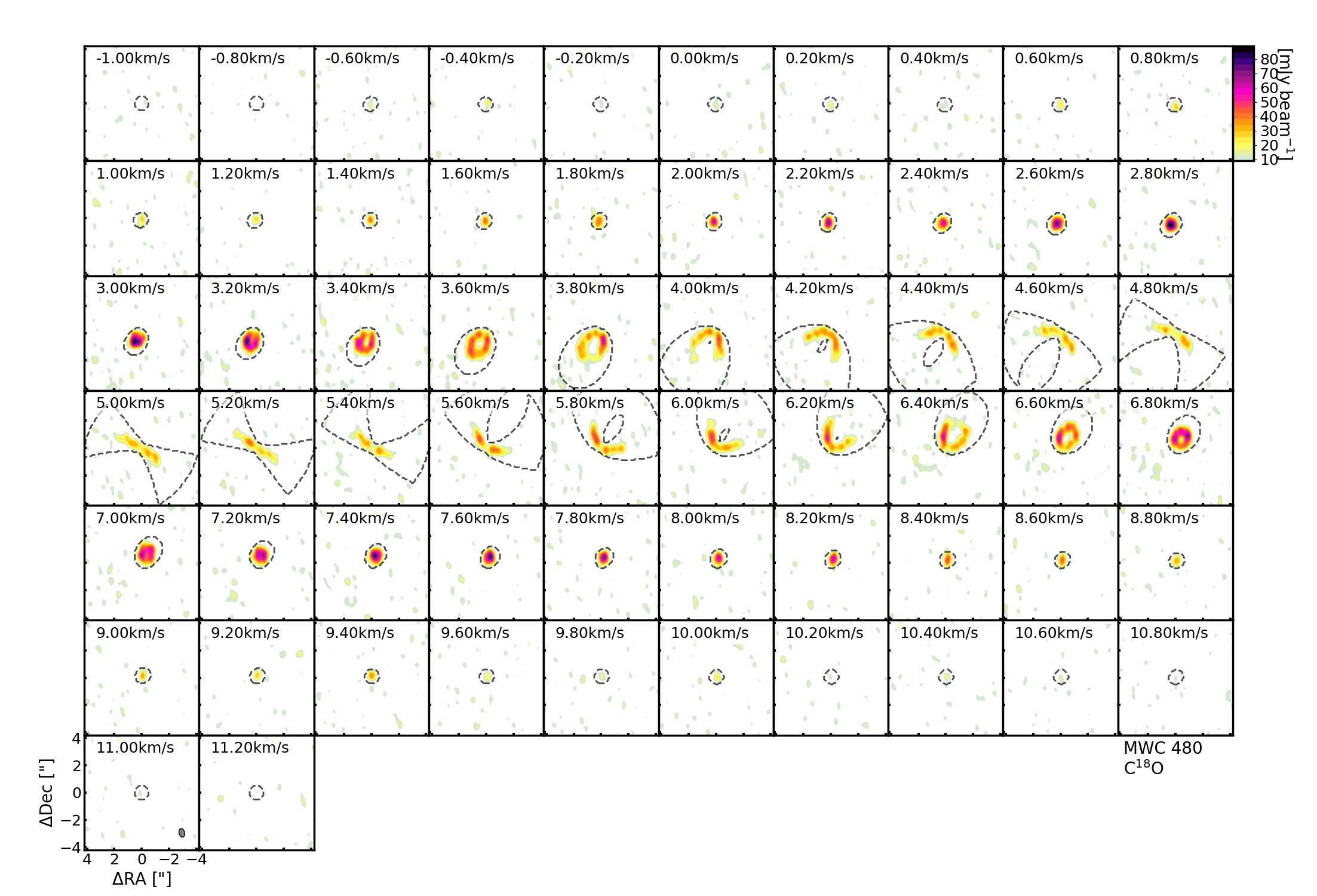}
\figsetgrpnote{C$^{18}$O towards MWC 480 above 2$\sigma$.}
\figsetgrpend

\figsetgrpstart
\figsetgrpnum{A2.11}
\figsetgrptitle{}
\figsetplot{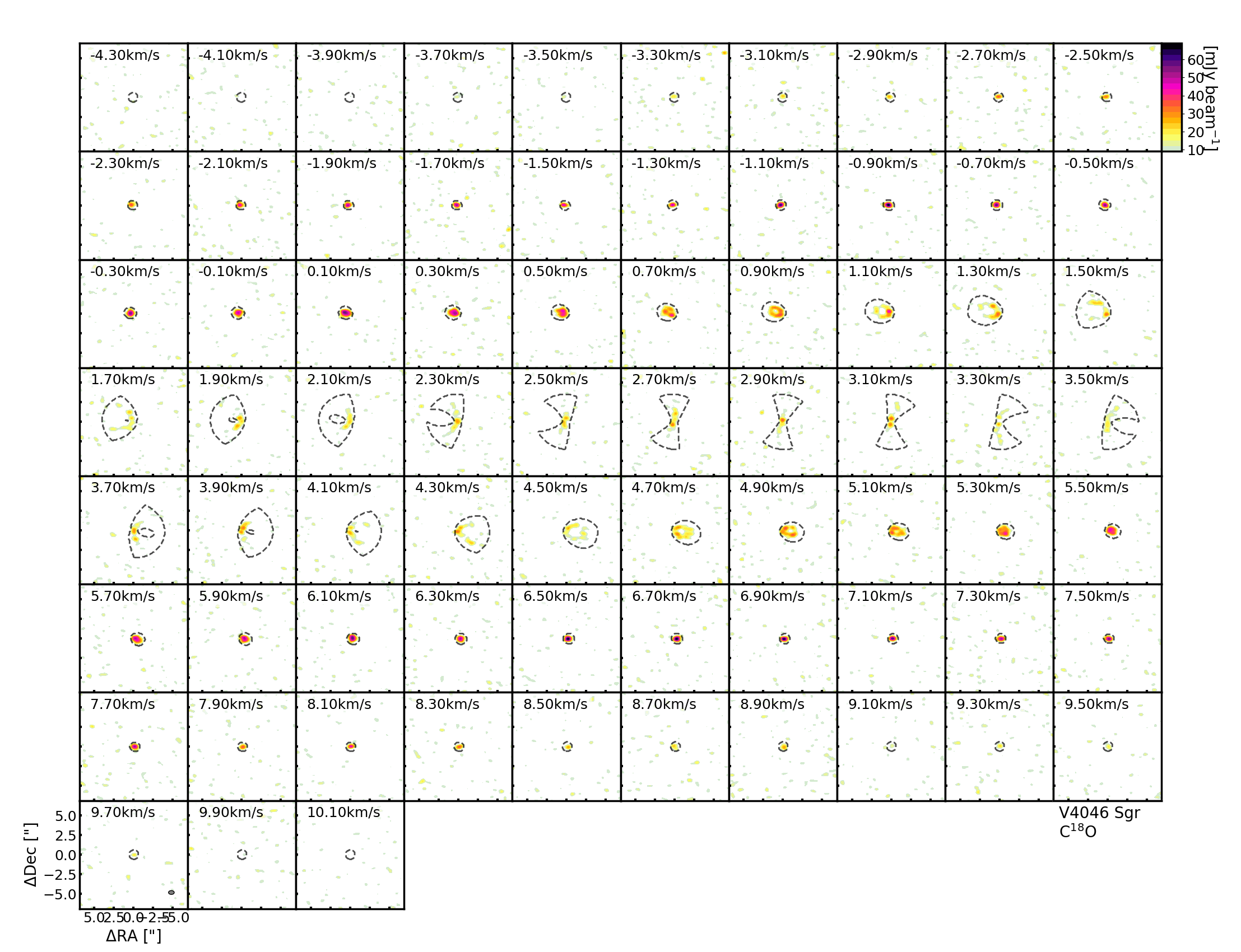}
\figsetgrpnote{C$^{18}$O towards V4046 Sgr above 2$\sigma$.}
\figsetgrpend

\figsetend


%

\figsetstart
\label{figset_h2coside}
\figsetnum{A3}
\figsettitle{Channel maps of all detected H$_2$CO lines (other than 3$_{03}$-2$_{02}$ and 4$_{04}$-3$_{03}$) above 2$\sigma$.}

\figsetgrpstart
\figsetgrpnum{A3.1}
\figsetgrptitle{}
\figsetplot{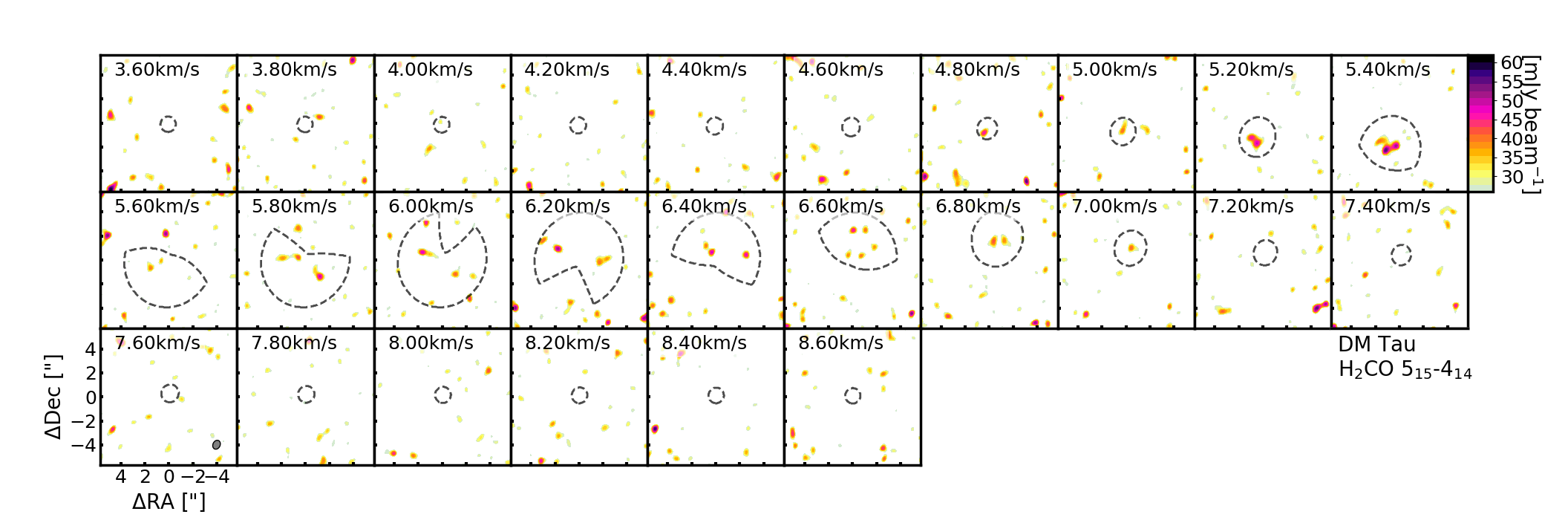}
\figsetgrpnote{H$_2$CO 5$_{15}$-4$_{14}$ towards DM Tau above 2$\sigma$.}
\figsetgrpend


\figsetgrpstart
\figsetgrpnum{A3.2}
\figsetgrptitle{}
\figsetplot{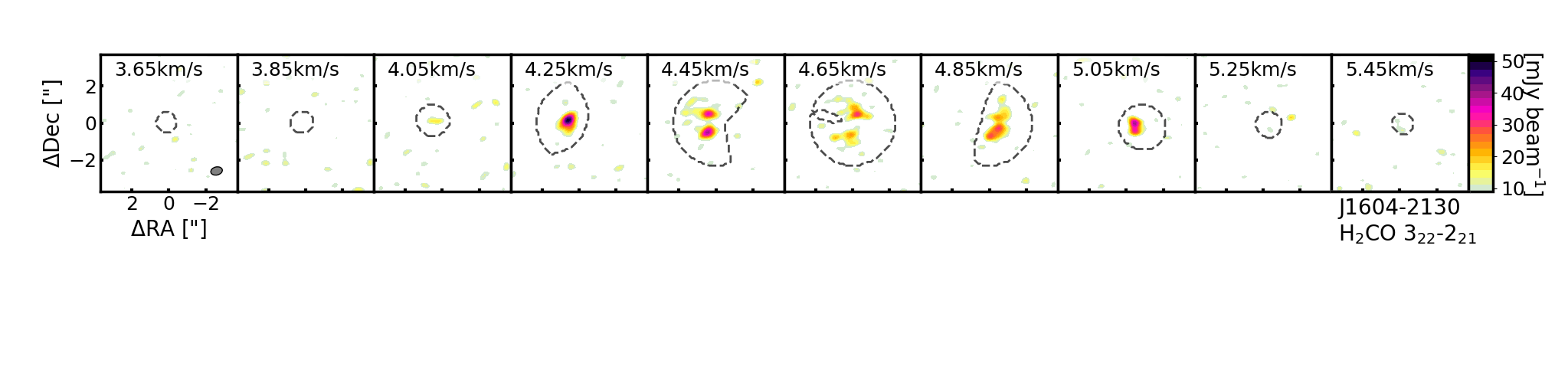}
\figsetgrpnote{H$_2$CO 3$_{22}$-2$_{21}$ towards J1604-2130 above 2$\sigma$.}
\figsetgrpend

\figsetgrpstart
\figsetgrpnum{A3.3}
\figsetgrptitle{H$_2$CO 3$_{21}$-2$_{20}$ towards J1604-2130 above 2$\sigma$.}
\figsetplot{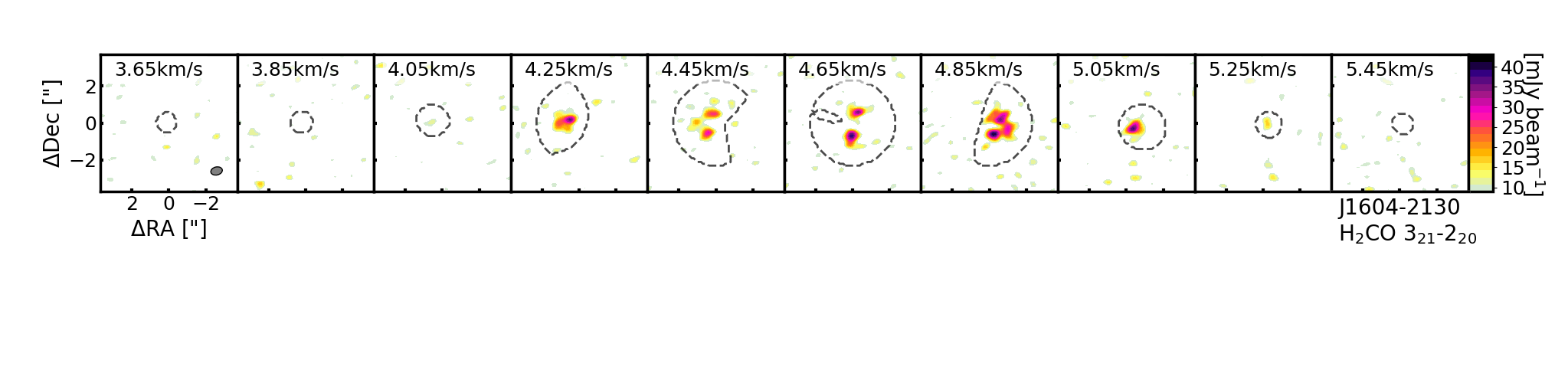}
\figsetgrpnote{ }
\figsetgrpend

\figsetgrpstart
\figsetgrpnum{A3.4}
\figsetgrptitle{}
\figsetplot{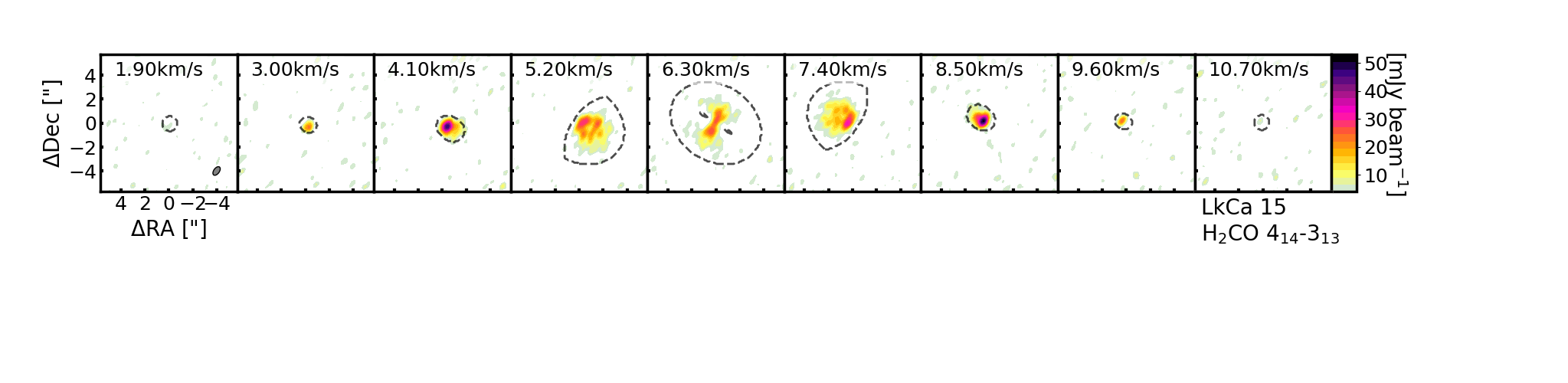}
\figsetgrpnote{H$_2$CO 4$_{14}$-3$_{13}$ towards LkCa 15 above 2$\sigma$.}
\figsetgrpend

\figsetgrpstart
\figsetgrpnum{A3.5}
\figsetgrptitle{}
\figsetplot{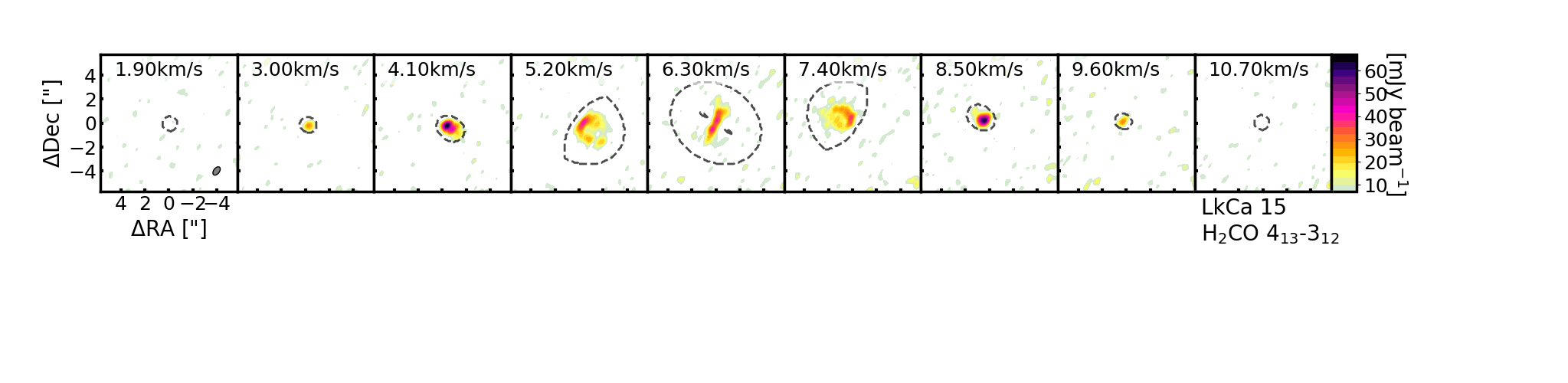}
\figsetgrpnote{H$_2$CO 4$_{13}$-3$_{12}$ towards LkCa 15 above 2$\sigma$.}
\figsetgrpend

\figsetgrpstart
\figsetgrpnum{A3.6}
\figsetgrptitle{}
\figsetplot{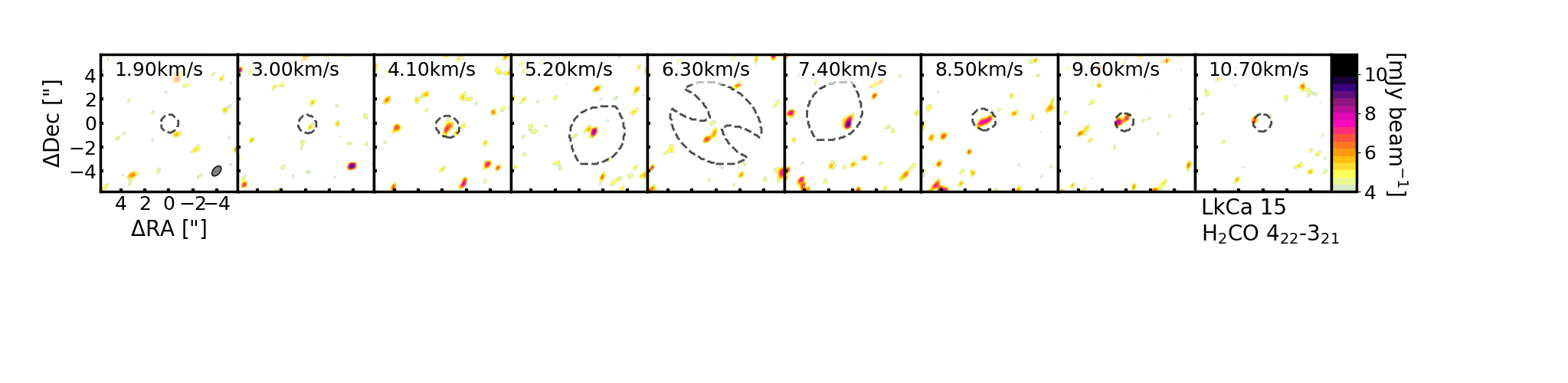}
\figsetgrpnote{H$_2$CO 4$_{22}$-3$_{21}$ towards LkCa 15 above 2$\sigma$.}
\figsetgrpend

\figsetgrpstart
\figsetgrpnum{A3.7}
\figsetgrptitle{}
\figsetplot{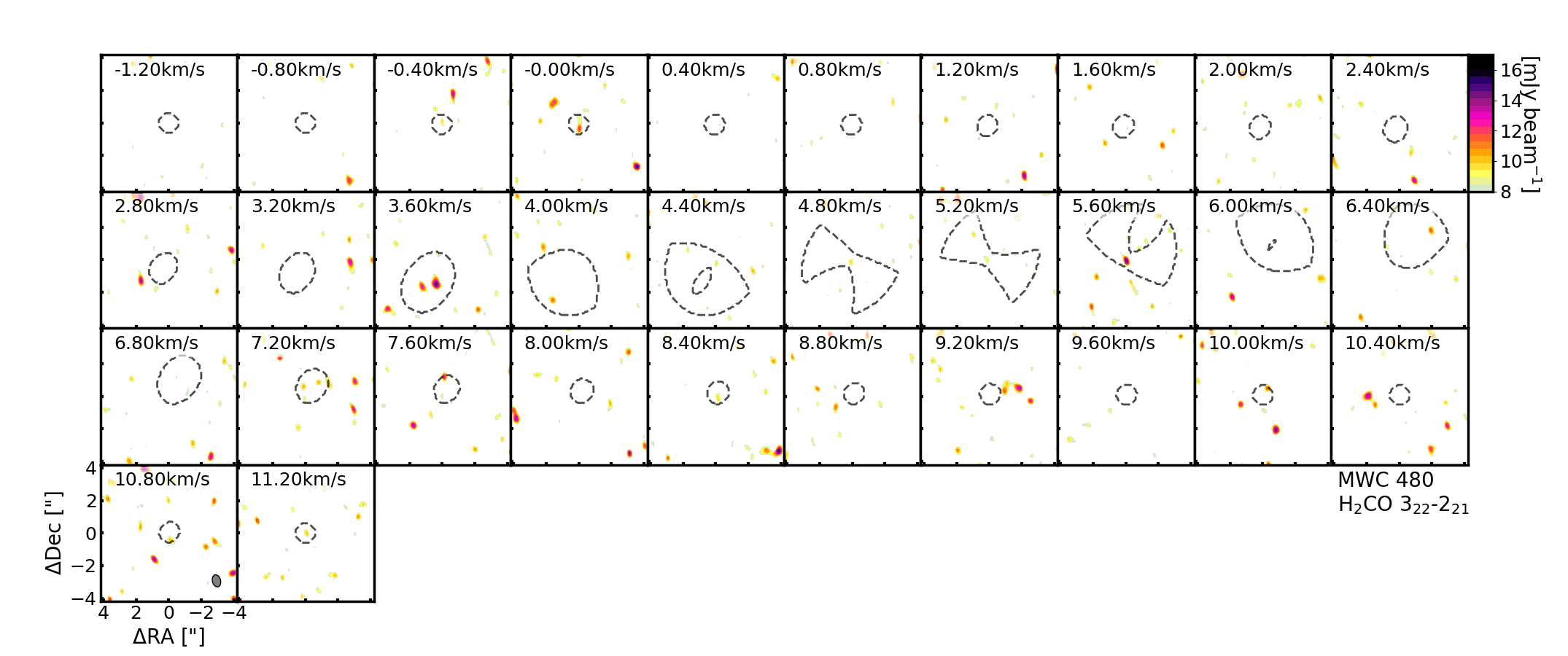}
\figsetgrpnote{H$_2$CO 3$_{22}$-2$_{21}$ towards MWC 480 above 2$\sigma$.}
\figsetgrpend

\figsetgrpstart
\figsetgrpnum{A3.8}
\figsetgrptitle{}
\figsetplot{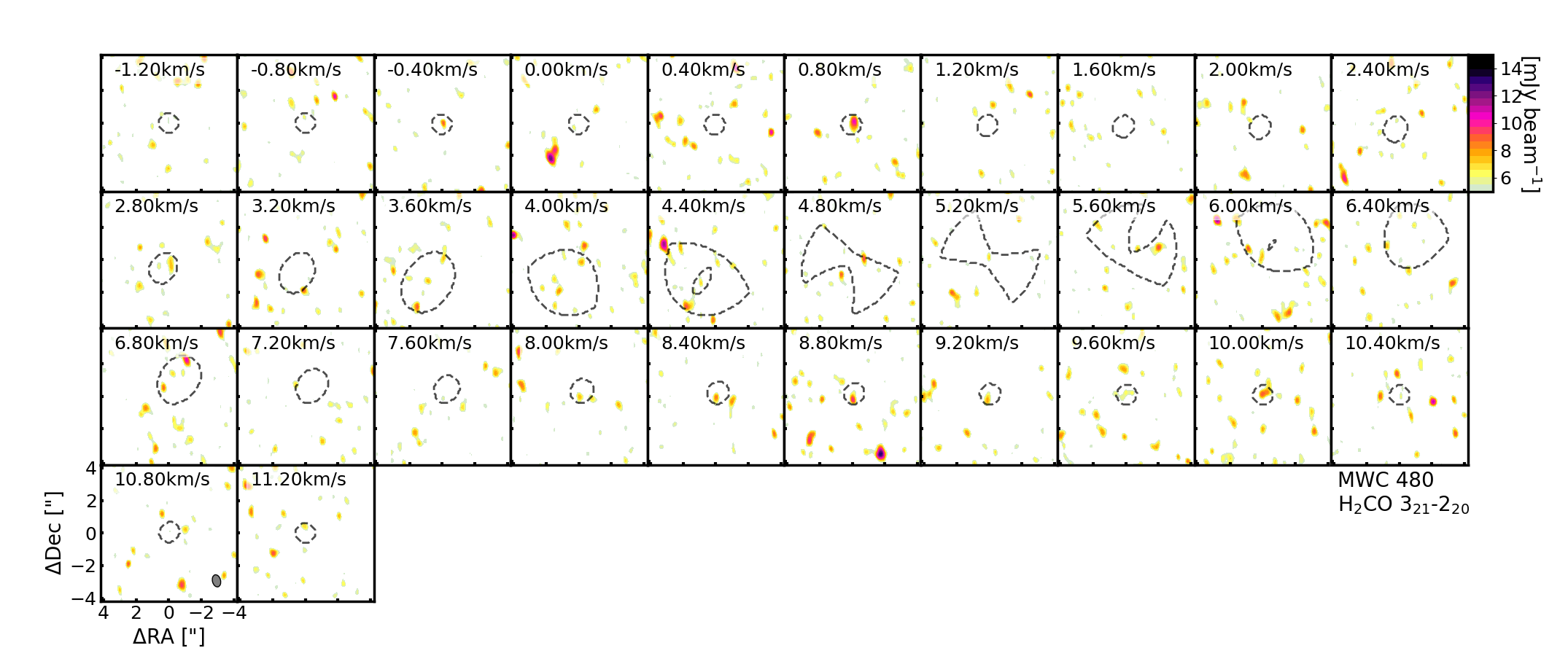}
\figsetgrpnote{H$_2$CO 3$_{21}$-2$_{20}$ towards MWC 480 above 2$\sigma$.}
\figsetgrpend

\figsetgrpstart
\figsetgrpnum{A3.9}
\figsetgrptitle{}
\figsetplot{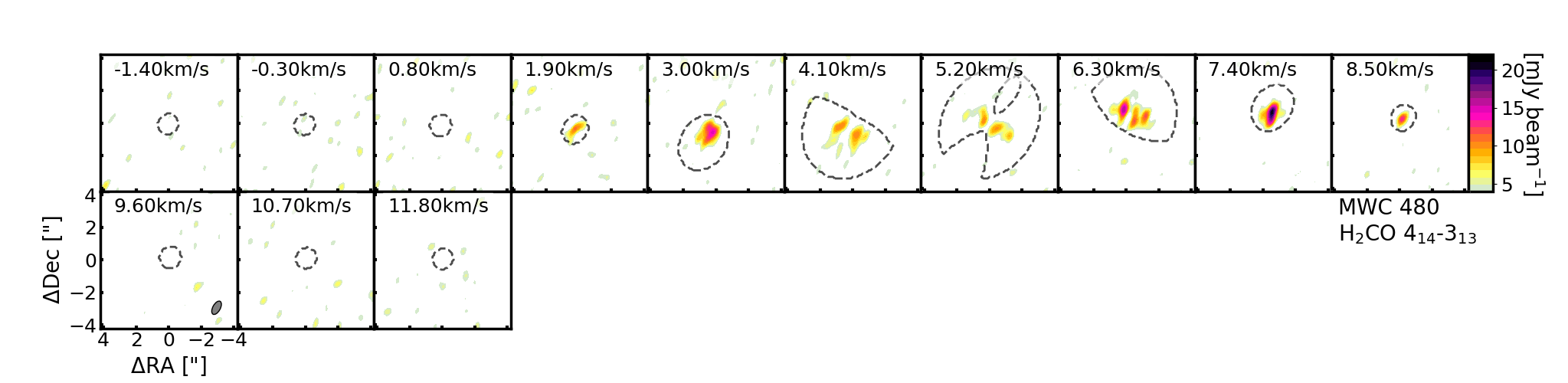}
\figsetgrpnote{H$_2$CO 4$_{14}$-3$_{13}$ towards MWC 480 above 2$\sigma$.}
\figsetgrpend

\figsetgrpstart
\figsetgrpnum{A3.10}
\figsetgrptitle{}
\figsetplot{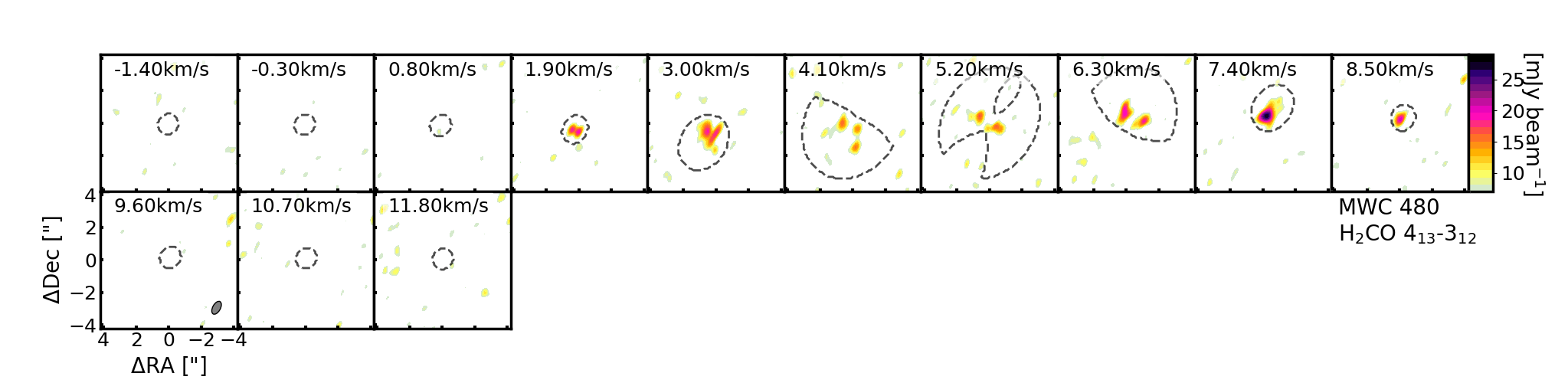}
\figsetgrpnote{H$_2$CO 4$_{13}$-3$_{12}$ towards MWC 480 above 2$\sigma$.}
\figsetgrpend

\figsetgrpstart
\figsetgrpnum{A3.11}
\figsetgrptitle{H$_2$CO 5$_{15}$-4$_{14}$ towards MWC 480 above 2$\sigma$.}
\figsetplot{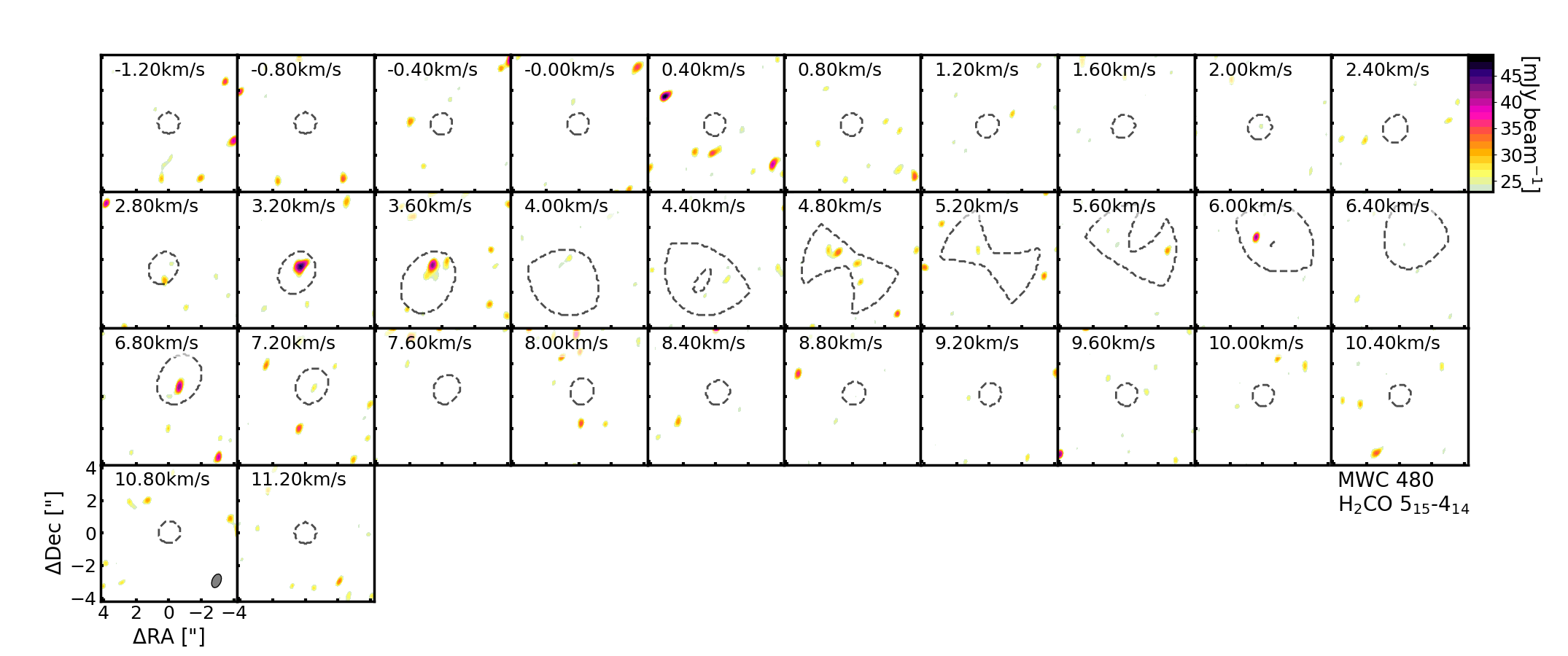}
\figsetgrpnote{H$_2$CO 5$_{15}$-4$_{14}$ towards MWC 480 above 2$\sigma$.}
\figsetgrpend

\figsetend


%

\clearpage

\setcounter{section}{2}
\setcounter{figure}{0}
\setcounter{table}{0}

\begin{deluxetable*}{llllllll}[!hbtp]
\centering
\tablewidth{0.75\textwidth}
\tablecaption{\textbf{Keplerian Mask Parameters.}
\centering
\label{table_kepmaskparams}}
\tablehead{
Disk	& Position           & Inclination           & H$_2$CO and Cont. & C$^{18}$O & Systemic  \\
	& Angle$^*$	&  Angle$^*$  & Mask Radius$^{**}$ & Mask Radius$^{**}$  & Velocity \\
	& {[}$^\circ${]} & {[}$^\circ${]} & {[}"{]} & {[}"{]}   & {[}km s$^{-1}${]}
}
\startdata
AS 209     & 6.6                 & 31.2                & 2.0                    & 2.9                  & 4.600      \\
CI Tau     & 260.4               & 41.6                & 3.4                    & 4.7                  & 5.840      \\
DM Tau     & 110.3               & 23.2                & 4.0                    & 4.1                  & 6.070      \\
DO Tau     & 113.4               & 18.5                & 3.5                    & 2.3                  & 5.900      \\
GM Aur     & 33.8                & 37.2                & 2.7                    & -                  & 5.550      \\
HD 143006  & 277.1               & 24.0                  & 1.7                    & 1.7                  & 7.755      \\
HD 163296  & 136.9               & 43.7                & 4.8                    & 4.1                  & 5.790      \\
IM Lup     & 305.3               & 46.2                & 3.4                    & 3.4                  & 4.400        \\
J1604-2130 & 194.6               & 6.2                 & 2.3                    & 2.0                    & 4.600        \\
J1609-1908 & 103.4               & 32.1                & 1.7                    & -                  & 3.800        \\
J1612-1859 & 151.9               & 30.9                & 0.8                    & -                  & 4.700        \\
J1614-1906 & 262.5               & 40.9                & 0.8                    & -                  & 3.800        \\
LkCa 15    & 33.1                & 37.8                & 4.0                      & 4.1                  & 6.300        \\
MWC 480    & 122                 & 37.4                & 3.6                    & 5.0                    & 5.100        \\
V4046 Sgr  & 191.9               & 29.3                & 5.6                    & 4.0                    & 2.900       
\enddata
\tablecomments{All Keplerian masks were generated assuming the combined thermal and turbulent line width is $\Delta v \sim v_0 (r_0 / \mathrm{100\mathrm{AU}})^{q}$~\citep{cite_yenetal2016}.  For all disks, we fixed $r_0$ and $q$ to be 100 AU and -0.3, respectively.  For most disks, we fixed $v_0$ to be 0.3km s$^{-1}$.  The exceptions were the broader H$_2$CO 4$_{14}$-3$_{13}$ and 4$_{13}$-3$_{12}$ lines, for which we fixed $v_0$ to be 0.6km s$^{-1}$, as well as the nearly face-on disk J1604-2130, for which we fixed $v_0$ to be 0.15km s$^{-1}$.  $^*$: The position and inclination angles were estimated using a grid-search algorithm and the fixed broadening parameters.  We stress that these estimates are not exact measurements of the disk angles.  They are parametric values used only to maximize the masked emission.  $**$: The Keplerian mask radii were set where the azimuthally-averaged emission intensities first reached zero.  For disks with multiple H$_2$CO lines observed, we used the same (largest) mask radius for all lines to maintain a consistent area of H$_2$CO emission.}
\end{deluxetable*}

\begin{figure*}[!htbp]
\centering
\resizebox{0.875\hsize}{!}{
    \includegraphics{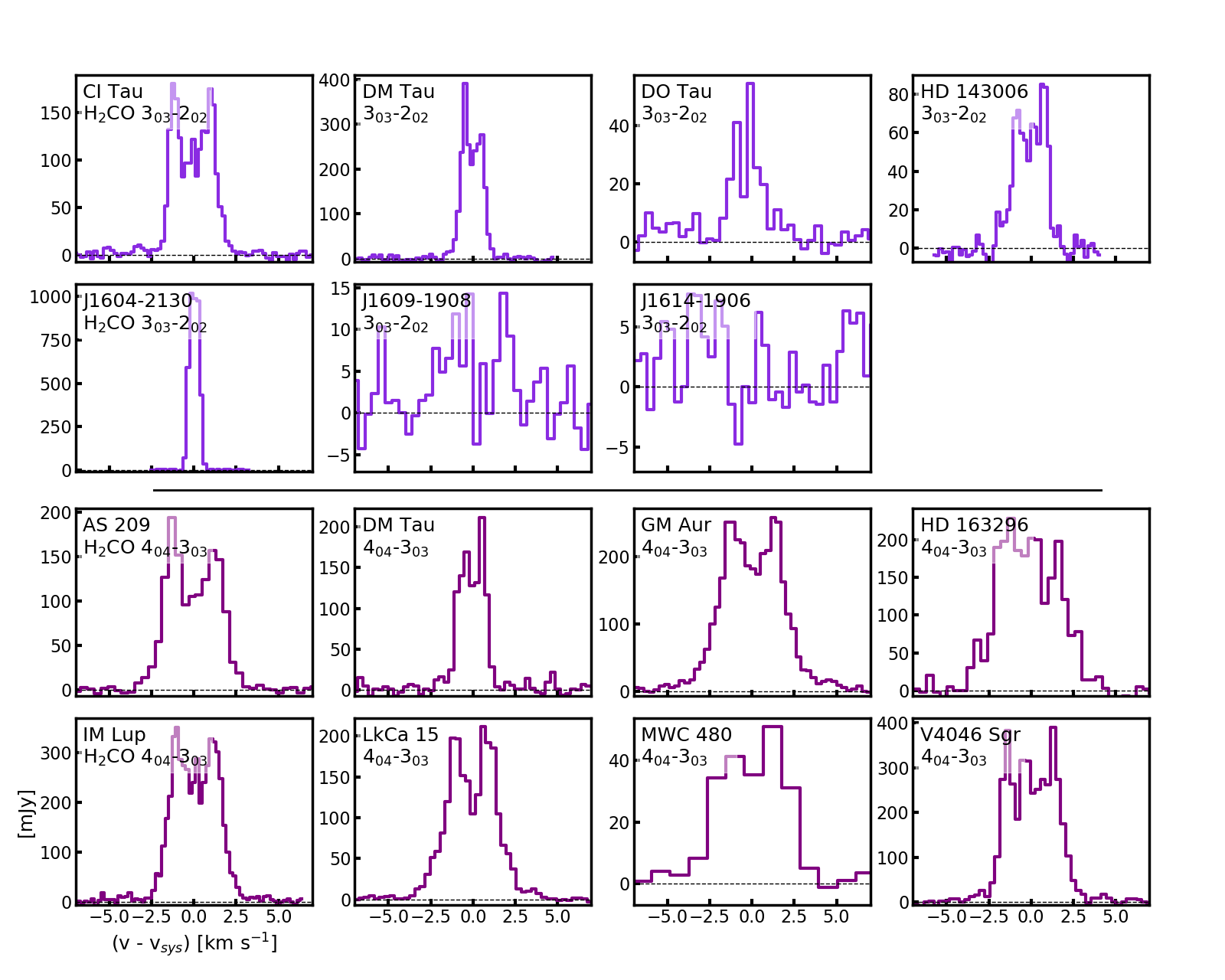}}
\caption{\textbf{H$_2$CO 3$_{03}$-2$_{02}$ and 4$_{04}$-3$_{03}$ Spectra.}  H$_2$CO 3$_{03}$-2$_{02}$ and 4$_{04}$-3$_{03}$ lines are depicted in light and dark purple, respectively, above and below the horizontal black line.  DM Tau appears twice because both lines are observed towards DM Tau.}
\label{fig_series32and43spec}
\end{figure*}

\clearpage

\setcounter{section}{3}
\setcounter{figure}{0}
\setcounter{table}{0}

\begin{deluxetable*}{lllllll}[!hbtp]
\tablecaption{\textbf{Dust Continuum Flux Measurements.}
\label{table_h2cocontfluxes}}
\tablehead{
Disk	& $\lambda$	& Size$^*$	& Sum$^{**}$ 	& Peak Em.$^{**}$   & rms ($\sigma$)$^\dagger$	& Beam Size (P.A.) \\
	& [mm] 	& [AU]  	& [mJy]	& [mJy beam$^{-1}$]	   & [mJy beam$^{-1}$]	& 
}
\startdata
AS 209                                                     & 1.0  & 66.5           & 297                 & 45               & 0.18                & 0.31" x 0.21" (-70.80 deg) \\
CI Tau                                                     & 1.3  & 206.7          & 161                 & 54               & 0.16                & 0.63" x 0.44" (30.36 deg)  \\
DM Tau                                                     & 1.3  & 159.5          & 99                  & 37               & 0.10                & 0.60" x 0.44" (35.26 deg)  \\
                                                           & 1.0  & 65.2           & 89                  & 8.7              & 0.15                & 0.18" x 0.15" (14.33 deg)  \\
                                                           & 0.9  & 159.5          & 224                 & 91               & 0.31                & 0.65" x 0.44" (-28.59 deg) \\
DO Tau                                                     & 1.3  & 98.0           & 127                 & 102              & 0.27                & 0.68" x 0.44" (29.50 deg)  \\
GM Aur                                                     & 1.0  & 79.5           & 255                 & 26               & 0.13                & 0.29" x 0.19" (-1.65 deg)  \\
HD 143006                                                  & 1.3  & 148.5          & 53                  & 21               & 0.12                & 0.69" x 0.47" (-75.15 deg) \\
HD 163296                                                  & 1.0  & 121.2          & 956                 & 219              & 0.87                & 0.44" x 0.33" (-84.74 deg) \\
IM Lup                                                     & 1.0  & 189.6          & 237                 & 78               & 0.35                & 0.40" x 0.33" (78.16 deg)  \\
J1604-2130                                                 & 1.3  & 178.8          & 69                  & 11               & 0.14                & 0.63" x 0.43" (-77.99 deg) \\
J1609-1908                                                 & 1.3  & 95.9           & 22                  & 18               & 0.14                & 0.63" x 0.43" (-76.80 deg) \\
J1612-1859                                                 & 1.3  & 96.6           & 3.6                 & 3.0              & 0.13                & 0.63" x 0.43" (-76.23 deg) \\
J1614-1906                                                 & 1.3  & 100.1          & 19                  & 17               & 0.14                & 0.62" x 0.43" (-76.99 deg) \\
LkCa 15                                                    & 1.0  & 94.8           & 281                 & 23               & 0.11                & 0.32" x 0.26" (-17.58 deg) \\
MWC 480                                                    & 1.0  & 209.3          & 482                 & 347              & 0.27                & 1.05" x 0.84" (-5.72 deg)  \\
                                                           & 0.9  & 161.0          & 848                 & 448              & 0.44                & 0.78" x 0.42" (-23.28 deg) \\
V4046 Sgr                                                  & 1.0  & 80.3           & 542                 & 75               & 0.18                & 0.50" x 0.39" (-70.09 deg)
\enddata
\tablecomments{{$*$: The radial extent of the dust continuum is defined as where the continuum emission is 5\% of its peak.  $**$: The fluxes and peak fluxes were measured within the same Keplerian mask radii used for H$_2$CO (Table~\ref{table_kepmaskparams}).  Errors are small relative to the fluxes.  $\dagger$:  The average rms across 1000 1"-by-1" random samples, extracted 3"-7" away from the disk center.}}
\end{deluxetable*}

\begin{deluxetable*}{llllllll}[!hbtp]
\tablecaption{\textbf{C$^{18}$O Emission Flux Measurements.}
\label{table_coemfluxes}}
\tablehead{
Disk	& Flux$^*$	& Peak Flux$^*$    	& Int. Vel.   & Chan.  & Chan. 	& Beam Size (P.A.) \\
	& [mJy	& [mJy beam$^{-1}$    & Range  & Width   & rms$^{**}$	& \\
	& $\times$ km s$^{-1}$]	& $\times$ km s$^{-1}$]    & [km s$^{-1}$]  & [km s$^{-1}$]   & [mJy beam$^{-1}$]	&
}
\startdata
AS 209     & 429 $\pm$ 30   & 53 $\pm$ 6.4        & 0.9 - 8.1        & 0.4              & 8.0            & 0.53" x 0.50" (37.32$^{\circ}$)  \\
CI Tau     & 549 $\pm$ 17   & 44 $\pm$ 3.3        & 2.6 - 9.0        & 0.2              & 5.4            & 0.67" x 0.46" (31.32$^{\circ}$)  \\
DM Tau     & 998 $\pm$ 17   & 66 $\pm$ 3.2        & 3.6 - 8.6        & 0.2              & 5.6            & 0.63" x 0.47" (36.01$^{\circ}$)  \\
DO Tau     & 210 $\pm$ 12   & 82 $\pm$ 3.6        & 3.2 - 8.8        & 0.2              & 5.3            & 0.72" x 0.46" (30.42$^{\circ}$)  \\
HD 143006  & 135 $\pm$ 10   & 43 $\pm$ 3.6        & 5.4 - 10.2       & 0.2              & 5.5            & 0.70" x 0.48" (-79.98$^{\circ}$) \\
HD 163296  & 5040 $\pm$ 31  & 355 $\pm$ 6.7       & -1.4 - 13.0      & 0.2              & 4.9            & 0.61" x 0.52" (62.26$^{\circ}$)  \\
IM Lup     & 1203 $\pm$ 14  & 80 $\pm$ 4.1        & 0.2 - 8.6        & 0.2              & 4.9            & 0.58" x 0.43" (-64.43$^{\circ}$) \\
J1604-2130 & 1267 $\pm$ 7.2 & 141 $\pm$ 1.7       & 3.7 - 5.5        & 0.2              & 5.4            & 0.65" x 0.45" (-80.59$^{\circ}$) \\
LkCa 15    & 619 $\pm$ 18   & 53 $\pm$ 3.7        & 2.3 - 10.3       & 0.2              & 4.7            & 0.55" x 0.43" (11.85$^{\circ}$)  \\
MWC 480    & 1648 $\pm$ 20  & 290 $\pm$ 11        & -1.0 - 11.2      & 0.2              & 4.6            & 0.64" x 0.41" (13.23$^{\circ}$)  \\
V4046 Sgr  & 1184 $\pm$ 18  & 385 $\pm$ 5.1       & -4.3 - 10.1      & 0.2              & 4.5            & 0.73" x 0.49" (89.68$^{\circ}$) 
\enddata
\tablecomments{{*: The velocity-integrated fluxes (Column 2) were measured within the bounds of the Keplerian masks (Table~\ref{table_kepmaskparams}).  The peak fluxes (Column 3) are the peaks of the velocity-integrated emission maps; note the difference in unit compared to the velocity-integrated fluxes.  The uncertainty in each peak flux is the standard deviation of the peaks across 1000 random samples.  Uncertainties do not include absolute flux uncertainties.  $**$: The channel rms was estimated as the standard deviation of 1000 1"-by-1" random samples.}}
\end{deluxetable*}

\clearpage

\setcounter{section}{4}
\setcounter{figure}{0}
\setcounter{table}{0}


\begin{figure*}[!htbp]
\centering
\resizebox{0.5\hsize}{!}{
    \includegraphics{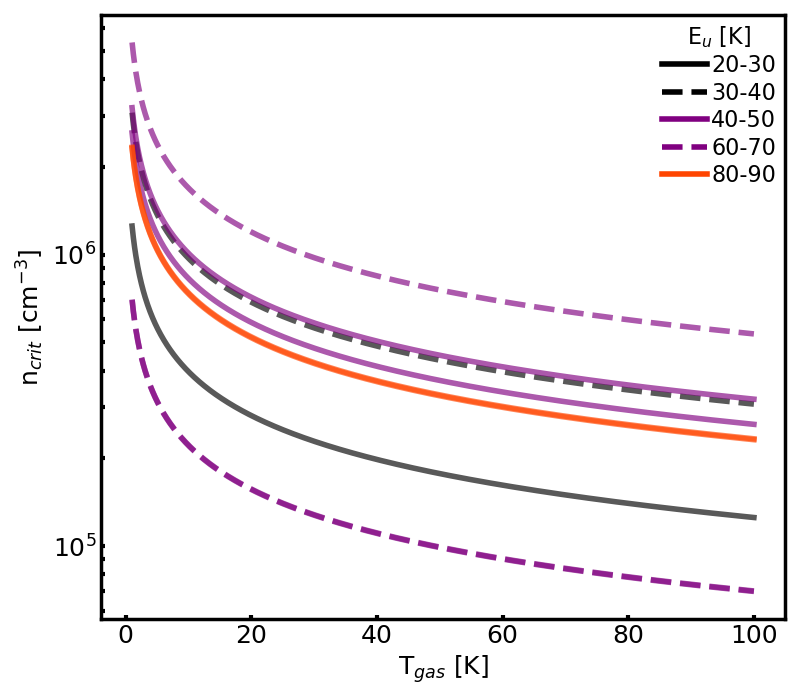}}
\caption{\textbf{H$_2$CO Critical Densities as a Function of Gas Temperature.}  The critical density $n_{crit}$ is approximately $n_{crit} \approx A_{ul}/(\sigma v)$~\citep[e.g.,][]{cite_textbook_ERA}, where $A_{ul}$ is the Einstein coefficient (Table~\ref{table_linechar}), $\sigma$ is the collisional cross section (assumed here to be $\sim 9.9 \times 10^{-15}$cm$^2$ for H$_2$CO), $v \approx \sqrt{3k_B T_{gas}/m}$ is the average gas velocity, $k_B$ is the Boltzmann constant, $T_{gas}$ is the gas temperature, and $m$ is the mass of the molecule.  The critical densities for all H$_2$CO lines in this survey (plotted above, grouped by upper energy E$_u$) exceed 10$^6$cm$^{-3}$ only at temperatures below 40K.  Based on the T Tauri protoplanetary disk modeling of~\cite{cite_walshetal2010}, we expect typical disk densities to be below 10$^6$cm$^{-3}$ only in the disk's atmospheric layer, where gas temperatures would be $\gtrsim$50K.}
\label{fig_critdens}
\end{figure*}


\bibliography{main}{}

\begin{thebibliography}{}
\expandafter\ifx\csname natexlab\endcsname\relax\def\natexlab#1{#1}\fi
\providecommand{\url}[1]{\href{#1}{#1}}

\bibitem[{{Aikawa} {et~al.}(2003){Aikawa}, {Momose}, {Thi}, {van Zadelhoff},
  {Qi}, {Blake}, \& {van Dishoeck}}]{cite_aikawaetal2003}
{Aikawa}, Y., {Momose}, M., {Thi}, W.-F., {et~al.} 2003, \pasj, 55, 11

\bibitem[{{Andrews} {et~al.}(2018){Andrews}, {Huang}, {P{\'e}rez}, {Isella},
  {Dullemond}, {Kurtovic}, {Guzm{\'a}n}, {Carpenter}, {Wilner}, {Zhang}, {Zhu},
  {Birnstiel}, {Bai}, {Benisty}, {Hughes}, {{\"O}berg}, \&
  {Ricci}}]{cite_andrewsetal2018}
{Andrews}, S.~M., {Huang}, J., {P{\'e}rez}, L.~M., {et~al.} 2018, \apj, 869,
  L41

\bibitem[{Atkinson {et~al.}(2006)Atkinson, Baulch, Cox, Crowley, Hampson,
  Hynes, Jenkin, Rossi, Troe, \& Subcommittee}]{cite_atkinsonetal2006}
Atkinson, R., Baulch, D.~L., Cox, R.~A., {et~al.} 2006, Atmospheric Chemistry
  and Physics, 6, 3625.
\newblock \url{https://www.atmos-chem-phys.net/6/3625/2006/}

\bibitem[{{Bergin} {et~al.}(2004){Bergin}, {Calvet}, {Sitko}, {Abgrall},
  {D'Alessio}, {Herczeg}, {Roueff}, {Qi}, {Lynch}, \&
  {Russell}}]{cite_berginetal2004}
{Bergin}, E., {Calvet}, N., {Sitko}, M.~L., {et~al.} 2004, \apj, 614, L133

\bibitem[{{Bergner} {et~al.}(2018){Bergner}, {Guzm{\'a}n}, {{\"O}berg},
  {Loomis}, \& {Pegues}}]{cite_bergneretal2018}
{Bergner}, J.~B., {Guzm{\'a}n}, V.~G., {{\"O}berg}, K.~I., {Loomis}, R.~A., \&
  {Pegues}, J. 2018, \apj, 857, 69

\bibitem[{{Bergner} {et~al.}(2019){Bergner}, {{\"O}berg}, {Bergin}, {Loomis},
  {Pegues}, \& {Qi}}]{cite_bergneretal2019}
{Bergner}, J.~B., {{\"O}berg}, K.~I., {Bergin}, E.~A., {et~al.} 2019, \apj,
  876, 25

\bibitem[{{Blake} {et~al.}(1987){Blake}, {Sutton}, {Masson}, \&
  {Phillips}}]{cite_blakeetal1987}
{Blake}, G.~A., {Sutton}, E.~C., {Masson}, C.~R., \& {Phillips}, T.~G. 1987,
  \apj, 315, 621

\bibitem[{{Boogert} {et~al.}(2008){Boogert}, {Pontoppidan}, {Knez}, {Lahuis},
  {Kessler-Silacci}, {van Dishoeck}, {Blake}, {Augereau}, {Bisschop}, \&
  {Bottinelli}}]{cite_boogertetal2008}
{Boogert}, A.~C.~A., {Pontoppidan}, K.~M., {Knez}, C., {et~al.} 2008, \apj,
  678, 985

\bibitem[{{Carney} {et~al.}(2017){Carney}, {Hogerheijde}, {Loomis}, {Salinas},
  {{\"O}berg}, {Qi}, \& {Wilner}}]{cite_carneyetal2017}
{Carney}, M.~T., {Hogerheijde}, M.~R., {Loomis}, R.~A., {et~al.} 2017, \aap,
  605, A21

\bibitem[{{Chapillon} {et~al.}(2012){Chapillon}, {Dutrey}, {Guilloteau},
  {Pi{\'e}tu}, {Wakelam}, {Hersant}, {Gueth}, {Henning}, {Launhardt},
  {Schreyer}, \& {Semenov}}]{cite_chapillonetal2012}
{Chapillon}, E., {Dutrey}, A., {Guilloteau}, S., {et~al.} 2012, \apj, 756, 58

\bibitem[{{Cleeves}(2016)}]{cite_cleeves2016}
{Cleeves}, L.~I. 2016, \apjl, 816, L21

\bibitem[{{Cleeves} {et~al.}(2011){Cleeves}, {Bergin}, {Bethell}, {Calvet},
  {Fogel}, {Sauter}, \& {Wolf}}]{cite_cleevesetal2011}
{Cleeves}, L.~I., {Bergin}, E.~A., {Bethell}, T.~J., {et~al.} 2011, \apj, 743,
  L2

\bibitem[{Condon \& Ransom(2016)}]{cite_textbook_ERA}
Condon, J.~J., \& Ransom, S.~M. 2016, Spectral Lines (The National Radio
  Astronomy Observatory)

\bibitem[{{Cridland} {et~al.}(2016){Cridland}, {Pudritz}, \&
  {Alessi}}]{cite_cridlandetal2016}
{Cridland}, A.~J., {Pudritz}, R.~E., \& {Alessi}, M. 2016, \mnras, 461, 3274

\bibitem[{{Cridland} {et~al.}(2017){Cridland}, {Pudritz}, {Birnstiel},
  {Cleeves}, \& {Bergin}}]{cite_cridlandetal2017}
{Cridland}, A.~J., {Pudritz}, R.~E., {Birnstiel}, T., {Cleeves}, L.~I., \&
  {Bergin}, E.~A. 2017, arXiv e-prints, arXiv:1705.02381

\bibitem[{{Drozdovskaya} {et~al.}(2019){Drozdovskaya}, {van Dishoeck}, {Rubin},
  {J{\o}rgensen}, \& {Altwegg}}]{cite_drozdovskayaetal2019}
{Drozdovskaya}, M.~N., {van Dishoeck}, E.~F., {Rubin}, M., {J{\o}rgensen},
  J.~K., \& {Altwegg}, K. 2019, \mnras, 490, 50

\bibitem[{{Drozdovskaya} {et~al.}(2014){Drozdovskaya}, {Walsh}, {Visser},
  {Harsono}, \& {van Dishoeck}}]{cite_drozdovskayaetal2014}
{Drozdovskaya}, M.~N., {Walsh}, C., {Visser}, R., {Harsono}, D., \& {van
  Dishoeck}, E.~F. 2014, \mnras, 445, 913

\bibitem[{{Dutrey} {et~al.}(1997){Dutrey}, {Guilloteau}, \&
  {Guelin}}]{cite_dutreyetal1997}
{Dutrey}, A., {Guilloteau}, S., \& {Guelin}, M. 1997, \aap, 317, L55

\bibitem[{{Endres} {et~al.}(2016){Endres}, {Schlemmer}, {Schilke}, {Stutzki},
  \& {M{\"u}ller}}]{cite_cdms2016}
{Endres}, C.~P., {Schlemmer}, S., {Schilke}, P., {Stutzki}, J., \&
  {M{\"u}ller}, H.~S.~P. 2016, Journal of Molecular Spectroscopy, 327, 95

\bibitem[{{Espaillat} {et~al.}(2010){Espaillat}, {D'Alessio}, {Hern{\'a}ndez},
  {Nagel}, {Luhman}, {Watson}, {Calvet}, {Muzerolle}, \&
  {McClure}}]{cite_espaillatetal2010}
{Espaillat}, C., {D'Alessio}, P., {Hern{\'a}ndez}, J., {et~al.} 2010, \apj,
  717, 441

\bibitem[{{Favre} {et~al.}(2018){Favre}, {Fedele}, {Semenov}, {Parfenov},
  {Codella}, {Ceccarelli}, {Bergin}, {Chapillon}, {Testi}, {Hersant},
  {Lefloch}, {Fontani}, {Blake}, {Cleeves}, {Qi}, {Schwarz}, \&
  {Taquet}}]{cite_favreetal2018}
{Favre}, C., {Fedele}, D., {Semenov}, D., {et~al.} 2018, \apjl, 862, L2

\bibitem[{{Fayolle} {et~al.}(2016){Fayolle}, {Balfe}, {Loomis}, {Bergner},
  {Graninger}, {Rajappan}, \& {{\"O}berg}}]{cite_fayolleetal2016}
{Fayolle}, E.~C., {Balfe}, J., {Loomis}, R., {et~al.} 2016, \apjl, 816, L28

\bibitem[{{Flaherty} {et~al.}(2015){Flaherty}, {Hughes}, {Rosenfeld},
  {Andrews}, {Chiang}, {Simon}, {Kerzner}, \& {Wilner}}]{cite_flahertyetal2015}
{Flaherty}, K.~M., {Hughes}, A.~M., {Rosenfeld}, K.~A., {et~al.} 2015, \apj,
  813, 99

\bibitem[{Fockenberg \& Preses(2002)}]{cite_fockenbergetal2002}
Fockenberg, C., \& Preses, J.~M. 2002, The Journal of Physical Chemistry A,
  106, 2924.
\newblock \url{http://dx.doi.org/10.1021/jp0141880}

\bibitem[{{Fuchs} {et~al.}(2009){Fuchs}, {Cuppen}, {Ioppolo}, {Romanzin},
  {Bisschop}, {Andersson}, {van Dishoeck}, \& {Linnartz}}]{cite_fuchsetal2009}
{Fuchs}, G.~W., {Cuppen}, H.~M., {Ioppolo}, S., {et~al.} 2009, \aap, 505, 629

\bibitem[{Féraud {et~al.}(2019)Féraud, Bertin, Romanzin, Dupuy, Le~Petit,
  Roueff, Philippe, Michaut, Jeseck, \& Fillion}]{cite_feraudetal2019}
Féraud, G., Bertin, M., Romanzin, C., {et~al.} 2019, ACS Earth and Space
  Chemistry

\bibitem[{{Gaia Collaboration} {et~al.}(2018){Gaia Collaboration}, {Brown},
  {Vallenari}, {Prusti}, {de Bruijne}, {Babusiaux}, \&
  {Bailer-Jones}}]{cite_gaiaetal2018}
{Gaia Collaboration}, {Brown}, A.~G.~A., {Vallenari}, A., {et~al.} 2018, ArXiv
  e-prints, arXiv:1804.09365

\bibitem[{{Gaia Collaboration} {et~al.}(2016{\natexlab{a}}){Gaia
  Collaboration}, {Prusti}, {de Bruijne}, {Brown}, {Vallenari}, {Babusiaux},
  {Bailer-Jones}, {Bastian}, {Biermann}, {Evans}, \&
  et~al.}]{cite_gaiaetal2016a}
{Gaia Collaboration}, {Prusti}, T., {de Bruijne}, J.~H.~J., {et~al.}
  2016{\natexlab{a}}, \aap, 595, A1

\bibitem[{{Gaia Collaboration} {et~al.}(2016{\natexlab{b}}){Gaia
  Collaboration}, {Brown}, {Vallenari}, {Prusti}, {de Bruijne}, {Mignard},
  {Drimmel}, {Babusiaux}, {Bailer-Jones}, {Bastian}, \&
  et~al.}]{cite_gaiaetal2016b}
{Gaia Collaboration}, {Brown}, A.~G.~A., {Vallenari}, A., {et~al.}
  2016{\natexlab{b}}, \aap, 595, A2

\bibitem[{{Gibb} {et~al.}(2004){Gibb}, {Whittet}, {Boogert}, \&
  {Tielens}}]{cite_gibbetal2004}
{Gibb}, E.~L., {Whittet}, D.~C.~B., {Boogert}, A.~C.~A., \& {Tielens},
  A.~G.~G.~M. 2004, \apjs, 151, 35

\bibitem[{{Goldsmith} \& {Langer}(1999)}]{cite_goldsmithetal1999}
{Goldsmith}, P.~F., \& {Langer}, W.~D. 1999, \apj, 517, 209

\bibitem[{{Guilloteau} {et~al.}(2013){Guilloteau}, {Di Folco}, {Dutrey},
  {Simon}, {Grosso}, \& {Pi{\'e}tu}}]{cite_guilloteauetal2013}
{Guilloteau}, S., {Di Folco}, E., {Dutrey}, A., {et~al.} 2013, \aap, 549, A92

\bibitem[{{Guzm{\'a}n} {et~al.}(2018){Guzm{\'a}n}, {{\"O}berg}, {Carpenter},
  {Le Gal}, {Qi}, \& {Pagues}}]{cite_guzmanetal2018}
{Guzm{\'a}n}, V.~V., {{\"O}berg}, K.~I., {Carpenter}, J., {et~al.} 2018, \apj,
  864, 170

\bibitem[{{Henning} \& {Semenov}(2013)}]{cite_henningetal2013}
{Henning}, T., \& {Semenov}, D. 2013, Chemical Reviews, 113, 9016

\bibitem[{{Herbst} \& {van Dishoeck}(2009)}]{cite_dishoecketal2009}
{Herbst}, E., \& {van Dishoeck}, E.~F. 2009, \araa, 47, 427

\bibitem[{{Hidaka} {et~al.}(2004){Hidaka}, {Watanabe}, {Shiraki}, {Nagaoka}, \&
  {Kouchi}}]{cite_hidakaetal2004}
{Hidaka}, H., {Watanabe}, N., {Shiraki}, T., {Nagaoka}, A., \& {Kouchi}, A.
  2004, \apj, 614, 1124

\bibitem[{Hiraoka {et~al.}(1994)Hiraoka, Ohashi, Kihara, Yamamoto, Sato, \&
  Yamashita}]{cite_hiraokaetal1994}
Hiraoka, K., Ohashi, N., Kihara, Y., {et~al.} 1994, Chemical Physics Letters,
  229, 408 .
\newblock
  \url{http://www.sciencedirect.com/science/article/pii/0009261494010668}

\bibitem[{{Hiraoka} {et~al.}(2002){Hiraoka}, {Sato}, {Sato}, {Sogoshi},
  {Yokoyama}, {Takashima}, \& {Kitagawa}}]{cite_hiraokaetal2002}
{Hiraoka}, K., {Sato}, T., {Sato}, S., {et~al.} 2002, \apj, 577, 265

\bibitem[{{Huang} {et~al.}(2017){Huang}, {{\"O}berg}, {Qi}, {Aikawa},
  {Andrews}, {Furuya}, {Guzm{\'a}n}, {Loomis}, {van Dishoeck}, \&
  {Wilner}}]{cite_huangetal2017}
{Huang}, J., {{\"O}berg}, K.~I., {Qi}, C., {et~al.} 2017, \apj, 835, 231

\bibitem[{Hunter(2007)}]{cite_matplotlib}
Hunter, J.~D. 2007, Computing In Science \& Engineering, 9, 90

\bibitem[{{Jensen} \& {Mathieu}(1997)}]{cite_jensenetal1997}
{Jensen}, E. L.~N., \& {Mathieu}, R.~D. 1997, \aj, 114, 301

\bibitem[{Jones {et~al.}(2001--)Jones, Oliphant, Peterson,
  {et~al.}}]{cite_scipy}
Jones, E., Oliphant, T., Peterson, P., {et~al.} 2001--, {SciPy}: Open source
  scientific tools for {Python}, , .
\newblock \url{http://www.scipy.org/}

\bibitem[{{Kastner} {et~al.}(2018){Kastner}, {Qi}, {Dickson-Vandervelde},
  {Hily-Blant}, {Forveille}, {Andrews}, {Gorti}, {{\"O}berg}, \&
  {Wilner}}]{cite_kastneretal2018}
{Kastner}, J.~H., {Qi}, C., {Dickson-Vandervelde}, D.~A., {et~al.} 2018, \apj,
  863, 106

\bibitem[{{Lee} {et~al.}(2019){Lee}, {Lee}, {Baek}, {Aikawa}, {Cieza}, {Yoon},
  {Herczeg}, {Johnstone}, \& {Casassus}}]{cite_leeetal2019}
{Lee}, J.-E., {Lee}, S., {Baek}, G., {et~al.} 2019, Nature Astronomy, 3, 314

\bibitem[{{Loomis} {et~al.}(2018){Loomis}, {Cleeves}, {{\"O}berg}, {Aikawa},
  {Bergner}, {Furuya}, {Guzman}, \& {Walsh}}]{cite_loomisetal2018}
{Loomis}, R.~A., {Cleeves}, L.~I., {{\"O}berg}, K.~I., {et~al.} 2018, ArXiv
  e-prints, arXiv:1805.01458

\bibitem[{{Loomis} {et~al.}(2015){Loomis}, {Cleeves}, {{\"O}berg}, {Guzman}, \&
  {Andrews}}]{cite_loomisetal2015}
{Loomis}, R.~A., {Cleeves}, L.~I., {{\"O}berg}, K.~I., {Guzman}, V.~V., \&
  {Andrews}, S.~M. 2015, \apjl, 809, L25

\bibitem[{{Marsh} \& {Mahoney}(1992)}]{cite_marshetal1992}
{Marsh}, K.~A., \& {Mahoney}, M.~J. 1992, \apj, 395, L115

\bibitem[{{Mathews} {et~al.}(2012){Mathews}, {Williams}, {M{\'e}nard},
  {Phillips}, {Duch{\^e}ne}, \& {Pinte}}]{cite_mathewsetal2012}
{Mathews}, G.~S., {Williams}, J.~P., {M{\'e}nard}, F., {et~al.} 2012, \apj,
  745, 23

\bibitem[{{Mathews} {et~al.}(2013){Mathews}, {Klaassen}, {Juh{\'a}sz},
  {Harsono}, {Chapillon}, {van Dishoeck}, {Espada}, {de Gregorio-Monsalvo},
  {Hales}, {Hogerheijde}, {Mottram}, {Rawlings}, {Takahashi}, \&
  {Testi}}]{cite_mathewsetal2013}
{Mathews}, G.~S., {Klaassen}, P.~D., {Juh{\'a}sz}, A., {et~al.} 2013, \aap,
  557, A132

\bibitem[{{Mordasini} {et~al.}(2008){Mordasini}, {Alibert}, {Benz}, \&
  {Naef}}]{cite_mordasinietal2008}
{Mordasini}, C., {Alibert}, Y., {Benz}, W., \& {Naef}, D. 2008, in Astronomical
  Society of the Pacific Conference Series, Vol. 398, Extreme Solar Systems,
  ed. D.~{Fischer}, F.~A. {Rasio}, S.~E. {Thorsett}, \& A.~{Wolszczan}, 235

\bibitem[{{M{\"u}ller} {et~al.}(2005){M{\"u}ller}, {Schl{\"o}der}, {Stutzki},
  \& {Winnewisser}}]{cite_cdms}
{M{\"u}ller}, H.~S.~P., {Schl{\"o}der}, F., {Stutzki}, J., \& {Winnewisser}, G.
  2005, Journal of Molecular Structure, 742, 215

\bibitem[{{Mumma} \& {Charnley}(2011)}]{cite_mummaetal2011}
{Mumma}, M.~J., \& {Charnley}, S.~B. 2011, \araa, 49, 471

\bibitem[{{Noble} {et~al.}(2012){Noble}, {Theule}, {Mispelaer}, {Duvernay},
  {Danger}, {Congiu}, {Dulieu}, \& {Chiavassa}}]{cite_nobleetal2012}
{Noble}, J.~A., {Theule}, P., {Mispelaer}, F., {et~al.} 2012, \aap, 543, A5

\bibitem[{{{\"O}berg} \& {Bergin}(2016)}]{cite_obergetal2016}
{{\"O}berg}, K.~I., \& {Bergin}, E.~A. 2016, \apjl, 831, L19

\bibitem[{{{\"O}berg} {et~al.}(2015{\natexlab{a}}){{\"O}berg}, {Furuya},
  {Loomis}, {Aikawa}, {Andrews}, {Qi}, {van Dishoeck}, \&
  {Wilner}}]{cite_obergetal2015b}
{{\"O}berg}, K.~I., {Furuya}, K., {Loomis}, R., {et~al.} 2015{\natexlab{a}},
  \apj, 810, 112

\bibitem[{{{\"O}berg} {et~al.}(2015{\natexlab{b}}){{\"O}berg}, {Guzm{\'a}n},
  {Furuya}, {Qi}, {Aikawa}, {Andrews}, {Loomis}, \&
  {Wilner}}]{cite_obergetal2015}
{{\"O}berg}, K.~I., {Guzm{\'a}n}, V.~V., {Furuya}, K., {et~al.}
  2015{\natexlab{b}}, \nat, 520, 198

\bibitem[{{{\"O}berg} {et~al.}(2010){{\"O}berg}, {Qi}, {Fogel}, {Bergin},
  {Andrews}, {Espaillat}, {van Kempen}, {Wilner}, \&
  {Pascucci}}]{cite_obergetal2010}
{{\"O}berg}, K.~I., {Qi}, C., {Fogel}, J.~K.~J., {et~al.} 2010, \apj, 720, 480

\bibitem[{{{\"O}berg} {et~al.}(2011){{\"O}berg}, {Qi}, {Fogel}, {Bergin},
  {Andrews}, {Espaillat}, {Wilner}, {Pascucci}, \&
  {Kastner}}]{cite_obergetal2011a}
---. 2011, \apj, 734, 98

\bibitem[{{{\"O}berg} {et~al.}(2017){{\"O}berg}, {Guzm{\'a}n}, {Merchantz},
  {Qi}, {Andrews}, {Cleeves}, {Huang}, {Loomis}, {Wilner}, {Brinch}, \&
  {Hogerheijde}}]{cite_obergetal2017}
{{\"O}berg}, K.~I., {Guzm{\'a}n}, V.~V., {Merchantz}, C.~J., {et~al.} 2017,
  \apj, 839, 43

\bibitem[{Oliphant(2006--)}]{cite_numpy}
Oliphant, T. 2006--, {NumPy}: A guide to {NumPy}, USA: Trelgol Publishing, , .
\newblock \url{http://www.numpy.org/}

\bibitem[{{Pinte} {et~al.}(2018){Pinte}, {M{\'e}nard}, {Duch{\^e}ne}, {Hill},
  {Dent}, {Woitke}, {Maret}, {van der Plas}, {Hales}, {Kamp}, {Thi}, {de
  Gregorio-Monsalvo}, {Rab}, {Quanz}, {Avenhaus}, {Carmona}, \&
  {Casassus}}]{cite_pinteetal2018}
{Pinte}, C., {M{\'e}nard}, F., {Duch{\^e}ne}, G., {et~al.} 2018, \aap, 609, A47

\bibitem[{{Podio} {et~al.}(2019){Podio}, {Bacciotti}, {Fedele}, {Favre},
  {Codella}, {Rygl}, {Kamp}, {Guidi}, {Bianchi}, {Ceccarelli}, {Coffey},
  {Garufi}, \& {Testi}}]{cite_podioetal2019}
{Podio}, L., {Bacciotti}, F., {Fedele}, D., {et~al.} 2019, \aap, 623, L6

\bibitem[{{Pontoppidan} {et~al.}(2004){Pontoppidan}, {van Dishoeck}, \&
  {Dartois}}]{cite_pontoppidanetal2004}
{Pontoppidan}, K.~M., {van Dishoeck}, E.~F., \& {Dartois}, E. 2004, \aap, 426,
  925

\bibitem[{{Qi} {et~al.}(2015){Qi}, {{\"O}berg}, {Andrews}, {Wilner}, {Bergin},
  {Hughes}, {Hogherheijde}, \& {D'Alessio}}]{cite_qietal2015}
{Qi}, C., {{\"O}berg}, K.~I., {Andrews}, S.~M., {et~al.} 2015, \apj, 813, 128

\bibitem[{{Qi} {et~al.}(2013){Qi}, {{\"O}berg}, \& {Wilner}}]{cite_qietal2013}
{Qi}, C., {{\"O}berg}, K.~I., \& {Wilner}, D.~J. 2013, \apj, 765, 34

\bibitem[{{Qi} {et~al.}(2019){Qi}, {{\"O}berg}, {Espaillat}, {Robinson},
  {Andrews}, {Wilner}, {Blake}, {Bergin}, \& {Cleeves}}]{cite_qietal2019}
{Qi}, C., {{\"O}berg}, K.~I., {Espaillat}, C.~C., {et~al.} 2019, \apj, 882, 160

\bibitem[{{Quast} {et~al.}(2000){Quast}, {Torres}, {de La Reza}, {da Silva}, \&
  {Mayor}}]{cite_quastetal2000}
{Quast}, G.~R., {Torres}, C.~A.~O., {de La Reza}, R., {da Silva}, L., \&
  {Mayor}, M. 2000, in IAU Symposium, Vol. 200, IAU Symposium, 28

\bibitem[{{Rosenfeld} {et~al.}(2013){Rosenfeld}, {Andrews}, {Hughes}, {Wilner},
  \& {Qi}}]{cite_rosenfeldetal2013a}
{Rosenfeld}, K.~A., {Andrews}, S.~M., {Hughes}, A.~M., {Wilner}, D.~J., \&
  {Qi}, C. 2013, \apj, 774, 16

\bibitem[{{Rosenfeld} {et~al.}(2012){Rosenfeld}, {Andrews}, {Wilner}, \&
  {Stempels}}]{cite_rosenfeldetal2012}
{Rosenfeld}, K.~A., {Andrews}, S.~M., {Wilner}, D.~J., \& {Stempels}, H.~C.
  2012, \apj, 759, 119

\bibitem[{{Salinas} {et~al.}(2017){Salinas}, {Hogerheijde}, {Mathews},
  {{\"O}berg}, {Qi}, {Williams}, \& {Wilner}}]{cite_salinasetal2017}
{Salinas}, V.~N., {Hogerheijde}, M.~R., {Mathews}, G.~S., {et~al.} 2017, \aap,
  606, A125

\bibitem[{{Schwarz} {et~al.}(2016){Schwarz}, {Bergin}, {Cleeves}, {Blake},
  {Zhang}, {{\"O}berg}, {van Dishoeck}, \& {Qi}}]{cite_schwarzetal2016}
{Schwarz}, K.~R., {Bergin}, E.~A., {Cleeves}, L.~I., {et~al.} 2016, \apj, 823,
  91

\bibitem[{{Thi} {et~al.}(2004){Thi}, {van Zadelhoff}, \& {van
  Dishoeck}}]{cite_thietal2004}
{Thi}, W.-F., {van Zadelhoff}, G.-J., \& {van Dishoeck}, E.~F. 2004, \aap, 425,
  955

\bibitem[{{van der Marel} {et~al.}(2015){van der Marel}, {van Dishoeck},
  {Bruderer}, {P{\'e}rez}, \& {Isella}}]{cite_mareletal2015}
{van der Marel}, N., {van Dishoeck}, E.~F., {Bruderer}, S., {P{\'e}rez}, L., \&
  {Isella}, A. 2015, \aap, 579, A106

\bibitem[{{van der Marel} {et~al.}(2014){van der Marel}, {van Dishoeck},
  {Bruderer}, \& {van Kempen}}]{cite_mareletal2014}
{van der Marel}, N., {van Dishoeck}, E.~F., {Bruderer}, S., \& {van Kempen},
  T.~A. 2014, \aap, 563, A113

\bibitem[{{van 't Hoff} {et~al.}(2018){van 't Hoff}, {Tobin}, {Trapman},
  {Harsono}, {Sheehan}, {Fischer}, {Megeath}, \& {van
  Dishoeck}}]{cite_hoffetal2018}
{van 't Hoff}, M. L.~R., {Tobin}, J.~J., {Trapman}, L., {et~al.} 2018, \apjl,
  864, L23

\bibitem[{{Walsh} {et~al.}(2010){Walsh}, {Millar}, \&
  {Nomura}}]{cite_walshetal2010}
{Walsh}, C., {Millar}, T.~J., \& {Nomura}, H. 2010, \apj, 722, 1607

\bibitem[{{Walsh} {et~al.}(2014){Walsh}, {Millar}, {Nomura}, {Herbst}, {Widicus
  Weaver}, {Aikawa}, {Laas}, \& {Vasyunin}}]{cite_walshetal2014}
{Walsh}, C., {Millar}, T.~J., {Nomura}, H., {et~al.} 2014, \aap, 563, A33

\bibitem[{{Walsh} {et~al.}(2016){Walsh}, {Loomis}, {{\"O}berg}, {Kama}, {van 't
  Hoff}, {Millar}, {Aikawa}, {Herbst}, {Widicus Weaver}, \&
  {Nomura}}]{cite_walshetal2016}
{Walsh}, C., {Loomis}, R.~A., {{\"O}berg}, K.~I., {et~al.} 2016, \apjl, 823,
  L10

\bibitem[{{Watanabe} \& {Kouchi}(2002)}]{cite_watanabeetal2002}
{Watanabe}, N., \& {Kouchi}, A. 2002, \apjl, 571, L173

\bibitem[{{Watanabe} {et~al.}(2004){Watanabe}, {Nagaoka}, {Shiraki}, \&
  {Kouchi}}]{cite_watanabeetal2004}
{Watanabe}, N., {Nagaoka}, A., {Shiraki}, T., \& {Kouchi}, A. 2004, \apj, 616,
  638

\bibitem[{{Yen} {et~al.}(2016){Yen}, {Koch}, {Liu}, {Puspitaningrum}, {Hirano},
  {Lee}, \& {Takakuwa}}]{cite_yenetal2016}
{Yen}, H.-W., {Koch}, P.~M., {Liu}, H.~B., {et~al.} 2016, \apj, 832, 204

\bibitem[{{Zhang} {et~al.}(2014){Zhang}, {Isella}, {Carpenter}, \&
  {Blake}}]{cite_zhangetal2014}
{Zhang}, K., {Isella}, A., {Carpenter}, J.~M., \& {Blake}, G.~A. 2014, \apj,
  791, 42

\end{thebibliography}

\end{document}